\documentclass{statsoc}
\pdfoutput=1 

\usepackage[a4paper]{geometry}
\usepackage{graphicx}
\usepackage[textwidth=8em,textsize=small]{todonotes}\usepackage{amsmath}
\usepackage{amssymb}
\usepackage{bm,bbm}
\usepackage{natbib}
\usepackage{enumitem}
\allowdisplaybreaks

\newtheorem{proposition}{Proposition}

\newtheorem{lemma}{Lemma}
\newtheorem{property}{Property}
\newtheorem{corollary}{Corollary}

\title[Post-cAIC inference]
{Post-selection inference for linear mixed model parameters using the conditional Akaike information criterion}
\author{Gerda Claeskens}
\address{KU Leuven, Belgium.}
\author{Katarzyna Reluga}
\address{University of Toronto, Canada.}
\email{katarzyna.reluga@utoronto.ca}
\author[Claeskens et al.]{Stefan Sperlich}
\address{University of Geneva,
	Switzerland.}

\begin{document}
	\begin{abstract}
		We investigate the issue of post-selection inference for a fixed and a mixed parameter in a linear mixed model using a conditional Akaike information criterion as a model selection procedure. Within the framework of linear mixed models we develop complete theory to construct confidence intervals for regression and mixed parameters under three frameworks: nested and general model sets as well as misspecified models. Our theoretical analysis is accompanied by a simulation experiment and a post-selection examination on mean income across Galicia’s counties. Our numerical studies confirm a good performance of our new procedure. Moreover, they reveal a startling robustness to the model misspecification of a naive method to construct the confidence intervals for a mixed parameter which is in contrast to our findings for the fixed parameters.
	\end{abstract}


\section{Introduction}\label{sec:Introduction}

Model or variable selection appears to be a routine practice in a great majority of statistical and machine learning data analyses. Despite the additional randomness coming from data-driven model selection procedures, both specialists and practitioners tend to disregard it in the subsequent steps of statistical inference. Instead, they often use classical theory to construct confidence intervals and testing procedure even though such theory might be invalid in their context. Many authors have stressed the need to account for selection uncertainty in the context of a classical regression \citep[e.g.,][and in a more recent surge of articles \citeauthor{berk2013valid}, \citeyear{berk2013valid};
\citeauthor{ferrari2015confidence}, \citeyear{ferrari2015confidence}; \citeauthor{charkhi2018asymptotic}, \citeyear{charkhi2018asymptotic}; \citeauthor{bachoc2019valid}, \citeyear{bachoc2019valid}]{hjort2003frequentist,leeb2003finite}. Moreover, the topic has been thoroughly discussed by, among others,  \cite{belloni2015uniform, lee2016exact, tibshirani2016exact} in the field of selective inference in which both the choice of the model and the target parameter are data-driven.

Despite the interest in the post-selection inference in many statistical domains, it remains a largely neglected problem in the field of linear mixed models (LMMs). The latter have been thoroughly studied, and are broadly applied for modelling clustered or longitudinal data \citep{verbekeMolenberghs2000,jiang2007linear} in, among others, ecology \citep{bolker2009generalized}, small area estimation \citep{rao2015small,morales2021sae} or medicine \citep{FrancqEtal2019}. Recently \cite{sugasawa2019observed} developed the first contribution towards the post-selection inference under LMM, and investigated a procedure based on a prediction error criterion for the area-level model of \cite{fay_herriot}. Their proposal involves a simultaneous estimation and a model selection of a mixed parameter, consisting of both fixed and random effects, and it is tailored to minimize the estimated mean squared error. For comparison with our method, we implement the observed best selective predictor (OBSP) of \cite{sugasawa2019observed} and study its performance in simulations in Section \ref{sec:simulations}. The issue of accounting for the model selection in the context of LMM has been also mentioned by \cite{cunen2020}, but the authors did not approach it in their article.

Our goal is thus to investigate post-selection inference under a classical, low-dimensional framework for fixed (regression) parameters and their linear combinations as well as a general mixed parameter. A precise estimation of the former is indispensable in any statistical analysis whereas the latter is essential in, among others, small area estimation (SAE). In particular, we address the construction of valid post-selection confidence intervals. Due to its practical relevance, we concentrate on the inference after the selection of covariates for fixed effects when random effects are present and the variance structure is not subject to the selection process. The selection of random effects involves a different strand of literature and methods. \cite{charkhi2018asymptotic} studied the asymptotic distribution of estimators after model selection using Akaike's information criterion (AIC), proposed by \cite{akaike1973information}, and applied this distribution to construct adjusted confidence intervals for fixed effects and linear combinations of them. Even though their general approach is suitable for any likelihood-based model, \cite{charkhi2018asymptotic} did not consider random effects. Within the mixed model setting, we can differentiate population and cluster foci, a distinction made already by \cite{harville1977maximum}. In LMMs, a classical AIC has a population focus and is obtained by integrating out random effects and using a marginal log-likelihood. Hence, it is often referred to as a marginal AIC (mAIC). Due to its main target, mAIC is not appropriate for the prediction of cluster-level parameters or mixed effects. We therefore use a criterion which selects covariates in terms of minimising the prediction errors with the focus on specific random effects. Considering this aspect, \cite{vaida2005conditional} proposed a conditional AIC under the assumption of known variance parameters. Since covariance matrices are usually unknown and need to be estimated, the assumption of \cite{vaida2005conditional} and later of \cite{liang2008} seems to be too stringent for a practical use. Therefore, in what follows we use a conditional AIC (cAIC) of \cite{kubokawa2011conditional} who extended the proposal of \cite{vaida2005conditional} and accounted for the estimation of the variability parameters. Our examination on the inference after cAIC-selection (henceforth we refer to it as post-cAIC inference) can thus be treated as a twofold extension of the theory of \cite{charkhi2018asymptotic}. First, we consider a different model selector; second, and more importantly, we focus not only on fixed effects, but also on mixed parameters consisting of both fixed and random effects. 

After the proposal of \cite{vaida2005conditional}, scholars have developed several extensions to the initial information criterion with a cluster focus \citep[see][for an extensive review of model selection techniques under LMM]{muller2013model}. Apart from cAIC of \cite{kubokawa2011conditional}, \cite{srivastava2010conditional} defined an alternative conditional Akaike information and investigated its unbiased estimator, which resulted in a modified cAIC. Furthermore, \cite{kawakubo2014modified} introduced a criterion which is appropriate to cover underspecified cases, that is, when the list of models does not include the true one \citep[in Section \ref{sec:misspec} we adopt the terminology of][and call it a misspecified setting]{charkhi2018asymptotic}. Another modification is to use information criteria with generalised degrees of freedom (GDF) combined with a marginal or a conditional likelihood as in \cite{greven2010behaviour} and \cite{you2016generalized}. On the other hand, \cite{lombardia2017mixed} employed GDF with the quasi-log-likelihood which focuses on random effects and the total variability, combining a conditional and a marginal log-likelihood. 
Hereinafter we consider only post-cAIC inference; the comparative study of the post-selection inference using different methods with a cluster focus might be a subject of possible future research.

For the sake of comparison, we use the framework and a similar notation of \cite{charkhi2018asymptotic} unless it is in conflict with ours.  In Section \ref{sec:inference_LMM} we present key concepts of LMM inference. Then we investigate three settings to construct post selection confidence intervals. We initialise with the set of nested models in Section \ref{sec:sec_nested} and then move towards any set of models in Section \ref{sec:sec_overlap}. In Section \ref{sec:misspec} we consider a post-selection inference for a set of misspecified models. In Section \ref{sec:simulations}, we outline the outcomes of the numerical study, whereas in Section \ref{sec:data_example} we apply post-cAIC inference in a study on mean income in the counties of Galicia. We conclude with a discussion in Section \ref{sec:ch4_discussion} while deferring certain technical details to Section \ref{sec:ch4_App} and the supplementary material (SM) in Section \ref{sec:sup_mat}.

\section{Inference in linear mixed models}\label{sec:inference_LMM}

We examine the inference under {individual cluster} LMM, {i.e., each observation belongs to one cluster, and clusters are independent.} To facilitate the exposition, we provide the definition for a full model with all possible fixed parameters included, that is
\begin{equation}\label{eq:LMM_vec}
	\bm{y}_{i}=\bm{X}_{i}\bm{\beta}+\bm{Z}_{i}\bm{u}_{i}+\bm{e}_{i},\quad i=1,\dots,n,
\end{equation}
where $\bm{y}_{i}\in \mathbbm{R}^{m_{i}}$ is a vector of target variables, $\bm{X}_{i} \in \mathbbm{R}^{m_i\times (a+K)}$ and $\bm{Z}_{i}\in\mathbbm{R}^{m_i\times q}$ are matrices of covariates, $\bm{\beta}\in\mathbbm{R}^{a+K}$ is a vector of fixed effects, $\bm{u}_{i}\in\mathbbm{R}^{q}$ is a vector of random effects and $\bm{\varrho}^t=(\bm{\beta}^t,\bm{u}^t)$, whereas $\bm{e}_{i}\in\mathbbm{R}^{m_i}$ is a vector of stochastic errors. Splitting the dimension of the covariates to the sum of $a+K$ is convenient for presenting our post-selection analysis, and is clarified in Section \ref{sec:sec_nested}. We assume that $\bm{u}_{i}\sim N_{q}{(\bm{0},\bm{G}_q( \bm{\theta} ))}$, $\bm{e}_{i}\sim N_{m_i}{(\bm{0},\bm{R}(\bm{\theta}) )}$ where $\bm{\theta}=(\theta_1, \dots, \theta_h)^t$ is an h-dimensional vector of variance parameters. Furthermore, we suppose that $\bm{G}=\bm{G}(\bm{\theta})$ and $\bm{R}=\bm{R}(\bm{\theta})$ are positive definite matrices known up to the vector $\bm{\theta}$. We denote the total number of clusters by $n$ and the total number of units by $m=\sum_{i=1}^n m_i$. Expression \eqref{eq:LMM_vec} can be rewritten in a succinct form
\begin{equation}\label{eq:LMM2}
	\bm{y}=\bm{X\beta}+\bm{Z u}+\bm{e},
\end{equation}
where $\bm{X}=(\bm{X}^t_1, \dots, \bm{X}^t_{n})^t$ is an ${m}\times(a+K)$ matrix of rank $a+K$, $\bm{Z}=\mathrm{diag}(Z_1, \dots, Z_{n})$ is an $m\times r$ matrix of rank $r=nq$, $\bm{u}=(\bm{u}^t_1,\dots,\bm{u}^t_n)^t$, $\bm{e}=(\bm{e}^t_1,\dots,\bm{e}^t_{n})^t$ and $\bm{G}=\mathrm{diag}_{n}(\bm{G}_q)$ is block-diagonal with $n$ blocks $\bm{G}_q$ on the diagonal. The marginal and conditional distributions of $\bm{y}$ are $\bm{y}\sim N_{m}(\bm{X\beta}, \bm{V}(\bm{\theta}))$ and $\bm{y}|\bm{u}\sim N_{m}(\bm{X\beta}+\bm{Zu}, \bm{R}(\bm{\theta}))$, respectively, where $\bm{V}=\bm{V}(\bm{\theta})=\bm{R}(\bm{\theta})+\bm{ZG}(\bm{\theta})\bm{Z}^t$. Twice negative marginal log-likelihood and extended log-likelihood functions for $\bm{y}$ modelled by equation \eqref{eq:LMM2} are:
\begin{eqnarray}
-2\ell_n^m&=&-2\ell_n^m(\bm{y}|\bm{\beta},\bm{\theta})=-2\log f(\bm{y})
\label{eq:marginal_loglik}\\
&=&m\log(2\pi)+\log|\bm{V}|+(\bm{y}-\bm{X\beta})^t\bm{V}^{-1}(\bm{y}-\bm{X\beta}),\nonumber\\
-2\ell_n^e&=&-2\ell_n^c(\bm{y}|\bm{u},\bm{\beta},\bm{\theta})-2\ell_r^u(\bm{u}|\bm{\theta})
=-2\log f(\bm{y}|\bm{u})-2\log f(\bm{u})\label{eq:joint_loglik2}\\
&=&m\log(2\pi)+\log|\bm{R}|+(\bm{y}-\bm{X\beta}-\bm{Z u})^t\bm{R}^{-1}(\bm{y}-\bm{X\beta}-\bm{Z u})\nonumber\\
&&+r\log(2\pi)+\log|\bm{G}|+\bm{u}^t\bm{G}^{-1}\bm{u}.\nonumber
\end{eqnarray}
Numerous methods have been established to estimate $\bm{\beta}$, $\bm{u}$ and $\bm{\theta}$. By far the most popular are two-stage techniques such as the best linear unbiased estimator and the best linear unbiased predictor (BLUP), mixed-model equations of \cite{henderson1950}, the Bayes estimation or the likelihood based inference \citep[see, for example, ][for all essential procedures]{verbekeMolenberghs2000,jiang2007linear}. In what follows we concentrate on the former. Regarding $\bm{\theta}$, we estimate it iteratively by maximizing \eqref{eq:marginal_loglik} or by using the restricted log-likelihood version. Both of them were discussed by \cite{laird1982random}. In addition, one can take a derivative of \eqref{eq:marginal_loglik} with respect to $\bm{\beta}$ to obtain $\tilde{\bm{\beta}}=\tilde{\bm{\beta}}(\bm{\theta})=(\bm{X}^t\bm{V}^{-1}\bm{X})^{-1}\bm{X}^t\bm{V}^{-1}\bm{y}$. Once $\bm{\theta}$ is estimated, we plug it into $\tilde{\bm{\beta}}$ to obtain $\hat{\bm{\beta}}=\hat{\bm{\beta}}(\hat{\bm{\theta}})$. An alternative analysis is required  if we tend to focus on the inferences with respect to random effects $\bm{u}$. \cite{henderson1950} used the extended likelihood in \eqref{eq:joint_loglik2}
to obtain the estimates of $\bm{\beta}$ and predictions of $\bm{u}$:
\begin{equation}\label{eq:mom_eq_mat}
		\begin{pmatrix}
			\tilde{\bm{\beta}}\\
			\tilde{\bm{u}}
		\end{pmatrix}
		=
		\begin{pmatrix}
			\bm{X}^t\bm{R}^{-1}\bm{X}    & \bm{X}^t\bm{R}^{-1}\bm{Z}  \\
			\bm{Z}^t\bm{R}^{-1}\bm{X}    & \bm{Z}^t\bm{R}^{-1}\bm{Z} + \bm{G}^{-1}  \\
		\end{pmatrix}^{-1}
		\begin{pmatrix}
			\bm{X}^t\bm{R}^{-1}\bm{y}\\
			\bm{Z}^T\bm{R}^{-1}\bm{y}
	\end{pmatrix},
\end{equation}
which results in the same expression for $\tilde{\bm{\beta}}$ as by using the marginal likelihood in \eqref{eq:marginal_loglik}. The mixed-model equations in \eqref{eq:mom_eq_mat} can be used to obtain a pseudo hat-matrix $\bm{H}$
\begin{equation}\label{eq:Hat_matrix}
	\bm{H}	=
	\begin{pmatrix}
		\bm{X}& \bm{Z}
	\end{pmatrix}
	\begin{pmatrix}
		\bm{X}^t\bm{X}    & \bm{X}^t\bm{Z}  \\
		\bm{Z}^t\bm{X}    & \bm{Z}^t\bm{Z} + \bm{G}^{-1}  \\
	\end{pmatrix}^{-1}
	\begin{pmatrix}
		\bm{X}^t\\
		\bm{Z}^T
	\end{pmatrix}
	\text{ and } \rho=\mathrm{\rm{tr}}(\bm{H}),
\end{equation}
where $\tilde{\bm{y}}=\bm{X}\tilde{\bm{\beta}}+\bm{Z}\tilde{\bm{u}}=\bm{H}\bm{y}$. We call $\rho$ the effective degrees of freedom \citep{hodges2001counting}. They are used as the main part of the penalty term in cAIC. It follows that $a+K\leqslant \rho \leqslant a+K+r$  \citep{vaida2005conditional} and we have $\hat{\rho}=\rho(\hat{\bm{\theta}})$. 

We focus on post-cAIC inference for (i) a fixed parameter in \eqref{eq:LMM_vec}, (ii) a linear combination $\bm{k}_i^t\bm{\beta}$, $\bm{k}_{i}\in \mathbbm{R}^{a+K}$ and (iii) a general mixed parameter
\begin{equation}\label{eq:mu}
\mu_{i}=\bm{k}^t_{i}\bm{\beta}+\bm{m}^t_{i}\bm{u}_{i},\quad\tilde{\mu}_{i}=\mu_{i}(\bm{\theta})=\bm{k}^t_{i}\tilde{\bm{\beta}}+\bm{m}^t_{i}\tilde{\bm{u}}_{i}\quad\text{and}\quad  \hat{\mu}_{i}=\mu_{i}(\hat{\bm{\theta}}),\quad i=1,\dots,n,
\end{equation}
where $\bm{m}_{i}\in \mathbbm{R}^{q}$, 
$\bm{c}_{i}=(\bm{k}_{i}^t,\bm{m}_{i}^t)^t$ and $\hat{\mu}_i$ is the EBLUP of $\mu_i$. 
The variability of regression parameters can be derived directly from the marginal log-likelihood in \eqref{eq:marginal_loglik}. Regarding the mixed effect, \cite{henderson} employed \eqref{eq:mom_eq_mat} to obtain a formula for the variance of $\tilde{\mu}_{i}$ in \eqref{eq:mu}. Due to the presence of a random effect, this variance is often referred to as mean squared error (MSE). We thus have $\mathrm{MSE}\{\tilde{\mu}_{i}(\bm{\theta})\}=\bm{c}^t_{i}\bm{K}^{-1} \bm{c}_{i}$, where $\bm{K}$ and $\bm{K}^{-1}$ are spelled out in the SM. Replacing $\bm{\theta}$ in $\mathrm{MSE}\{\tilde{\mu}_{i}(\bm{\theta})\}$ with $\hat{\bm{\theta}}$ results in an estimator
\begin{equation}\label{eq:naive_mse}
	\mathrm{mse}_{1}(\hat{\mu}_{i})=\bm{c}^t_{i}\hat{\bm{K}}^{-1} \bm{c}_{i}=g_{1i}(\hat{\bm{\theta}})+g_{2i}(\hat{\bm{\theta}}),
\end{equation}
which is called the first-order correct MSE estimator in the SAE literature. 
On the contrary, an analytical second-order correct estimator is given by \begin{equation}\label{eq:MSE_lin}
	\mathrm{mse}_{2}(\hat{\mu}_{i} )= g_{1i}(\hat{\bm{{\theta}}}) + g_{2i}(\hat{\bm{\theta}}) + 2 g_{3i}(\hat{\bm{\theta}}).
\end{equation}
The exact expressions for $g_{1i}$, $g_{2i}$ and $g_{3i}$ can be found in, for example, \cite{rao2015small} and our SM. We use $\mathrm{mse}_{2}$ in \eqref{eq:MSE_lin} to construct naive confidence intervals which do not account for the selection uncertainty. Finally, the cAIC of \cite{kubokawa2011conditional}
\begin{equation}\label{eq:c_AIC_Kubo}
	\mathrm{cAIC}=-2\ell^c_n(\hat{\bm{\beta}}_{m})+2\rho(\hat{\bm{\theta}})+2b(\hat{\bm{\theta}}),
\end{equation}
is an asymptotically unbiased estimator of the  conditional Akaike information (cAI)
$\mathrm{cAI}(\bm{\theta})=-2\int \int \int \log \{f(\bm{y}^*|\hat{\bm{u}}, \hat{\bm{\beta}},\hat{\bm{\theta}} ) \} f(\bm{y}^*|{\bm{u}}, {\bm{\beta}},{\bm{\theta}} ) f(\bm{y}|\bm{u}, \bm{\beta},\bm{\theta}) f(\bm{u}|\bm{\theta}) \mathrm{d}\bm{y}^*
\mathrm{d}\bm{y} \mathrm{d}\bm{u}$,
where $\bm{y}^*$ is a future variable distributed according the same normal distribution as $\bm{y}$, $\rho(\hat{\bm{\theta}})$ is an estimated version of the effective degrees of freedom in \eqref{eq:Hat_matrix}, $b(\hat{\bm{\theta}})$ is the additional penalty accounting for the estimation of variance parameter $\bm{\theta}$. Since the exact form of cAIC for a general LMM is complex, we defer it to our SM \citep[cf.][for 
the derivation]{kubokawa2011conditional}.

\section{Selection properties of the cAIC in nested models}\label{sec:sec_nested}

We investigate the nested sequence of $K+1$ likelihood models $M_0\subseteq M_1 \dots \subseteq M_K$ which depend on the parameter vector $\bm{\beta}=(\bm{\beta}^t_a,\beta_{a+1},\dots,\beta_{a+K})\in \mathbb{B} \subseteq \mathbbm{R}^{a+K}$. More specifically, model $M_0$ contains $a$ covariates, in model $M_1$ we employ $a+1$ covariates, etc. The largest model $M_K$ contains a full vector $\bm{\beta}\in \mathbbm{R}^{a+K}$. The parameter that is common to all models and thus not subject to the selection procedure is denoted by $\bm{\beta}_a\in \mathbbm{R}^a$. Without loss of generality, we assume that $M_{i}$ adds one covariate to $M_{i-1}$. Furthermore, there exists a single minimal true model $M_{p_0}$ in the set of general models $\mathcal{M}_{nest}=\{M_i:i=0,\dots, K \}$, that is, $p_0$ is the smallest model order for which all non-zero components of the true vector $\bm{\beta}_0$ are included. Models with $i<p_0$ are underparametrised and with $i>p_0$ overparametrised. In addition, $\bm{v}^s(k)=(v_1,\dots,v_{a+k})^t$ denotes a subvector of $\bm{v}$ which corresponds to model $M_k$. Furthermore, in model $M_i$ we define  $\bm{\beta}^{s}(i)=(\bm{\beta}_{a}^t,\dots,\beta_i)\in\mathbbm{R}^{a+i}$, its counterparts $\hat{\bm{\beta}}^{s}_{m}(i)\in\mathbbm{R}^{a+i}$ and $\hat{\bm{\beta}}^{s}_{c}(i)\in\mathbbm{R}^{a+i}$ estimated using maximum marginal $\ell^m_n$ and conditional log-likelihoods $\ell^c_n$ as defined in equations \eqref{eq:marginal_loglik} and \eqref{eq:joint_loglik2} respectively. In addition, let $\hat{\bm{\beta}}_{m}(i)=[\{\hat{\bm{\beta}}^{s}_{m}(i)\}^t, \bm{0}^t_{K-i}]\in\mathbbm{R}^{a+K}$ and $\hat{\bm{\beta}}_{c}(i)=[\{\hat{\bm{\beta}}^{s}_{c}(i)\}^t, \bm{0}^t_{K-i}]\in\mathbbm{R}^{a+K}$. Last but not least, we can distinguish $\bm{\beta}_{0}=\bm{\beta}_{0}(p_0)$, which is the true value with $\beta_{0j}=0$ for $j>p_0$, whereas  $\bm{\beta}^s_{0}(p_0)$ is composed of non-zero elements of $\bm{\beta}_{0}(p_0)$. If no confusion is possible, we omit the dependence on $p_0$ in $\bm{\beta}_{0}(p_0)$.

The conditional Akaike information criterion for model $M_j$ in the set of models $\mathcal{M}_{nest}$ is formally given as $\mathrm{cAIC}(M_j)=-2\ell^c_n\{\hat{\bm{\beta}}_{m}(j)\}+2\hat{\rho}_j+2\hat{b}_{j}$, where $\ell^c_n$ is the conditional likelihood defined in equation \eqref{eq:joint_loglik2}, $\hat{\bm{\beta}}_{m}(j)$ is a vector of estimated covariates using the marginal likelihood, $\hat{\rho}_j$ and $\hat{b}_j$ are estimated penalty terms. 
The index of the selected model is $\hat{p}_0=\min\{j:\mathrm{cAIC}(M_j)=\min_{0\leqslant i\leqslant K} \mathrm{cAIC}(M_i)\}$. To continue with the post-selection inference 
we need to rewrite the cAIC-based selection procedure using a set of inequalities which impose geometrical restrictions on the support of the normally distributed random variables. First, we redefine $\hat{p}_0=\min\{j \in \{ 0,\dots,K\}: j=\arg \max_{0,\dots,K} \mathrm{cAIC}'(M_j)\}$, with
\begin{eqnarray*} \mathrm{cAIC}'(M_j) &=& 2\left[\ell^c_{n,j}\big\{\hat{\bm{\beta}}_{m}(j)\big\}-\ell_n^c(\bm{\beta}_0)\right]+2(\rho_0-\hat{\rho}_j)+2(b_0-\hat{b}_j)
\\&=& 2\ell'^{c}_{n,j}+2(\rho_0-\hat{\rho}_j)+2(b_0-\hat{b}_j),
\end{eqnarray*}
where $\rho_0+b_0-\hat{\rho}_j -\hat{b}_j$ can be treated as an effective difference between the degrees of freedom imposed on the true model and on the selected model. 
The probability of underselection using cAIC is asymptotically zero (see Lemma \ref{lemma:oversel} in Section \ref{sec:sec_overlap} and its proof in our SM), which implies that $p\geqslant p_0$. A similar result was demonstrated for AIC by \cite{woodroofe1982model} and generalised by \cite{charkhi2018asymptotic}. 
If we condition on $\hat{p}_0=p$,  $\mathrm{cAIC}'(M_p)-\mathrm{cAIC}'(M_j)>0$ for $j=p_0,\dots,p-1$ and $\mathrm{cAIC}'(M_p)-\mathrm{cAIC}'(M_j)\geqslant0$ for $j=p_0,\dots,K$. For a full model $M_K$, denote by $\mathcal{I}^m=\mathcal{I}^m(M_K)$ the Fisher information matrix in a marginal setting with all parameters, and by $\mathcal{J}^c=\mathcal{J}^c_{n}(M_K)$ the negative Hessian calculated from the conditional likelihood (their precise definitions are given in Section \ref{sec:assump}). 
Unless the model is correctly specified, we have $\mathcal{I}^m\neq\mathbb{E}(\mathcal{J}^m)$. Furthermore, we define $\bm{\Sigma}=\bm{\Sigma}(M_K)=(\mathcal{I}^m)^{-1/2}\mathcal{J}^c(\mathcal{I}^m)^{-1/2}$. Consequently, a submatrix of $\bm{\Sigma}$ which corresponds to model $M_i$ is denoted by $\bm{\Sigma}(M_i)$ and refers only to the covariates from the considered model $M_i$. 
Let the diagonal and the off-diagonal elements of $\bm{\Sigma}$ be $\Sigma_i$ and $\Sigma_{ij}$, $i,j=1, \dots, a+K$. Furthermore, let $\mathcal{I}^m(i)$ and $\bm{K}(i)$ be submatrices of $\mathcal{I}^m$ and $\bm{K}$, respectively,  corresponding to the model $M_i$. In addition, we denote by $id(\cdot)$ an indicator function. Let $\rho_{jp}=\rho_j-\rho_p$ and $b_{jp}=b_{j}-b_{p}$ and $\hat{\rho}_{jp}$ and $\hat{b}_{jp}$ their empirical versions. Consider the sequence of nested models $\mathcal{M}_{nest}$. It follows that the selection region for a fixed parameter $\bm{\beta}$ is defined as follows: 

$(a)$ for $p=p_0$ we have  $\mathcal{A}_{p}(\mathcal{M}_{nest})= $
\begin{eqnarray}\label{eq:nested_region_1}
\bigcap\limits_{j=p+1, \dots, K} \left\{\bm{w}\in \mathbbm{R}^{a+K}:
\sum_{i=p+1}^{j}w^2_{a+i}\Sigma_{a+i}+2\sum_{i=p+1}^{j}\sum_{k=1}^{j-1}w_{a+i}w_{k}\Sigma_{(a+i)k} < 2\rho_{jp}+2b_{jp} \right\},
\end{eqnarray}
$(b)$ for $p>p_0$ we have  $\mathcal{A}_{p}(\mathcal{M}_{nest})=\mathcal{B}_{1,p}\cap \mathcal{B}_{2,p}$ where
\begin{equation}\label{eq:nested_region_2}
	\begin{split}
		\mathcal{B}_{1,p}&=\bigcap\limits_{j=p_0+1, \dots, p}
		\left\{\bm{w}\in \mathbbm{R}^{a+K}: \sum_{i=j}^{p}w^2_{a+i}\Sigma_{a+i}+2\sum_{i=j}^{p}\sum_{k=1}^{p-1}w_{a+i}w_{k}\Sigma_{(a+i)k}\geqslant
		2\rho_{jp}+2b_{jp}  \right\}, \\
		\mathcal{B}_{2,p}&=\bigcap\limits_{j=p+1, \dots, K} \left\{\bm{w}\in \mathbbm{R}^{a+K}:\sum_{i=p+1}^{j}w^2_{a+i}\Sigma_{a+i}+2\sum_{i=p+1}^{j}\sum_{k=1}^{j-1}w_{a+i}w_{k}\Sigma_{(a+i)k} <  2\rho_{jp}+2b_{jp} \right\}.
	\end{split}
\end{equation}
In other words, $\mathcal{A}_{p}$ describes constraints on the domain of multidimensional normal random variables. The specific form of $\mathcal{A}_{p}^{\mu}$ is defined by the set of inequalities coming from the asymptotic analysis of $\{\hat{\bm{\varrho}}(M)-\bm{\varrho}_0\}$. Since the random effects are not subject to the selection procedure, the selection region for a mixed parameter $\mathcal{A}^{\mu}_{p}(\mathcal{M}_{nest})$ has almost the same form for $p=p_0$ and $p>p_0$. In fact, one would need to only replace $\mathbbm{R}^{a+K}$ with $\mathbbm{R}^{a+K+r}$ in \eqref{eq:nested_region_1} and \eqref{eq:nested_region_2}.

We illustrate the allowable domains for normal random variables  $W_1$, $W_2$ and $W_3$ using the restrictions imposed by the cAIC selection. The domains of the random effects are not affected by the geometrical restrictions. Consider $K=2$, $a=1$, $\mathcal{M}_{nest}$, and suppose that $M_0$ is a true model containing only $\bm{\beta}_1$, that is, $M_0=\{\bm{\beta}_1\}$. Moreover,  $M_1=\{\bm{\beta}_1,\bm{\beta}_2\}$ whereas $M_2=\{\bm{\beta}_1,\bm{\beta}_2,\bm{\beta}_3\}$. To be able to plot the domains, we need to fix or estimate the values of $\bm{\Sigma}$, $\rho_j$ and $b_j$, $j=1,2,3$. We thus constructed a simulated dataset using a simplified setting from Section \ref{sec:simulations}, the details can be found in our SM.
\begin{figure}
	\centering
	\includegraphics[width=\textwidth]{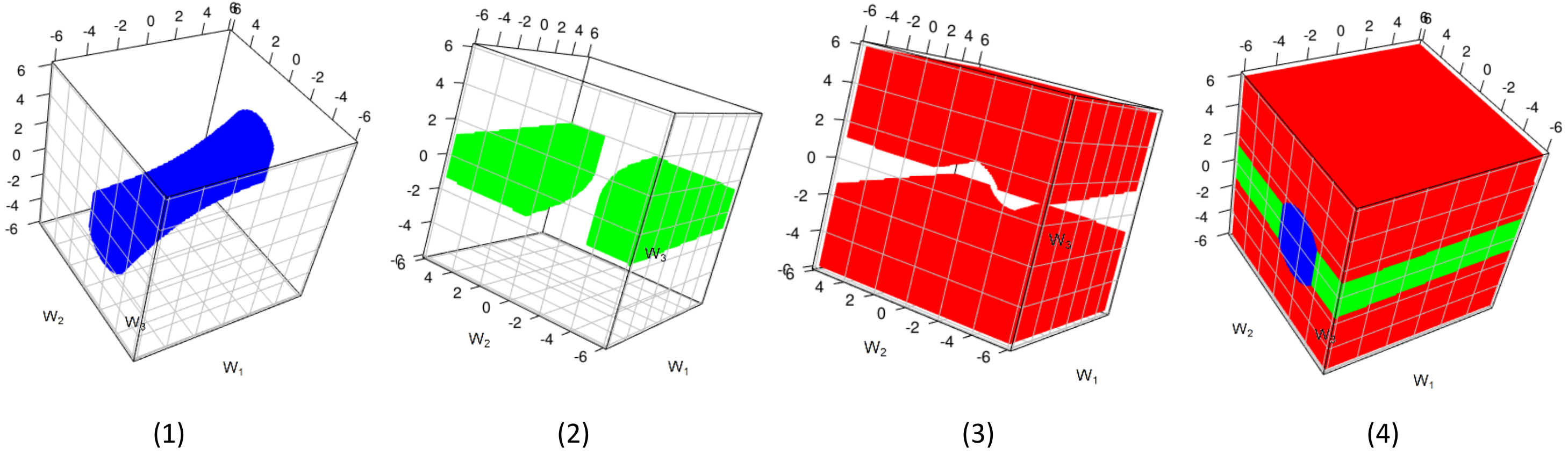}
	\vspace{-0.3cm}
	\caption{\label{fig:nest_single} Allowable domains of $W_1$, $W_2$ and $W_3$ for nested model selection when cAIC selects: (1) $M_0$ with $\bm{\beta}_1$, (2) $M_1$ with $\{\bm{\beta}_1,\bm{\beta}_2\}$, (3) $M_2$ with $\{\bm{\beta}_1,\bm{\beta}_2,\bm{\beta}_3\}$ and (4) $M_0$ or $M_1$ or $M_2$.}
	\vspace{-0.5cm}
\end{figure}
Figure \ref{fig:nest_single} depicts geometrical regions which restrict the domains of $W_1$, $W_2$ and $W_3$. The regions are defined by the appropriate equations in \eqref{eq:nested_region_1} and \eqref{eq:nested_region_2}. 
Once $M_{cAIC}=M_0$, we have $p=p_0$ and use \eqref{prop:prop_nested1} to derive
\begin{eqnarray*}
\mathcal{A}_{M_0}(\mathcal{M}_{nest})& = & \{\bm{w}\in \mathbbm{R}^3:
		w_2^2\Sigma_2+2w_1w_2\Sigma_{12}<2(\rho_{21}+b_{21}),\\
&&w_2^2\Sigma_2 +w_3^2\Sigma_3+2w_1w_2 \Sigma_{12}+2w_1w_3\Sigma_{13}+2w_2w_3\Sigma_{23}<2(\rho_{31}+b_{31}) \},
\end{eqnarray*}
which corresponds to the left panel of Figure \ref{fig:nest_single}. One obtains similar sets of equations if $M_{cAIC}=M_1$ or $M_2$ (exact calculations are worked out in the SM). We conclude that a data-driven model selection heavily influences the domain of asymptotic distributions of the parameters that are subject to the selection process. The last panel of Figure  \ref{fig:nest_single} shows the partition of the space composed of $W_1$, $W_2$ and $W_3$. The following proposition describes the asymptotic distribution of a fixed and mixed effect after cAIC selection.

\begin{proposition} \label{prop:prop_nested1}	Suppose that Assumptions $(a)-(d)$ from Section \ref{sec:assump} are satisfied. Conditionally on $\mathcal{A}_{p}(\mathcal{M}_{nest})$ the limiting distribution for a post-cAIC fixed parameter is
\begin{equation*}
\begin{split}
F_p(\bm{t})&=\lim\limits_{n\rightarrow \infty} P\left[ n^{1/2}\{\hat{\bm{\beta}}_{m}(p)-\bm{\beta}_{0}\}\leqslant \bm{t}\mid \hat{p}_0=p, \mathcal{M}_{nest}\right]\\
&=P\left[ \{\mathcal{I}^{m}(p)\}^{-1/2}\bm{W}^s(p)\leqslant \bm{t}^s(p) \mid\bm{W}\in \mathcal{A}_{p}(\mathcal{M}_{nest})\right]id(\bm{t}\in \mathcal{T}_p),
\end{split}
\end{equation*}
where $\mathcal{T}_p\in\mathbbm{R}^{a+p}\times (\mathbbm{R}^+)^{K-p}$, $\bm{W}\sim N(0,\bm{I}_{a+K})$, $\mathcal{I}^m(p)$ is a submatrix whereas $\bm{t}^s(p)$ and $\bm{W}^s(p)$ are subvectors corresponding to a selected model $M_p$. In addition, conditionally on $\mathcal{A}^{\mu}_{p}(\mathcal{M}_{nest})$, the limiting distribution for a post-cAIC mixed parameter is
\begin{equation*}
	\begin{split}
		F_p(t)&=\lim\limits_{n\rightarrow \infty} P\left[	n^{1/2}\{\hat{\mu}_{i}(p)-\mu_{i}\}\leqslant t \mid \hat{p}_0=p, \mathcal{M}_{nest}\right]\\
		&\approx P\left[  \{\bm{c}^s_{i}(p)\}^t\bm{K}^{-1/2}(p) \bm{W}_{\mu}^{s}(p)\leqslant t\mid \bm{W}_{\mu}\in \mathcal{A}^{\mu}_{p}(\mathcal{M}_{nest})    \right],\\
	\end{split}
\end{equation*}
where $\bm{W}_{\mu}\sim N(0,\bm{I}_{a+K+r})$, $\bm{c}^{s}_{i}(p)$ and $\bm{W}_{\mu}^{s}(p)$ are subvectors and $\bm{K}(p)$ a submatrix corresponding to a selected model $M_p$. 
\end{proposition}

Proposition \ref{prop:prop_nested1} leads to the following corollary on the asymptotic post-selection density of fixed effects. In the SM we illustrate the effect of the selection on the densities.
\begin{corollary}\label{cor:density_beta}
	Under the assumptions of Proposition \ref{prop:prop_nested1}, the post-cAIC density of
	$n^{1/2}\{\hat{\bm{\beta}}_{m}(p)-\bm{\beta}_{0}\}$
	with $\hat{p}_0=p$ from $\mathcal{M}_{nest}$ is $f_p(\bm{t})=\phi_p [\bm{t}^s(p)|\mathcal{A}_{p}(\mathcal{M}_{nest});  \{\mathcal{I}^{m}(p)\}^{-1}]id(\bm{t}\in \mathcal{T}_p).$ When cAIC selects the true model, $\hat{p}_0=p_0$, then $f_{p_0}(\bm{t})=\phi_{p_0} \{\bm{t}^s(p_0)\}id(\bm{t}\in \mathcal{T}_p)$.
\end{corollary}

Proposition \ref{prop:prop_nested1} can be used to construct a post-cAIC confidence interval (CI) for a mixed parameter or components of a fixed effect. Using the same ideas as \cite{charkhi2018asymptotic}, we first focus on the latter. In fact, under the assumptions of Proposition \ref{prop:prop_nested1}, the asymptotic $1-\alpha$ quantiles of the marginal distributions of $\beta_j$, $j = 1, \dots, a+K$ satisfy $\int_{\mathcal{R}_{1-\alpha}}f_p(\bm{\bm{t}})d\bm{t}=1-\alpha$, where $\mathcal{R}_{1-\alpha}=\mathbbm{R}^{j-1}\times[-c_{j}(\alpha/2),c_{j}(\alpha/2)]\times\mathbbm{R}^{a+p-j}\times(\mathbbm{R}^+)^{K-p}$. Regarding a mixed effect, let $S^{p}_{0i}=\hat{\mu}_{i}(p)-\mu_{i}$ and $c^{\mu}_{i}(1-\alpha)= \inf\{s\in \mathbbm{R}: P(S^p_{0i}\leqslant s)\geqslant 1 -\alpha\}$, where we keep the dependence on $p$ to stress that $\hat{\mu}_{i}(p)$ is calculated after cAIC selection of covariates. Post-cAIC CI for $\beta_{j}$, $j = 1, \dots, a+K$, and $\mu_i$, $ i\in i = 1, \dots, n$, are regions defined as
\begin{equation} \label{eq:cAIC_int}
	\mathcal{I}^{\beta}_{j} =  \{\hat{\beta}_{j}(p) \pm c_{j}(\alpha/2) \}
	\quad \text{and} \quad
	\mathcal{I}^{\mu}_{i} =  \left\{\hat{\mu}_{i} \pm c^{\mu}_{i}(\alpha/2) \right\}.
\end{equation}
To retrieve critical values, we need to approximate the distribution of $n^{1/2}\{\hat{\bm{\beta}}_{m}(p)-\bm{\beta}_{0}\}$ and $S^p_{0i}$ using selection regions $\mathcal{A}_{p}(\mathcal{M}_{nest})$ and  $\mathcal{A}^{\mu}_{p}(\mathcal{M}_{nest})$. A detailed computational procedure involving Monte Carlo sampling is described in Section \ref{sec:simulations}. We can use classical results to construct $(1-\alpha)$-CI which do not account for the selection uncertainty
\begin{equation}\label{eq:naive_CI}
\mathcal{I}_{j}^{\beta, N}=\{\beta_j\pm \Phi^{-1}(\alpha/2)\times\hat{\sigma}(\hat{\beta}_{j})\}
\quad \text{and} \quad
\mathcal{I}^{\mu, N}_{i} = \left\{\hat{\mu}_{i} \pm \Phi^{-1}(\alpha/2) \times \hat{\sigma}(\hat{\mu}_{i})\right\},
\end{equation}
$j=1, \dots, a+K$, $i= 1, \dots, n$, where $\Phi$ is a normal cumulative distribution function. We refer to intervals in \eqref{eq:naive_CI} as naive CI. A high quantile from the normal distribution is sometimes replaced in \eqref{eq:naive_CI} by a bootstrap based or analytically derived quantity which results in the second order correct CI \citep[see, for example, a monograph of][for a detailed discussion of the second-order correctness for a mixed parameter]{rao2015small}.

\section{Selection properties of cAIC in general models}\label{sec:sec_overlap}

\begin{figure}
\centering
\includegraphics[width=0.75\textwidth]{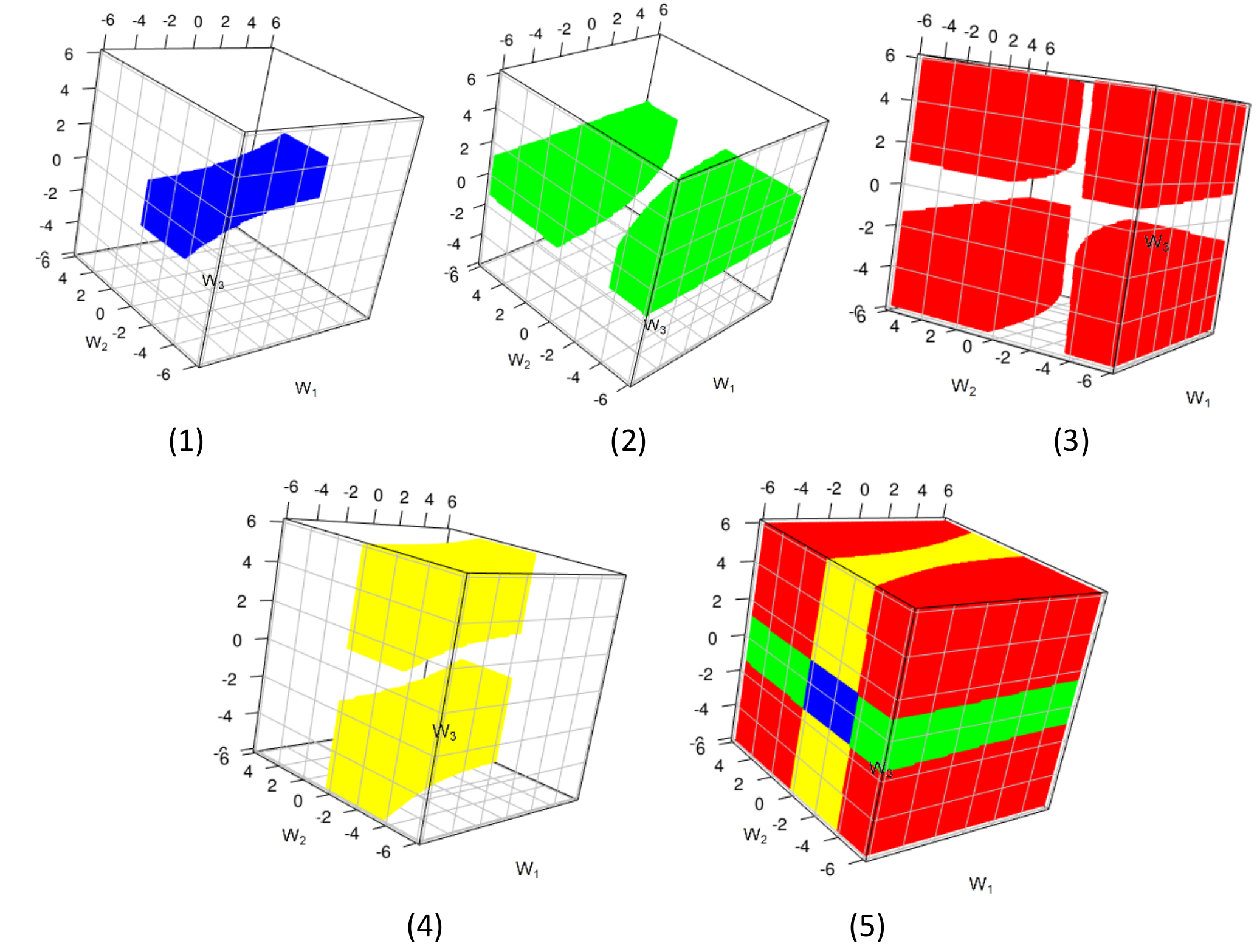}
\vspace{-0.3cm}
\caption{\label{fig:all_single}
Allowable domains of $W_1$, $W_2$ and $W_3$ for nested model selection when cAIC selects: (1) $M_0$, (2) $M_1$, (3) $M_2$, (4) $M_3$ and (5) $M_0$ or $M_1$ or $M_2$ or $M_3$.}
\vspace{-0.3cm}
\end{figure}

The set of candidate models $\mathcal{M}$ substantially influences asymptotic post-selection inference (see Figures \ref{fig:nest_single}, \ref{fig:all_single} as well as the discussion accompanying them). Suppose that $\mathcal{M}=\mathcal{M}_{all}$ is a set composed of all possible submodels of a largest model. Second, let $\mathcal{M}_{o}\subset\mathcal{M}_{all}$ be the set of overparametrised models including the true model. It immediately follows that the models in $\mathcal{M}_{o}$ are overlapping, according to the definition in \cite{vuong1989}. Lemma \ref{lemma:oversel} is an equivalent of Lemma 1 in \cite{charkhi2018asymptotic} for cAIC. As one would expect, cAIC also exhibits an overselection property.
\begin{lemma}\label{lemma:oversel}
	Consider a set of models $\mathcal{M}$ that contains at least one overparametrised candidate model and $\mathrm{cAIC}$ as a model selection criterion. Under assumptions $(a)-(d)$ in Section \ref{sec:assump}, an underparametrised model is selected with a probability converging to zero asymptotically.
\end{lemma}
The proof is deferred to our SM. Under this generalised modelling framework, the estimator of $\bm{\beta}_0$ in model $M$ is denoted by $\hat{\bm{\beta}}_m(M)$. Furthermore, let $\hat{\mu}_{i}(M)=\hat{\mu}_{i}\{\hat{\bm{\beta}}_m(M),\hat{\bm{\theta}}\}$ and $\hat{\mu}_{i}(M)=\hat{\mu}_{i}\{\bm{\beta}_0(M),\bm{\theta}\}$ for model $M\in \mathcal{M}$. In addition, $\bm{v}^s(M)$, $\mathcal{I}^m(M)$, $\bm{K}(M)$ denote a subvector and submatrices of $\mathcal{I}^m$ and  $\bm{K}$, respectively, corresponding to model $M$. If the orthogonality assumption (e) in Section \ref{sec:assump} holds, we obtain a simplified set of constraints given in \eqref{eq:gen_inequalities}. Otherwise, we follow the approach of \cite{charkhi2018asymptotic} for overlapping models. Define matrix $\bm{E}$ composed of two blocks. The first, $\bm{E}_1$, is a block diagonal matrix with  $\bm{E}(M_i,M_j)=\{\mathcal{I}^m(M_i)\}^{-1}\mathcal{I}^c_{ij}\{\mathcal{I}^{m}(M_i)\}^{-1}$ on a $(i,j)$th block. The second, $\bm{E}_2$, is the unitary matrix $\bm{I}_r$. The former corresponds to the covariates selected by cAIC. We define an extended selection matrix which indicates the diagonal and off-diagonal elements of $\bm{\Sigma}$. 
This matrix 
is necessary to construct a region similar to $\mathcal{A}_p$ in \eqref{eq:nested_region_1} and  \eqref{eq:nested_region_2}. Let $\bm{P}_{m}$ be a $|m|\times \left(a+K+K_b\right)$ projection matrix that selects the elements of $\bm{\Sigma}$ which belong to model $m$, $K_b=\binom{a+K}{2}$, $|m|$ the number of covariates in model $m$. The extended selection matrix $\bm{\upsilon}_{\mathcal{M}}$ of dimension $|\mathcal{M}| \times \left(a+K+K_b\right)$ is a matrix composed of $\{0,1\}$ such that $\bm{\upsilon}_{\mathcal{M}}=(\bm{1}^t_{(a+K+K_b)}\bm{P}_{1}^t\bm{P}_{1},\bm{1}^t_{(a+K+K_b)}\bm{P}_{M}^t\bm{P}_{M})$, where $|\mathcal{M}|$ is the number of considered models and $\bm{P}_{1},\dots,\bm{P}_{M}$ the projection matrices. 

If assumption (e) from Section \ref{sec:assump} holds, the selection region for a fixed parameter under model $M$ is
\begin{equation}\label{eq:gen_inequalities}
\begin{split}
	\mathcal{A}_{M}(\mathcal{M}_{o})=&\left\{ \bm{w}\in \mathbbm{R}^{a+K}:
	\left( \bm{1}_{|\mathcal{M}_o-1| } \otimes \left(\bm{1}^t_{a+K+K_b}\right) \bm{P}_{M}^t\bm{P}_{M} -\bm{\upsilon}_{\mathcal{M}_o\setminus M} \right) \right.\\
	&\times\left. \left(\Sigma_1w^2_1,\dots, \Sigma_{a+K}w^2_{a+K}, \Sigma_{12}w_1w_2, \dots, \right. \right.\\
	& \left. \left.
	\Sigma_{(a+K)(a+K)}w_{a+K}w_{a+K-1}   \right)^t\geqslant
	2\bm{\rho}_{M,|\mathcal{M}_o-1|} + 2\bm{b}_{M,|\mathcal{M}_o-1|} \right\},
\end{split}
\end{equation}
where  $\bm{\rho}_{M,|\mathcal{M}_o-1|} =\bm{\rho}_M-\bm{\rho}_{|\mathcal{M}_o-1|}$, $\bm{b}_{M,|\mathcal{M}_o-1|}=\bm{b}_M-\bm{b}_{|\mathcal{M}_o-1|}$,  $\bm{\rho}_M=\bm{1}_{|\mathcal{M}_o-1|}\rho_M$,  $\bm{\rho}_{|\mathcal{M}_o-1|}=(\rho_1,\dots, \rho_{|\mathcal{M}_o-1|})^t$,
$\bm{b}_M=\bm{1}_{|\mathcal{M}_o-1|}b_M$,  $\bm{b}_{|\mathcal{M}_o-1|}=(b_1,\dots, b_{|\mathcal{M}_o-1|})^t$. Similarly as in Section \ref{sec:sec_nested}, one needs to replace $\mathbbm{R}^{a+K}$ by $\mathbbm{R}^{a+K+r}$ to obtain the region $\mathcal{A}^{\mu}_{M}(\mathcal{M}_{o})$ for a mixed parameter. If the orthogonality condition (e) from Section \ref{sec:assump} does not hold, define $e=\sum_{M\in\mathcal{M}_o}^{}|M|$. Consider $\bm{B}_{cAIC,i}$ and $\bm{A}_{cAIC,i}$ as defined in \eqref{eq:region_overparm} and \eqref{eq:matrix_A_gen}. Let $\mathcal{M}^c_{o}=\mathcal{M}_{o} \setminus M_{\mathrm{cAIC}}$ and $\rho b_{M_{\mathrm{cAIC}},M_i}=\rho_{M_{\mathrm{cAIC}},M_i}+b_{M_{\mathrm{cAIC}},M_i}$. The selection regions are 
\begin{eqnarray}
\label{eq:gen_inequalities2_fixed}
&\mathcal{A}_{M}(\mathcal{M}_{o})=\{\bm{w}\in \mathbbm{R}^{e}:
\bm{w}^t(\bm{J}^m_{o})^{-1/2}\bm{B}_{\mathrm{cAIC},i}(\bm{J}^{m}_{o})^{-1/2}\bm{w}\geqslant
2\rho b_{M_{\mathrm{cAIC}},M_i}, M_i\in \mathcal{M}^c_{o} \},\\
\label{eq:gen_inequalities2_mixed}
&\mathcal{A}^{\mu}_{M}(\mathcal{M}_{o})=\{\bm{w}\in \mathbbm{R}^{e+r}:\bm{w}^t\bm{E}^{1/2}\bm{A}_{\mathrm{cAIC},i}\bm{E}^{1/2}\bm{w} \geqslant
2\rho b_{M_{\mathrm{cAIC}},M_i}, M_i\in \mathcal{M}^c_{o} \}.
\end{eqnarray}

We follow up with the example from Section \ref{sec:sec_nested}. Nevertheless, hereinafter we consider $\mathcal{M}_{all}=\{M_0, M_1, M_2, M_3 \}$ with $M_0$, $M_1$, $M_2$ as in the framework of the nested models and $M_3=(\bm{\beta}_1,\bm{\beta}_3)$. 
Assuming $\mathcal{M}_{all}$, our restrictions are defined by 4 inequalities -- in contrast to 3 inequalities for $\mathcal{M}_{nest}$ --  which naturally affect the domain for random variables. Once $M_{cAIC}=M_0$, we have
\begin{equation*}
	\begin{split}
		&\mathcal{A}_{M_0}(\mathcal{M}_{nest})=\{\bm{w}\in \mathbbm{R}^3:
		w_2^2\Sigma_2+2w_1w_2\Sigma_{12}<2(\rho_{M_1, M_0}+b_{M_1, M_0}),\\&w_2^2\Sigma_{2}+w_3^2\Sigma_{3}+2w_1w_2\Sigma_{12}+2w_1w_3\Sigma_{13}+2w_2w_3\Sigma_{23}<
		2(\rho_{M_2, M_0}+b_{M_2, M_0})\\
		& w_3^2\Sigma_{3}+2w_1w_3\Sigma_{13}<
		2(\rho_{M_3, M_0}+b_{M_3, M_0})
		\},
	\end{split}
\end{equation*}
which is illustrated in panel (1) of Figure \ref{fig:all_single}. Similar equations can be derived for $M_{cAIC}=M_1$ in panel (2), $M_{cAIC}=M_2$ in panel (3) and $M_{cAIC}=M_3$ in panel (4) (see our SM for exact expressions).
It is crucial to emphasise that even though we select the same model, the initial set, in our case $\mathcal{M}_{nest}$ or $\mathcal{M}_{all}$, influences allowable domains. This phenomenon is clearly visible if we compare Figures \ref{fig:nest_single} and \ref{fig:all_single}. Consider for example $M_{cAIC}=M_2$. The allowable domains assuming $\mathcal{M}_{nest}$ and $\mathcal{M}_{all}$ are shown in panel (3) of Figures \ref{fig:nest_single} and \ref{fig:all_single}, respectively. We immediately conclude that the domains differ significantly. 
The choice of $\mathcal{M}$ is of paramount importance -- it affects the distribution of all parameters, even those which are common to all models. 
Similarly to Figure \ref{fig:nest_single}, panel (5) of Figure \ref{fig:all_single} presents the partition of the 3-dimensional space.

The following proposition describes the asymptotic distribution of a regression parameter and a mixed parameter after cAIC selection from a general set of models.
\begin{proposition}\label{prop:prop_overparam}
(I)	Suppose that Assumptions $(a)-(e)$ from Section \ref{sec:assump} are satisfied. A limiting distribution for the post-cAIC fixed parameter is
\begin{equation*}
	\begin{split}
		F_M(\bm{t})&=\lim\limits_{n\rightarrow \infty}  P\left[
		n^{1/2}\{ \hat{\bm{\beta}}_{m}(M)-\bm{\beta}_{0}\}
		\leqslant \bm{t}\mid M_{\mathrm{cAIC}}=M,\mathcal{M}_{all}\right]\\
		&=P\left[ \{\mathcal{I}^{m}(M)\}^{-1/2}\bm{W}^s(M)\leqslant \bm{t}^s(M)\mid \bm{W}\in \mathcal{A}_{M}(\mathcal{M}_{o})\right]id(\bm{t}\in \mathcal{T}_M) , \\
	\end{split}
\end{equation*}
where $\mathcal{T}_M=\mathbbm{R}^{|M|}\times (\mathbbm{R}^+)^{K-|M|}$, $\bm{W}\sim N(0, \bm{I}_{a+K})$, $\mathcal{A}_{M}(\mathcal{M}_o)$ defined in \eqref{eq:gen_inequalities},  $\mathcal{I}^m(M)$ is a submatrix and $\bm{t}^s(M)$, $\bm{W}^s(M)$ subvectors corresponding to a selected model $M$. In addition, the limiting distribution for a post-cAIC mixed parameter is
\begin{equation*}
\begin{split}
F_M(t)&=\lim\limits_{n\rightarrow \infty} P\left[	n^{1/2}\{\hat{\mu}_{i}(M)-\mu_{i}\}\leqslant t\mid M_{\mathrm{cAIC}}=M,\mathcal{M}_{all}\right]\\
	&\approx P\left[    \{\bm{c}^{s}_{i}(M)\}^t\bm{K}^{-1/2}(M) \bm{W}_{\mu}^{s}(M)<t \mid \bm{W}_{\mu}\in \mathcal{A}^{\mu}_{p}(\mathcal{M}_{o})    \right] ,
\end{split}
\end{equation*}
where $\bm{W}_{\mu}\sim N(0, \bm{I}_{a+K+r})$, $\bm{c}^{s}_{i}(M)$ and $\bm{W}_{\mu}^{s}(M)$ are subvectors, whereas $\bm{K}(M)$ is a submatrix corresponding to model $M$.

(II)	
Suppose that Assumptions $(a)-(d)$ from Section \ref{sec:assump} are satisfied. The limiting distribution for a post-cAIC fixed parameter is
\begin{equation*}
F_M(\bm{t})
=P\left[ \{\mathcal{I}^{m}(M)\}^{-1/2}\bm{W}^s(M)\leqslant \bm{t}^s(M)\mid \bm{W}\in \mathcal{A}_{M}(\mathcal{M}_{o})\right]id(\bm{t}\in \mathcal{T}_M),\\
\end{equation*}
where $\mathcal{T}_M=\mathbbm{R}^{|M|}\times (\mathbbm{R}^+)^{e-|M|}$, $\bm{W}\sim N(0, \bm{I}_{e})$ and $\mathcal{A}_{M}(\mathcal{M}_o)$, defined in \eqref{eq:gen_inequalities2_fixed}. In addition, the limiting distribution for a post-cAIC mixed parameter is
\begin{equation*}
	F_M(t)\approx P\left[ \{\bm{c}^{s}_{i}(M)\}^t\bm{K}^{-1/2}(M) \bm{W}_{\mu}^{s}(M)<t\mid \bm{W}_{\mu}\in \mathcal{A}^{\mu}_{M}(\mathcal{M}_{o})    \right],
\end{equation*}
where $\bm{W}\sim N(0, \bm{I}_{e+r})$ and  $\mathcal{A}^{\mu}_{M}(\mathcal{M}_{o})$ defined in \eqref{eq:gen_inequalities2_mixed}.
\end{proposition}

Similarly as in Section \ref{sec:sec_nested}, Proposition \ref{prop:prop_nested1} leads to a corollary on the asymptotic post-selection density of fixed effects.
\begin{corollary}\label{cor:density_beta_gen}
	Under the assumptions of Proposition \ref{prop:prop_nested1}, the limiting post-cAIC density of 	$n^{1/2}\{\hat{\bm{\beta}}^s_{m}(M)-\bm{\beta}^s_{0}(M)\}$
	with $M_{cAIC}=M$ from $\mathcal{M}_{o}$ is $f_M(\bm{t}^s(M))=\phi_M [\bm{t}^s(M)\mid  \mathcal{A}_{M}(\mathcal{M}_{o}); \{\mathcal{I}^m(M)\}^{-1}]$, where $\bm{\beta}_{m}(M)$, $\bm{\beta}_{0}(M)$ are $|M|$-vectors of non-zero values.
\end{corollary}
One employs the density in Corollary \ref{cor:density_beta_gen} to construct confidence intervals for post-cAIC elements of ${\bm{\beta}}_{m}(M_{cAIC})$. The asymptotic $1-\alpha$ quantile satisfies  $\int_{\mathcal{R}_{1-\alpha}}f_M(\bm{t}^s(M))d\bm{t}^s(M)=1-\alpha$, where $\mathcal{R}_{1-\alpha}\in \mathbbm{R}^{|M|}$ imposes the restrictions $[-c_{j}(\alpha/2),c_{j}(\alpha/2)]$ on the j$th$ components. The form of the confidence intervals is almost identical as in \eqref{eq:cAIC_int} -- we only need to replace $\hat{\bm{\beta}}_{m}(p)$ with $\hat{\bm{\beta}}_{m}(M)$. The same applies to the CI for a fixed parameter. Proposition \ref{prop:prop_overparam} leads us to the result on a linear combination $\bm{l}^t{\bm{\beta}}$. We have
\begin{equation*}
\begin{split}
	F(t)&=\lim\limits_{n\rightarrow \infty}  P\left[
	n^{1/2}\{\bm{k}^s(M)\}^t\{ \hat{\bm{\beta}}^s_{m}(M)-\bm{\beta}^s_{m}(M)\}
	\leqslant t\mid M_{\mathrm{cAIC}}=M,\mathcal{M}_{all}\right]\\
	&=P\left[ \{\bm{k}^s(M)\}^t\{\mathcal{I}^{m}(M)\}^{-1/2}\bm{W}^s(M)\leqslant t\mid \bm{W}\in \mathcal{A}_{M}(\mathcal{M}_{o})\right].
\end{split}	
\end{equation*}

\section{Selection properties of cAIC in misspecified models}\label{sec:misspec}

In this section we provide some uniformly valid results which do not require the assumption of the existence of the true model. To do so, we need to extend the misspecification framework of \cite{charkhi2018asymptotic} to account for clustered data and cAIC model selection. In a series of papers \cite{leeb2003finite,leeb2006can,leeb2008can} proved that uniform results for post-selection estimators are not available for the traditional quantities $\hat{\bm{\beta}}(M)-\bm{\beta}_0(M)$, $M\in \mathcal{M}$ which we considered in Sections \ref{sec:sec_nested} and \ref{sec:sec_overlap}. These results are general and apply to various selection procedures such as LASSO or AIC \citep[see][]{tibshirani2018uniform,charkhi2018asymptotic}. Nevertheless, under a misspecified setting \citep{charkhi2018asymptotic}, considering nonstandard targets \citep{berk2013valid} or modified pivots \citep{tibshirani2018uniform}, uniform results are attainable.

In our misspecified setting, the true parameter vector does not exist, because all models are misspecified or the true density is not a member of a parametric family. To be able to prove the uniform convergence, we use a framework with asymptotics based on a pseudo triangular array 
adapted for dependent data. In practice we collect one data sample. Thus our construction serves only to demonstrate the theoretical results. If we had a possibility to collect different samples, we assume that the observed vectors $\bm{y}$ might be represented in an extended, vector based triangular array  $\{\bm{Y}_{ni}: i=1,\dots, n; n\in \mathbbm{N}\}$, that is, we suppose that $\bm{Y}_{nk}$ and  $\bm{Y}_{nl}$ were independent for $k\neq l$ and for different samples. 
Let $g^m_n$ and $G^m_n$ as well as $g^e_n$ and $G^e_n$ be the true joint marginal and joint extended density and distribution of $\{\bm{Y}_{n1},\dots,\bm{Y}_{n}\}$. In what follows, all probabilities are computed with respect to the true distributions $P=P_{G_m}$ and $P=P_{G_e}$. Within this framework, the estimation of $\bm{\beta}$ and $\bm{\varrho}$ often requires the same conditions imposed on marginal and extended loglikelihoods. If no confusion is possible, we state them using $(\cdot)$ which stands for $m$ or $e$. Since the likelihood might be misspecified, we use White's \citeyearpar{white_1994} quasi-likelihood framework for modelling. Models can be thus represented as
\begin{equation*}
	M^{m}_{n,j}=\left\{\prod_{i=1}^{n}f^{m}_{j,i}(\bm{y}_{i};\bm{\beta}_j):\bm{\beta}_j\in \Theta^m_j \subset \mathbbm{R}^{v^m_j}\right\},\;
	M^{e}_{n,j}=\left\{\prod_{i=1}^{n}f^{e}_{j,i}(\bm{y}_{i};\bm{\varrho}_j):\bm{\varrho}_j\in \Theta^e_j \subset \mathbbm{R}^{v^e_j}\right\},
\end{equation*}
with $v^{(\cdot)}_j$ the number of parameters in $M^{(\cdot)}_{n,j}$, and $\Theta^{(\cdot)}_j$ a compact set. The collection of all models is denoted by $\mathcal{M}^{(\cdot)}_n=\{M^{(\cdot)}_{n,1},\dots, M^{(\cdot)}_{n,J}\}$. Following Definition 2.2 in \cite{white_1994}, the true class of distribution $\mathcal{G}^{(\cdot)}_n$ is defined by $\int g^{(\cdot)}_n(\bm{y}) \log g^{(\cdot)}_n(\bm{y}) d\bm{y} < \infty$ for each $n\in \mathbbm{N}$. When no confusion is possible, we skip the subscript $n$. Furthermore, for each $i\in \mathbbm{N}$ and each $j\in 1\dots, J$, $f^m_{j,i}(\cdot;\bm{\beta}_j)$ and $f^e_{j,i}(\cdot;\bm{\varrho}_j)$ are measurable for all $\bm{\beta}_j\in \Theta^m_j$, $\bm{\varrho}_j\in \Theta^e_j$. We suppose that $f^{(\cdot)}_{j,i}(\bm{y}_{i};\cdot)$ is almost surely continuous and continuously differentiable on $\Theta^{(\cdot)}_j$. The existence of the marginal and extended likelihood estimators follows from the extension of Lemma 2.1 in \cite{gallant1988unified}, that is we adapt their results to account for modelling independent vectors, rather than independent scalars. The ideas of the proof are general enough to be applied in this setting. We therefore assume that there exist estimators $\hat{\bm{\beta}}_{m,j}$, $\hat{\bm{\varrho}}_{e,j}$ maximising $\prod_{i=1}^{n}f^m_{j,i}(\bm{y}_{i};\bm{\beta}_j)$ and
$\prod_{i=1}^{n}f^e_{j,i}(\bm{y}_{i};\bm{\varrho}_j)$
over $\Theta^{(\cdot)}_j$.  Furthermore, we call the pseudo-true values $\bm{\beta}'_{0,n}(M_j)$ and $\bm{\varrho}'_{0,n}(M_j)$
the maximisers of
\begin{equation*}
	\mathbb{E}_{G^m_n}\left\{ n^{-1}\sum_{i=1}^{n}
	\log f^{m}_{j,i}(\bm{y}_{i};\bm{\beta}_j)\right\}\quad\text{and}\quad \mathbb{E}_{G^e_n}\left\{ n^{-1}\sum_{i=1}^{n}
	\log f^{e}_{j,i}(\bm{y}_{i};\bm{\varrho}_j)\right\},
\end{equation*}
if such exists. These maximisers depend on the sample size $n$, the true joint density and the model densities. Denote with $v'^{(\cdot)}=\sum_{j=1}^{J}v_j^{(\cdot)}$. For the marginal likelihood we have  $\bm{\beta}'_{0,\mathcal{M}}=\{\bm{\beta}'_{0,n}(M_1)^t,\dots,\bm{\beta}'_{0,n}(M_K)^t\}^t$, $\hat{\bm{\beta}}_{m,\mathcal{M}}=\{\hat{\bm{\beta}}_m(M_1)^t,\dots,\hat{\bm{\beta}}_m(M_K)^t\}^t$, that is, vectors of length $v'^{m}$. On the other hand, for the extended parameters the vectors  $\bm{\varrho}'_{0,\mathcal{M}}=\{\bm{\varrho}'_{0,{n}}(M_1)^t,\dots,\bm{\varrho}'_{0,{n}}(M_K)^t\}^t$, $\hat{\bm{\varrho}}_{e,\mathcal{M}}=\{\hat{\bm{\varrho}}_e(M_1)^t,\dots,\hat{\bm{\varrho}}_{e}(M_K)^t\}^t$ are of length $v'^{e}$.

Lemma \ref{lemma: misspec} refers only to the extended vector of parameters due to our mixed parameter focus. An equivalent statement is valid for fixed parameters estimated using the marginal loglikelihood. In addition, recall that the estimating equations for the fixed parameters using marginal and extended loglikelihood result in the same expression (see Section \ref{sec:inference_LMM} and references therein for more details). Even though the extended likelihood is not a proper likelihood as it includes non-observable random effects, the general results of Lemma 2.1 in \cite{gallant1988unified} are applicable in this setting.
Therefore,
\begin{lemma}\label{lemma: misspec}
	Define $\mathcal{H}_{n}\sim N _{v'}\{0,\bm{D}(\bm{\varrho}'_{0,\mathcal{M}})\}$ where $\bm{D}(\bm{\varrho}'_{0,\mathcal{M}})$ is a $v'^{e}\times v'^{e}$ block matrix with $(i,j)$th block equal to $\mathcal{J}_{M_i}^{-1}(\bm{\varrho}'_{0,M_i})\mathcal{I}_{ij}(\bm{\varrho}'_{0,M_i},\bm{\varrho}'_{0,M_j})\mathcal{J}_{M_j}^{-1}(\bm{\varrho}'_{0,M_j})$. We thus have
	\begin{equation}\label{eq:unif2}
		\lim \limits_{{n}\rightarrow \infty }\sup_{\bm{t}\in \mathbbm{R}^{v'^{e}} }
		\sup_{G^{e}_{n}\in \mathcal{G}^{e}_{n} }|P\{{n}^{1/2}(\hat{\bm{\varrho}}_{e,\mathcal{M}}-\bm{\varrho}'_{0,\mathcal{M}})\leqslant \bm{t}\} -P (\mathcal{H}_{n}\leqslant\bm{t})|=0,
	\end{equation}
	where $\mathcal{J}_{M_i}$ and $\mathcal{I}_{ij}$ as defined in Section \ref{sec:assump}.
\end{lemma}
We assume that there exists an estimator $\hat{\bm{D}}$ of $\bm{D}$ such that
$\lim\limits_{{n}\rightarrow \infty }\sup_{G^{e}_{n}\in \mathcal{G}^{e}_{n} }P(||\hat{\bm{D}}-\bm{D}||>\varepsilon)=0$,
where $||A||$ is the Euclidean matrix norm operator and we suppose that $\mathcal{W}_{v'^{e}}\sim N_{v'}(0,\bm{I}_{v'^{e}})$ \citep[for a discussion about the existence of such estimators see][ \textsection 8.3]{white_1994}. The uniform convergence result \eqref{eq:unif2} is also valid for a pivotal statistic:
\begin{equation*}
\lim \limits_{{n}\rightarrow \infty }\sup_{\bm{t}\in \mathbbm{R}^{v'^{e}} }
\sup_{G^{e}_{n}\in \mathcal{G}^{e}_{n} }|P\{\hat{\bm{D}}^{-1/2}{n}^{-1/2}(\hat{\bm{\varrho}}_{e,\mathcal{M}}-\bm{\varrho}'_{0,\mathcal{M}})\leqslant \bm{t}\} -P (\mathcal{W}_{v'}\leqslant\bm{t})|=0.
\end{equation*}
\subsection{Post-selection inference for misspecified models}
As we do above, we define a selection region using pairwise comparisons between misspecified models from a set $\mathcal{M}$. Define $\ell^c_{n,M_j}(\bm{y},\bm{\beta}_j)=\sum_{i=1}^{n}\log f^c_{j,i}(\bm{y}_{i}|\bm{u}_{i};\bm{\beta}_j)$. Model $M_{\mathrm{cAIC}}$ is selected if
$	2[\ell^c_{n,M_{\mathrm{cAIC}}}\{\bm{y},\hat{\bm{\beta}}_m(M_\mathrm{cAIC})\}-\ell^c_{n,M}\{\bm{y},\hat{\bm{\beta}}_m(M)\}]\geqslant 2(\rho_{M_{\mathrm{cAIC}}, M}+b_{M_{\mathrm{cAIC}}, M})$,
for each $M\in \mathcal{M} \setminus M_{\mathrm{cAIC}}$. As it was stated in Section \ref{sec:sec_nested}, this is equivalent with $	2l'^{c}_{n,M_{\mathrm{cAIC}}}-2l'^{c}_{n,M}\geqslant 2(\rho_{M_{\mathrm{cAIC}}, M}+b_{M_{\mathrm{cAIC}}, M})$, where in this section
\begin{equation*}
	2l'^{c}_{n,M_{\mathrm{cAIC}}}=2(\ell^c_{n,M_{\mathrm{cAIC}}}\{\bm{y},\hat{\bm{\beta}}_m(M_\mathrm{cAIC})\}-\ell^c_{n,M_{\mathrm{cAIC}}}\{\bm{y},\bm{\beta}'_{0,n}(M_{\mathrm{cAIC}})\}),
\end{equation*}
and $2l'^{c}_{n,M}$ is defined in an analogous way with $M_{\mathrm{cAIC}}$ replaced by $M$. As we showed in Sections \ref{sec:sec_nested} and \ref{sec:sec_overlap}, when both models are correctly specified, the difference of the conditional log-likelihoods can be described using scaled chi-squared random variables. \cite{vuong1989} investigated the conditions under which the difference of marginal likelihoods converges assuming model misspecification. 
A full characterization of the asymptotic distribution is possible only in case of the similarity of likelihoods. Since in our selection procedure we only consider fixed effects, 
similar arguments, that is, Taylor expansions of the conditional likelihoods around the true value can be used to prove divergence in our setting. We thus focus on the misspecified setting in the case of the similarity of the conditional likelihoods, that is $\ell^c_{{n},M_k}\{\bm{y},\bm{\beta}'_{0,{n}}(M_k)\}=\ell^c_{{n},M_l}\{\bm{y},\bm{\beta}'_{0,{n}}(M_l)\}$ for $k,l=1,\dots,K$. Following \cite{charkhi2018asymptotic}, we consider a general set of models $\mathcal{M}$ and suppose that it includes the smallest model $M_{s}=M_{pars}$ nested in all other models. Our strategy is to compare all models with the smallest one and then determine the final regions using, as before, pairwise comparisons. For each $M\in \mathcal{M} \setminus M_{\mathrm{s}}$, we have
\begin{equation*}
	\begin{split}
		&2[\ell^c_{{n},M}\{\bm{y},\hat{\bm{\beta}}_m(M)\}-\ell^c_{{n},M}
		\{\bm{y},\bm{\beta}'_{0,{n}}(M)\}]
		=2n^{1/2}\{\hat{\bm{\beta}}_m(M)-\bm{\beta}'_{0,{n}}(M)\}^t\mathcal{R}_{M}^c\left\{\bm{\beta}'_{0,{n}}(M)\right\}\\
		&+n
		\{\hat{\bm{\beta}}_m(M)-\bm{\beta}'_{0,{n}}(M)\}^t\mathcal{J}_{M}^c
		\{\bm{\beta}'_{0,{n}}(M)\}\{\hat{\bm{\beta}}_m(M)-\bm{\beta}'_{0,{n}}(M)\}+o_P(1)\\
		&=n\{\hat{\bm{\beta}}_m(M)-\bm{\beta}'_{0,{n}}(M)\}^t\mathcal{J}_{M}^c\{\bm{\beta}'_{0,{n}}(M)\}\{\hat{\bm{\beta}}_m(M)-\bm{\beta}'_{0,{n}}(M)\}+o_P(1),
	\end{split}
\end{equation*}
where $\mathcal{R}_{M}^c$ and $\mathcal{J}_{M}^c$ as defined in Section \ref{sec:assump}, the third line is a direct result of $\mathcal{R}_{{n}}^c(\bm{\beta}'_{0,{n}}(M))=0$. The expansion for $M_s$ follows along the same lines, we just need to replace $M$ with $M_s$. Once we compare them we obtain
\begin{equation*}
	\begin{split}
		2l'^{c}_{{n},M}-2l'^{c}_{{n},M_s}
		&=n\{\hat{\bm{\beta}}_m(M)-\bm{\beta}'_{0,{n}}(M)\}^t\mathcal{J}_{M}^c\{\bm{\beta}'_{0,{n}}(M)\}\{\hat{\bm{\beta}}_m(M)-\bm{\beta}'_{0,{n}}(M)\}\\
		&-n\{\hat{\bm{\beta}}_m(M_s)-\bm{\beta}'_{0,{n}}(M_s)\}^t\mathcal{J}_{M_s}^c\{\bm{\beta}'_{0,{n}}(M_s)\}\{\hat{\bm{\beta}}_m(M_s)-\bm{\beta}'_{0,{n}}(M_s)\}+o_p(1)\\
		&=n(\hat{\bm{\beta}}_{m,\mathcal{M}}-\bm{\beta}'_{0,\mathcal{M}})^t \bm{B}_{M,M_s}(\hat{\bm{\beta}}_{m,\mathcal{M}}-\bm{\beta}'_{0,\mathcal{M}}) +o_p(1).
	\end{split}
\end{equation*}
In addition, $\bm{B}_{M,M_s}$ is a diagonal matrix with blocks $\mathcal{J}^c_{M}\{\bm{\beta}'_{0,{n}}(M)\}$ and $-\mathcal{J}^c_{M_s}\{\bm{\beta}'_{0,{n}}(M_s)\}$ corresponding to models $M$ and $M_{s}$. Following the same reasoning as in the proof of Proposition \ref{prop:prop_overparam} in Section \ref{sec:sec_overlap}, we obtain the asymptotic selection event for model $M_{\mathrm{cAIC}}$.
\begin{proposition}\label{prop:uniform_conv} The selection region for mixed parameter assuming a set of misspecified models $\mathcal{M}$ is
	\begin{equation*}
		\begin{split}
			\mathcal{A}_{M_{\mathrm{cAIC}}}(\mathcal{M})&=\left\{w\in \mathbbm{R}^{v'+r}:\bm{w}^t\bm{E}^{1/2}(\bm{A}_{\mathrm{cAIC},M_s}-\bm{A}_{M,M_s})\bm{E}^{1/2}\bm{w} \right.\\
			&\geqslant\left. 
			2(\rho_{M_{\mathrm{cAIC}, M}} +b_{M_{\mathrm{cAIC},M}}),
			M\in \mathcal{M}\setminus M_{cAIC}	 \right\}.
		\end{split}
	\end{equation*}
	Suppose that the assumptions from Lemma \ref{lemma: misspec} hold. Then we have
	\begin{equation}\label{eq:unif_conv_mu}
		\begin{split}
			&\lim \limits_{{n}\rightarrow \infty }\sup_{G^e_{n}\in \mathcal{G}^e_{n} } \sup_{t\in \mathbbm{R} }\left|
			P\left[
			{n}^{1/2}\{
			\hat{\mu}_{i}(M_{cAIC})-\mu_{i}\}<t|M_{\mathrm{cAIC}}
			\}<t|M_{\mathrm{cAIC}}\right]\right.\\
			&\left.-P\left[ \{\bm{c}^{s}_{i}(M_{cAIC})\}^t\bm{K}^{-1/2}(M_{cAIC}) \bm{W}^{s}(M_{cAIC})<t|\mathcal{A}^{\mu}_{M_{\mathrm{cAIC}}}   \right]\right|=0.
		\end{split}
	\end{equation}
\end{proposition}
Proposition \ref{prop:uniform_conv} guarantees 
uniform convergence over 
$t \in \mathbbm{R}$ 
and $G^e_{n}$ in a class $\mathcal{G}^e_{n}$ in contrast to the pointwise convergence which is valid only over $t \in \mathbbm{R}$. 
The construction of the cluster-wise uniformly valid post-selection intervals follows in the same way as in \eqref{eq:cAIC_int}. 

\section{Simulation study}\label{sec:simulations}

We carried out an empirical simulation study to assess the performance of post-cAIC CI for a regression parameter, a linear combination of its components and a mixed parameter. In our analysis, we compare post-cAIC CI in \eqref{eq:cAIC_int} with naive intervals in \eqref{eq:naive_CI}. In case of the mixed parameter, we construct them using the first- and second-order correct MSE estimators in \eqref{eq:naive_mse} and \eqref{eq:MSE_lin}, respectively. It is well known that mixed-parameters are quite robust to misspecification of the shape of random effects \citep{mcculloch2011misspecifying}. We investigate the performance of our new method as well as the robustness of naive CI to model misspecification for fixed and mixed effects. The literature offers us a benchmark when it comes to the post-selection inference for mixed parameters under LMM. More specifically,, we compare our post-cAIC intervals with post-OBSP intervals constructed using OBSP for area-level parameters and post-selected MSE developed by \cite{sugasawa2019observed}. Since the authors focused on the area-level model only and did not consider the construction of the intervals, we somewhat extend their work regarding these two aspects.

The data generation process was inspired by \citeauthor{charkhi2018asymptotic}'s \citeyearpar{charkhi2018asymptotic} procedure. Namely, we assume a nested error regression model (NERM) $y_{ij}=\sum_{d=1}^{5}\beta_dx_{dij}+u_{i}+e_{ij}$ with a true vector of fixed parameters $\bm{\beta}=(2.25, -1.1, 2.43,0,0)$, $e_{ij}\sim N(0,\sigma^2_e)$ and $u_{i}\sim N(0,\sigma^2_u)$, $i=1,\dots, {n}$, $j=1,\dots, m_i$. We consider two settings for $\sigma^2_e$ and $\sigma^2_u$; under setting 1 (S1)  $(\sigma^2_e,\sigma^2_u)=(1,1)$ whereas under setting 2 (S2) $(\sigma^2_e,\sigma^2_u)=(1,0.5)$. We wish to mimic two types of asymptotic regimes. In the first case, we assume that ${n}\rightarrow \infty $ with $m_i$ fixed such that ${n}:m_i=\{(15:5),(30:5),(60:5),(90:5)\}$.  Then, in the second case we suppose that ${n}$ is fixed and $m_i\rightarrow\infty $ such that ${n}:m_i=\{(30:5),(30:10),(30:20)\}$. The former scenario is popular in SAE \citep{rao2015small} and longitudinal studies \citep{verbekeMolenberghs2000}, whereas the latter in repeated cross-sectional studies. Further, $x_{1ij}=1$ and $x_{2ij},\dots,x_{5ij}\sim N(0_4,\bm{\Omega})$, where $\bm{\Omega}$ is a positive definite matrix with 1 on the diagonal and 0.25 elsewhere. In case of post-cAIC inference for a linear combination of fixed effects $\bm{k}_i^t\bm{\beta}$, we computed $n$ linear combinations in each simulation, and set $\bm{k}_i = \bar{\bm{x}}_i$, that is, vectors $\bm{k}_i$ were means of cluster covariates. Under NERM, the computation of the second term of the penalty in cAIC defined in \eqref{eq:c_AIC_Kubo} is simplified \citep[a spelled out formula can be found in][Section 4.3]{kubokawa2011conditional}. We consider three different model sets. Denote by $\bm{\upsilon}^i_{all}$ the extended selection matrix when the first $i$ parameters are present in all models. In our empirical study we examine $\bm{\upsilon}^2_{all}$ which is a $2^3\times 15$ matrix (5 covariates and 10 covariance terms), $\bm{\upsilon}^3_{all}$ (a $2^2\times 15$ matrix) and $\bm{\upsilon}^4_{all}$ (a $2\times 15$ matrix). Since all model sets led to the same conclusions, the results under $\bm{\upsilon}^3_{all}$ and $\bm{\upsilon}^4_{all}$ are deferred to the SM. We run our simulations until model $M$ with parameters $\beta_1, \beta_2, \dots, \beta_5$ had been selected $I=1000$ times. In each simulation run, we estimate the matrix $\bm{\Sigma}$ defined in Section \ref{sec:sec_nested} for a full model. Its submatrix $\bm{\Sigma}(M)$ corresponds to the model selected in a particular simulation. We apply a result derived from Proposition \ref{prop:prop_overparam} (I) to calculate the confidence intervals. Observe that one should employ (II) if the orthogonality condition from Section \ref{sec:assump} does not hold. Nevertheless, following the practice of \cite{charkhi2018asymptotic}, we use (I) which leads to good numerical outcomes. In the SM, we describe a procedure to obtain post-cAIC CI as a practical algorithm. Furthermore, we provide some practical guidance on sampling from a multivariate truncated normal.

\begin{table}
\caption{\label{tab:PoSI_beta_1} Coverage probabilities and average lengths (in parenthesis) of post-cAIC and naive confidence intervals for $\beta_{j}$. Nominal coverage probability: 95\%, selection matrix: $\bm{\upsilon}^2_{all}$.}
\centering
\hspace{-0.1cm}
\setlength{\tabcolsep}{1.6pt}
\fbox{%
\begin{tabular}{ccccccccccc}
&         &           &  & $(15:5)$     & $(30:5)$     & $(60:5)$     & $(90:5)$     &  & $(30:10)$    & $(30:20)$    \\
S & Meth.   & $\beta_j$ &  & CP (L)    & CP (L)    & CP (L)    & CP (L)    &  & CP (L)    & CP (L)    \\ \cline{1-3} \cline{5-8} \cline{10-11}
S1  & p.-cAIC & $\beta_1$ &  & 93.2 (1.144) & 94.4 (0.818) & 95.9 (0.582) & 96.2 (0.467) &  & 95.9 (0.831) & 94.1 (0.768) \\
&         & $\beta_5$     &  & 94.8 (0.806) & 96.2 (0.531) & 95.8 (0.379) & 96.5 (0.320) &  & 96.8 (0.363) & 96.0 (0.251) \\
& N       & $\beta_1$    &  & 92.9 (1.099) & 93.8 (0.796) & 95.0 (0.560)    & 95.8 (0.456) &  & 93.9 (0.748) & 93.0 (0.727) \\
&         & $\beta_5$     &  & 66.8 (0.569) & 72.3 (0.377) & 69.6 (0.267) & 70.4 (0.218) &  & 69.2 (0.256) & 66.9 (0.175) \\[2mm]
S2  & p.-cAIC & $\beta_1$    &  & 92.4 (0.849) & 93.6 (0.622) & 95.0 (0.439)   & 97.1 (0.355) &  & 94.3 (0.589) & 94.3 (0.540) \\
&         & $\beta_5$    &  & 93.1 (0.807) & 93.1 (0.521) & 95.7 (0.378) & 96.7 (0.323) &  & 97.2 (0.371) & 96.4 (0.254) \\
& N       & $\beta_1$    &  & 92.3 (0.837) & 93.5 (0.618) & 94.9 (0.435) & 97.1 (0.353) &  & 93.1 (0.558) & 93.7 (0.527) \\
&         & $\beta_5$    &  & 64.5 (0.562) & 68.3 (0.373) & 70.8 (0.264) & 67.2 (0.216) &  & 68.1 (0.255) & 66.7 (0.175)
\end{tabular}}
\vspace{-0.3cm}
\end{table}

Table \ref{tab:PoSI_beta_1} presents coverage probabilities (CP) and lengths (L) for post-cAIC (p.-cAIC) and naive (N) CI for the components of fixed parameters $\beta_j$. CP and L were calculated as an average over simulation runs. The superiority of post-cAIC is unquestionable as its coverage always oscillates around the nominal level. In contrast, the naive CI for $\beta_5$ never surpasses $71\%$, which is a consequence of treating a chosen model as given. Our results are in alignment with those in \cite{charkhi2018asymptotic}, in which post-AIC CI are studied for fixed parameters in a modelling setting without random effects.
Let us investigate the effect of including covariates to our parameter of interest. 
Table \ref{tab:PoSI_betax} displays coverage probabilities and average lengths for linear combinations of the components of fixed parameters. We present two randomly selected linear combinations and $\bar{\bm{k}}^t\hat{\bm{\beta}}$ which stands for the average over $n$ parameters. While we can observe an improvement of the performance of the naive CI in comparison with Table \ref{tab:PoSI_beta_1}, the undercoverage still persist. In contrast, post-cAIC CI perform better overall with a coverage close to the nominal level.  
\begin{table}
\caption{\label{tab:PoSI_betax} Coverage probabilities and average lengths (in parenthesis) of post-cAIC and naive confidence intervals for $\bm{k}_i^t\hat{\bm{\beta}}$ and $\bar{\bm{k}}^t\hat{\bm{\beta}}$. Nominal coverage probability: 95\%, selection matrix: $\bm{\upsilon}^2_{all}$.}
\setlength{\tabcolsep}{1.6pt}
\hspace{-0.1cm}
\fbox{%
\begin{tabular}{ccccccccccc}
	&         &                  &  & $(15:5)$     & $(30:5)$     & $(60:5)$     & $(90:5)$     &  & $(30:10)$    & $(30:20)$    \\
	S & Meth.   &   Par.     &  & CP (L)    & CP (L)    & CP (L)    & CP (L)    &  & CP (L)    & CP (L)    \\
	\cline{1-3} \cline{5-8} \cline{10-11}
	S1  & p.-cAIC & $\bm{k}_1^t\hat{\bm{\beta}}$                &  & 95.4 (1.411) & 97.7 (0.990)  & 94.7 (0.711) & 98.6 (0.577) &  & 96.6 (0.891) & 93.9 (0.751) \\
	&         & $\bm{k}_2^t\hat{\bm{\beta}}$                &  & 96.6 (1.631) & 94.5 (0.923) & 96.3 (0.669) & 98.0 (0.575) &  & 96.2 (0.870) & 94.5 (0.793) \\
	&         & $\bar{\bm{k}}^t\hat{\bm{\beta}}$ &  & 95.3 (1.374) & 95.8 (0.998) & 96.7 (0.711) & 97.1 (0.572) &  & 96.5 (0.891) & 94.5 (0.786) \\
	& N       & $\bm{k}_1^t\hat{\bm{\beta}}$                 &  & 89.6 (1.155) & 94.2 (0.820)  & 91.0 (0.608) & 95.5 (0.467) &  & 92.9 (0.765) & 93.6 (0.731) \\
	&         & $\bm{k}_2^t\hat{\bm{\beta}}$                 &  & 88.9 (1.186) & 91.9 (0.837) & 94.1 (0.596) & 94.2 (0.480) &  & 93.3 (0.752) & 93.3 (0.728) \\
	&         & $\bar{\bm{k}}^t\hat{\bm{\beta}}$ &  & 92.0 (1.159) & 92.3 (0.844) & 93.4 (0.594) & 94.1 (0.487) &  & 93.5 (0.758) & 93.3 (0.731) \\[2mm]
	S2  & p.-cAIC & $\bm{k}_1^t\hat{\bm{\beta}}$                &  & 94.1 (1.155) & 97.3 (0.864) & 94.1 (0.534) & 97.0 (0.451) &  & 96.8 (0.672) & 94.2 (0.568) \\
	&         & $\bm{k}_2^t\hat{\bm{\beta}}$                &  & 95.0 (0.997) & 96.6 (1.053) & 96.7 (0.538) & 97.8 (0.625) &  & 96.0 (0.665) & 94.5 (0.548) \\
	&         & $\bar{\bm{k}}^t\hat{\bm{\beta}}$ &  & 95.2 (1.117) & 97.0 (0.844)   & 97.0 (0.597) & 97.6 (0.485) &  & 96.0 (0.661) & 94.5 (0.560) \\
	& N      &$\bm{k}_1^t\hat{\bm{\beta}}$                &  & 87.4 (0.910) & 94.0 (0.697)   & 91.7 (0.460) & 95.1 (0.406) &  & 91.5 (0.571) & 93.3 (0.537) \\
	&         & $\bm{k}_2^t\hat{\bm{\beta}}$                &  & 92.3 (0.841) & 86.9 (0.769) & 94.7 (0.479) & 87.2 (0.437) &  & 93.1 (0.580) & 93.9 (0.530) \\
	&         & $\bar{\bm{k}}^t\hat{\bm{\beta}}$ &  & 91.0 (0.912) & 92.3 (0.677) & 92.9 (0.477) & 93.9 (0.390) &  & 93.1 (0.571) & 93.5 (0.532)
\end{tabular}}	
\vspace{-0.3cm}
\end{table}
\begin{table}
	\caption{\label{tab:PoSI_1} Coverage probabilities and average lengths (in parenthesis) of post-cAIC, post-OBSP and naive confidence intervals for $\mu_{i}$. Nominal coverage probability: 95\%, selection matrix: $\bm{\upsilon}^2_{all}$.}
	\setlength{\tabcolsep}{2pt}
	\hspace{-0.1cm}
	\fbox{%
		\begin{tabular}{*{10}{c}}
			&                    &            & $(15:5)$     & $(30:5)$     & $(60:5)$     & $(90:5)$     &  & $(30:10)$                           & $(30:20)$    \\
			S & Meth.              &            & CP (L)    & CP (L)    & CP (L)    & CP (L)    &  & CP (L)                           & CP (L)    \\ \cline{1-2} \cline{4-7} \cline{9-10}
			S1                  & {\small p.-cAIC}           &            & 94.9 (1.646) & 95.5 (1.660) & 95.4 (1.645) & 95.5 (1.641) &  & 95.3 (1.213)        & 95.1 (0.871) \\
			& \multicolumn{2}{c}{N1} & 93.6 (1.564) & 94.8 (1.607) & 95.1 (1.615) & 95.2 (1.619) &  & 94.9 (1.192)                        & 94.9 (0.864) \\
			& \multicolumn{2}{c}{N2} & 95.1 (1.648) & 95.2 (1.636) & 95.3 (1.629) & 95.3 (1.629) &  & 95.1 (1.202)                        & 95.0 (0.868) \\
			& {\small p.-OBSP}    &            & 94.8 (1.628) & 95.2 (1.632) & 95.2 (1.626) & 95.3 (1.627) &  & 95.1 (1.200)                        & 94.3 (0.867) \\[2mm]
			S2                & {\small p.-cAIC}        &            & 93.4 (1.518) & 94.9 (1.550) & 95.5 (1.539) & 95.5 (1.529) &  & 95.2 (1.161)  & 95.0 (0.849) \\
			& \multicolumn{2}{c}{N1} & 91.7 (1.422) & 93.7 (1.481) & 94.9 (1.502) & 95.1 (1.503) &  & 94.7 (1.140)                        & 94.9 (0.843) \\
			& \multicolumn{2}{c}{N2} & 96.4 (1.669) & 95.5 (1.567) & 95.5 (1.537) & 95.5 (1.526) &  & 95.3 (1.165)                        & 95.1 (0.851) \\
			& {\small p.-OBSP}    &            & 93.5 (1.516) & 94.5 (1.524) & 95.3 (1.523) & 95.3 (1.516) &  &  95.1 (1.154) & 94.3 (0.848)
	\end{tabular}}	
	\vspace{-0.3cm}
\end{table}
Finally, we study the performance of CI for mixed effects which are linear combinations of fixed and random effects; the latter are partly intended to smooth model misspecifications. Table \ref{tab:PoSI_1} shows coverage probabilities and lengths for post-cAIC (p.-cAIC), post-OBSP (p.-OBSP) and naive confidence intervals constructed using the first-(N1) and the second-(N2) order correct MSE estimators 
under selection matrix $\bm{\upsilon}^2_{all}$. CP and L were calculated as an average over the simulation runs and mixed parameters. Regarding post-cAIC confidence intervals, they attain a nominal coverage or suffer from a minor undercoverage when a sample size is small. The performance of post-OBSP intervals is similar. In addition, in both cases the intervals are very often narrower than in the case of a naive method N2. Yet, the most striking feature is a surprisingly good performance of the second-order naive intervals. They almost always reach the nominal level and they are only slightly wider than the post-selection intervals. Although the naive CI are not theoretically valid, because they ignore the selection step in their asymptotic distributions, it seems that the are extremely robust to this misspecification. 

\section{Post-cAIC inference of income data in Galicia}\label{sec:data_example}
We illustrate our post-cAIC procedure by constructing confidence intervals for the average of rescaled household incomes in 52 counties of Galicia in north-western Spain. We make use of the 2015 Structural Survey of Homes in Galicia (SSHG) with 9203 households, yet with certain areas where the number of units is small with $m_i<20$, see \cite{reluga2021simultaneous} for a detailed study about household income on the original scale using the same data set. Galicia is subdivided into 53 counties (\textit{comarcas}), but the data were not collected in county Quiroga. The SSHG contains covariates on different sources of income, personal characteristics (for example, age, education level) as well as information on the household status (such as number of household members, mortgage situation, etc.). The originally observed $y_{dj}$ is the yearly family household income per capita. It is well known that income data are right skewed, and our dependent variable exhibits this feature too. As our theory relies on normality, we follow a standard practice in the SAE literature and transform it by $y_{l,dj} = \log(y_{dj}+c)$, where constant $c>0$ minimises the Fisher skewness of the model residuals with $y_{l,dj}$ as a response. Constant $c$ is selected from a grid within the range of household incomes $[\min(y_{dj}), \max(y_{dj})]$ \cite[the same approach was used, among others, by][]{marhuenda2017poverty}. We analyse two estimators for the cluster-level means of the household income which are popular in SAE: EBLUP of a mixed parameter and a linear combination of the estimated regression parameters in \eqref{eq:mu}. The latter is called the regression-synthetic estimator in the survey statistic and SAE literature \citep[][]{rao2015small}. More specifically, we consider $\hat{\mu}_i=	\hat{\bar{\bm{X}}}^{dir}_{i}\bm{\beta}+\hat{\bm{u}}_{i}$ and  $\hat{\mu}^{F}_i=	\hat{\bar{\bm{X}}}^{dir}_{i}\bm{\beta}$, where $\hat{\bar{\bm{X}}}^{dir}_{i}$ is the official estimate of covariate means which we calculate from the SSHG (the details of the calculations are deferred to the SM). As was illustrated in Sections \ref{sec:sec_nested} and \ref{sec:sec_overlap}, the initial set of models is crucial in the post-cAIC inference. We therefore did not use all possible covariates in the SSHG. In contrast, we selected eight covariates which are the most correlated to the outcome variable. We then constructed 16 models, and each of them contained an intercept and a subset of four covariates with the highest correlation  (correlation coefficients, the inclusion of covariates in considered models and the results of the selection criteria can be found in our SM). cAIC selected Model 1 with an intercept and four covariates whereas OBSP of \cite{sugasawa2019observed} Model 12 with an intercept and seven covariates.

\begin{figure}
\centering
\includegraphics[width=0.9\textwidth]{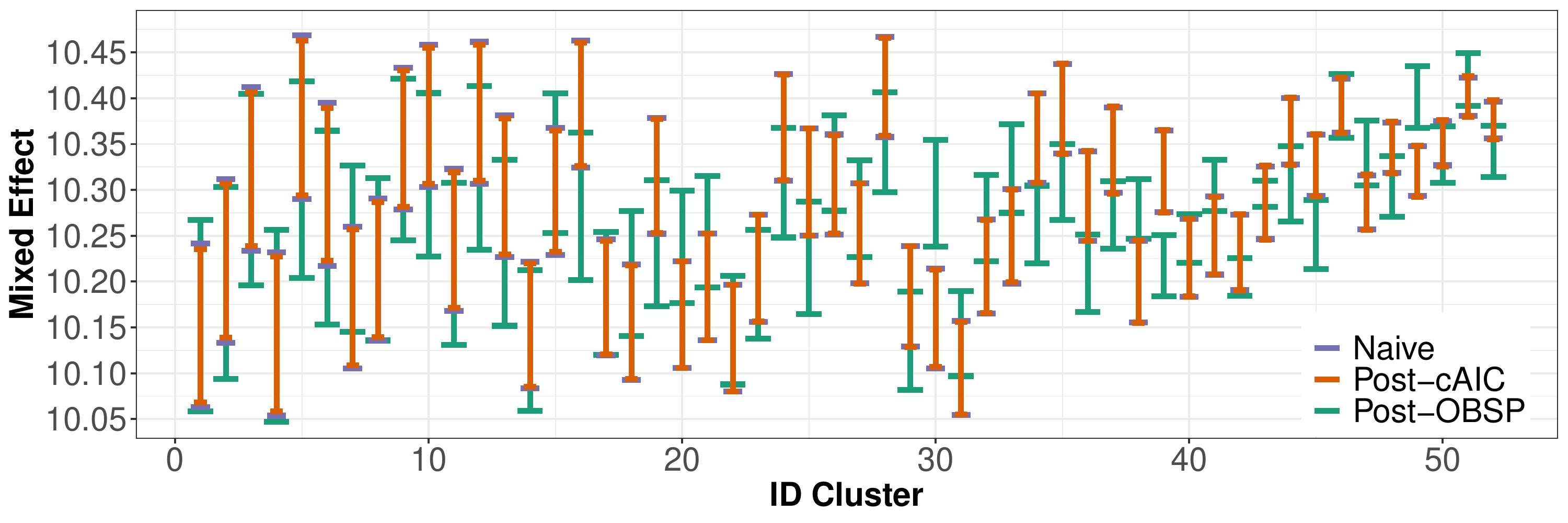}
\vspace{-0.3cm}
\caption{\label{fig:PoSI_int_data} Post-cAIC, post-OBSP and naive confidence intervals for the EBLUPs of the county-level averages of transformed household income in Galicia.}
\vspace{-0.48cm}
\end{figure}

\begin{figure}
\centering
\includegraphics[width=0.9\textwidth]{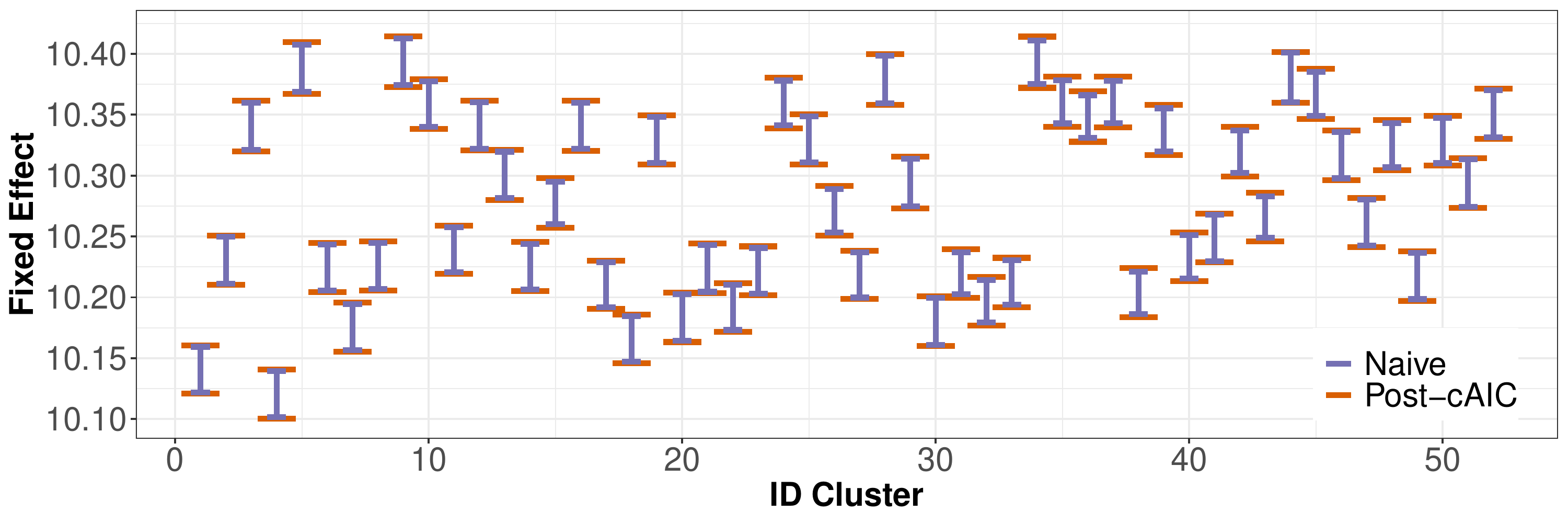}
\vspace{-0.3cm}
\caption{\label{fig:fixed} Post-cAIC and naive confidence intervals for  the regression-synthetic estimates of the county-level averages of transformed household income in Galicia.}
\vspace{-0.5cm}
\end{figure}

Figure \ref{fig:PoSI_int_data} presents naive confidence intervals constructed using the second-order correct MSE, post-cAIC and post-OBSP confidence intervals for the mixed parameter. We did not plot naive CI with first-order correct MSE because they were indistinguishable from the second-order intervals. First, some of the post-OBSP confidence intervals do not overlap with naive or  post-cAIC intervals, because distinct models were selected by cAIC and OBSP. In the majority of counties, the post-cAIC CI are narrower than their naive and post-OBSP counterparts. This conclusion is confirmed by the descriptive statistics shown in our SM and in accordance with our simulation findings.
Different widths of naive and post-cAIC intervals are related to the sample size of each county.
Figure \ref{fig:fixed} shows naive and post-cAIC confidence intervals for the synthetic-regression estimates. In contrast to Figure \ref{fig:PoSI_int_data}, post-cAIC CI are wider than naive CI.  Even though the difference between post-cAIC and naive intervals in this study is minor, the latter have a tendency to undercover because they do not account for the model selection, cf. Table \ref{tab:PoSI_betax}.

\section{Discussion}\label{sec:ch4_discussion}

We developed the asymptotic theory for post-cAIC inference. We employed our theoretical derivation to construct post-cAIC confidence intervals for mixed and fixed parameters under LMM. To the best of our knowledge, this is the first contribution which addresses post-selection inference under a general LMM framework. We tested finite sample properties of our proposal in simulations and a data example.  
In simulation scenarios, our post-cAIC CI performed well in terms of the coverage probability and average length. 
In contrast, the naive intervals performed very poorly in the numerical analysis of fixed parameters. Surprisingly, though, naive intervals for mixed parameters yielded satisfactory results. This demonstrates their robustness to possible model misspecifications and may justify their usage among practitioners. Nevertheless, we believe that theoretically valid methods, which are generally applicable, should always be preferred if they perform equally good as naive methods and they are not too intricate to implement. In follow-up studies, more extensive simulations will be needed to thoroughly examine the startling feature of naive intervals.

Finally, we developed post-cAIC after model selection with cluster focus, using the cAIC of \cite{kubokawa2011conditional}. The post-cAIC methodological advancements might be put forward in a similar way for other conditional Akaike information criteria, because the majority of them is composed of twice the conditional log-likelihood and a penalty 
function. Consider, for example, the cAIC of \cite{srivastava2010conditional}, that is
$	\mathrm{cAIC}_{SK}=-2\ell^c_n\{\hat{\bm{\beta}}_{m}\}+
(2n[ \mathrm{\rm{tr}}\{ (\bm{X}, \bm{Z}) \hat{\bm{H}}\}+1])/
(n -\mathrm{rank}\{(\bm{X}, \bm{Z})\}-2)$.
To derive the selection region and hence post-$\mathrm{cAIC}_{SK}$ CI, we could follow analogous steps as those for the cAIC of \cite{vaida2005conditional} and modify the penalty function. A comprehensive account of the conditional Akaike criteria for which post-selection analysis is similar to ours is included in the review paper of \cite{muller2013model}.

\section{Technical details}\label{sec:ch4_App}

\subsection{Assumptions}\label{sec:assump}

We denote by  $\mathcal{B}_K(\varepsilon)$ an $(a+K)$-dimensional sphere centred at $\bm{\beta}_0$ with radius $\varepsilon$, and by $\mathcal{B}^c_K(\varepsilon)$ its complement. In addition, $(\cdot)$ stands for $c$ or $m$ which refer to a conditional or a marginal framework.
\begin{enumerate}[itemsep=1.5pt, nosep]
		\item For each $\varepsilon_{(\cdot)}>0$, as $ n\rightarrow \infty$, $\sup_{\bm{\beta}\in \mathcal{B}^c_K(\varepsilon_{(\cdot)}) } \{ \ell_n^{(\cdot)}(\bm{\beta})-\ell_n^{(\cdot)}(\bm{\beta}_0) \}\rightarrow -\infty$ in probability.
		\item There exists $\varepsilon_{(\cdot)0}>0$ such that $\ell_n^{(\cdot)}(\bm{\beta})$ is twice continuously differentiable in  $\mathcal{B}_K(\varepsilon_{0(\cdot)})$ for all large $n$. We define the score vector $\mathcal{R}_{n}^{(\cdot)}(\bm{\beta})=(\partial/\partial \bm{\beta})\ell_n^{(\cdot)}(\bm{\beta})$ and the negative Hessian matrix  $\mathcal{J}_{n}^{(\cdot)}(\bm{\beta})=-(\partial^2/\partial \bm{\beta} \partial \bm{\beta}^t)\ell_n^{(\cdot)}(\bm{\beta})$.
		\item For some $0<\varepsilon_{(\cdot)1}<\varepsilon_{(\cdot)0}$, as $n\rightarrow \infty$, there exist nonrandom positive-definite continuous matrices $\mathcal{J}^{(\cdot)}(\bm{\beta})$ such that for $\bm{\beta}$ in $\mathcal{B}_K(\varepsilon_{(\cdot)1})$ $\sup_{\bm{\beta}\in \mathcal{B}_K(\varepsilon_{(\cdot)1}) } tr\{\mathcal{J}_{n}^{(\cdot)}(\bm{\beta})/n-\mathcal{J}^{(\cdot)}(\bm{\beta})\}\rightarrow 0$ in probability.
		\item $\sqrt{n}\mathcal{R}_{n}^m(\bm{\beta}) \rightarrow N\{0,\mathcal{I}^{m}(\bm{\beta}_0)\}$ in distribution once $n\rightarrow\infty$.
		\item Orthogonality of the models under $\mathrm{cAIC}$: for $M_i, M_j\in\mathcal{M}_o$, $i\neq j$, we have that $\mathcal{I}^c_{ij}=\mathbb{E}( \{\partial/\partial \bm{\beta}(M_i)\}[\ell_{c}\{\bm{\beta}(M_i)\}]  \{\partial/\partial \bm{\beta}(M_j)\}[\ell_{c}\{\bm{\beta}(M_j)^t\}]  )=\bm{0}_{|M_i|\times |M_j|}$, where the expectation is taken with respect to the true model.
	\end{enumerate}

The derivation of the cAIC of \cite{kubokawa2011conditional}, and first- and second-order correct MSE estimators require additional regularity conditions. Since we do not use them explicitly in the following derivations, they are deferred to the SM together with algebraic derivations, the proof of Lemma \ref{lemma:oversel}, and details on the structure of matrices $\bm{K}$ and $\bm{K}^{-1}$.

\subsection{Asymptotic post-selection derivations}\label{sec:ch4_App_PoSI}

\subsubsection{Statement and proof of Lemma \ref{lemma:joint_con}}
We need to guarantee a joint convergence of estimators which is obtained in Lemma \ref{lemma:joint_con}.
\begin{lemma} \label{lemma:joint_con}
Suppose that Assumptions $(a)-(b)$ in Section \ref{sec:assump} are valid. For any fixed ordering of $\mathcal{M}_o$, we denote by $o=|\mathcal{M}_o|$ the size of $\mathcal{M}_o$. It follows that $n^{1/2}(\hat{\bm{\beta}}_{m,\mathcal{M}_o}-\bm{\beta}_{0,\mathcal{M}_o})=n^{1/2}\{\hat{\bm{\beta}}_{m}(M_1)^t-\bm{\beta}_{0}(M_1)^t, \dots,\hat{\bm{\beta}}_{m}(M_o)^t-\bm{\beta}_{0}(M_o)^t\}^t\rightarrow N(\bm{0},\bm{T}_{\mathcal{M}_o})$ in distribution, where $\bm{T}_{\mathcal{M}_o}$ is partitioned such that $\mathcal{I}_{ij}\{\bm{\beta}_0(M_i),\bm{\beta}_0(M_j)\}$ is the $(i,j)$th block.
\end{lemma}
The proof follows from the Taylor expansion applied to $\ell^m_n$ as 
in \cite{charkhi2018asymptotic}, that is
$0=n^{-1/2}\mathcal{R}_{n}^m\{\bm{\beta}_0(M_i)\}+\mathcal{J}_{n}^m\{\bm{\beta}_0(M_i)\}n^{1/2}\{\hat{\bm{\beta}}_{m}(M_i)^t-\bm{\beta}_{0}^t\}+o_p(1)$,\quad $M_i\in \mathcal{M}_o$.
The asymptotic distribution of the estimators is immediate using the multivariate central limit theorem
$n^{-1/2}\left[\mathcal{R}_{n}^{mt}\left\{\bm{\beta}_0(M_1)\right\},\dots,\mathcal{R}_{n}^{mt}\left\{\bm{\beta}_0(M_o)\right\} \right]^t\xrightarrow[] {n\rightarrow\infty}N(\bm{0},\bm{T}_{\mathcal{M}_o})$.
\vspace{-0.4cm}
\subsubsection{Proof of Proposition \ref{prop:prop_nested1}} \label{sec:prop_nested1} By Lemma \ref{lemma:joint_con}, there is a joint convergence of estimators $\hat{\bm{\beta}}_m$ from different models. Geometrical regions can be defined by pairwise comparisons of the $\mathrm{cAIC}'$ values. Therefore, for $j=p_0,\dots, a+K$ we write
\begin{eqnarray}\label{eq:l_c_T}
\ell^c_n\{\bm{\beta}_0(j)\} &=& \ell^c_n\{\hat{\bm{\beta}}_{m}(j)\}+
n^{1/2}\{\bm{\beta}_0(j)-\hat{\bm{\beta}}_{m}(j)\}^t\mathcal{R}_{n,j}^c\{\hat{\bm{\beta}}_{m}(j)\}
\\&& +
\frac{n}{2}\{\bm{\beta}_0(j)-\hat{\bm{\beta}}_{m}(j)\}^t\mathcal{J}_{j}^c
\{\hat{\bm{\beta}}_{m}(j)\}\{\bm{\beta}_0(j)-\hat{\bm{\beta}}_{m}(j)\}+o_p(1).\nonumber
\end{eqnarray}
We focus on the second term on the right-hand side in the first line of \eqref{eq:l_c_T}. Considering the model with a full set of parameters and by condition $(b)$ from Section \ref{sec:assump}, we have
$\mathcal{R}_{n,K}^c(\bm{\beta})=\frac{\partial \ell_{n,K}^c}{\partial \bm{\beta}}=\bm{X}^t\bm{R}^{-1}\bm{y}
	-\bm{X}^t\bm{R}^{-1}\bm{X}\bm{\beta}-\bm{X}^t\bm{R}^{-1}\bm{Zu},$\;
$	\mathcal{J}_{n,K}^c(\bm{\beta})=\frac{\partial^2 \ell_{n,K}^c}{\partial \bm{\beta}\partial \bm{\beta}^t}=\bm{X}^t\bm{R}^{-1}\bm{X}$.
We thus obtain $\bm{\beta}_c=(\bm{X}^t\bm{R}^{-1}\bm{X})^{-1}(\bm{X}^t\bm{R}^{-1}\bm{y}-\bm{X}^t\bm{R}^{-1}\bm{Zu})$ and $\tilde{\bm{\beta}}_c(\bm{\theta})$ with $\bm{u}$ replaced with $\tilde{\bm{u}}$. On the other hand, it follows that
\begin{equation}\label{eq:beta_dist}
	n^{1/2}\{\bm{\beta}_0(j)-\hat{\bm{\beta}}_{m}(j)\}\rightarrow
	N[\bm{0},\{\mathcal{I}_{j}^{m}(\bm{\beta}_0)\}^{-1}] \text{ in distribution},
\end{equation}
\begin{equation}\label{eq:l_c_l_ml}
		\mathcal{R}_{n,j}^c\{\hat{\bm{\beta}}_{m}(j)\}=
		\mathcal{R}_{n,j}^c\{\hat{\bm{\beta}}_{c}(j)\}+
		n^{1/2}\{\hat{\bm{\beta}}_{m}(j)-\hat{\bm{\beta}}_{c}(j)\}^t\mathcal{J}_{j}^c\{\hat{\bm{\beta}}_{c}(j)\}+o_p(1).
\end{equation}
Since we used maximum likelihood to estimate $\hat{\bm{\beta}}_{c}(j)$, the first term on the right-hand side of \eqref{eq:l_c_l_ml} is $\mathcal{R}_{n,j}^c\{\hat{\bm{\beta}}_{c}(j)\}=0$. Similarly the second term is also 0 which is a consequence of algebraic derivations in the SM. Moreover, $\mathcal{J}_{n,j}^c$ does not include the random effect. Since $\hat{\bm{\theta}}-\bm{\theta}=O_p(n^{-1/2})$, the convergence in probability of  $\rho(\hat{\bm{\theta}})$, $b(\hat{\bm{\theta}})$ to $\rho(\bm{\theta})$,  $b(\bm{\theta})$ follows from the continuous mapping theorem, and the rate $O_p(n^{-1/2})$ was proven in Theorem 2.3 of \cite{kubokawa2011conditional}. Combining results of \eqref{eq:l_c_T}, \eqref{eq:beta_dist} and \eqref{eq:l_c_l_ml}, it follows that
\begin{equation}\label{eq:l_c_T2}
\begin{split}
2l'^{c}_{n,j}&=-2[\ell_n^c\{\bm{\beta}_0(j)\}-\ell_n^c\{\hat{\bm{\beta}}_{m}(j)\}]\\&=
n\{\bm{\beta}_0(j)-\hat{\bm{\beta}}_{m}(j)\}^t\mathcal{J}_{j}^c\{\hat{\bm{\beta}}_{m}(j)\}\{\bm{\beta}_0(j)-\hat{\bm{\beta}}_{m}(j)\}+o_p(1)\\
&\xrightarrow[]{n\rightarrow \infty}\bm{W}^{st}_j\bm{\Sigma}(M_j)\bm{W}^{s}_j,
\end{split}
\end{equation}
where $\bm{W}^{s}_j$, $\bm{\Sigma}(M_j)$ as defined in Section \ref{sec:sec_nested}. Furthermore, observe that $\mathrm{cAIC}'(M_p)-\mathrm{cAIC}'(M_j)\geqslant0$ is equivalent to $2(l'^{c}_{n,p}-l'^{c}_{n,j})\geqslant2(\hat{\rho}_{pj}+\hat{b}_{pj})$. On the other hand, applying Lemma \ref{lemma:joint_con} and the reasoning above we have a joint convergence of $(2l'^{c}_{n,p_0},\dots, 2l'^{c}_{n,K})$ to $(\bm{W}^{st}_{p_0}\bm{\Sigma}(M_{p_0})\bm{W}^{s}_{p_0}\dots,\bm{W}^{st}_K\bm{\Sigma}(M_K)\bm{W}^{s}_K)$, that is, $2l'^{c}_{n,(\cdot)}$ converges to a scaled, chi-square distribution, where $\bm{W}^{s}_{(\cdot)}$ is distributed according to a multivariate normal distribution, and $(\cdot)$ stands for $p_0, \dots, K$. As a result, the difference between two conditional likelihoods corresponds to the difference between these two chi-square random variables, and can be written using a summation sign. We use these sums to define the selection region $\mathcal{A}_p(M_{nest})$, and by continuous mapping theorem we have $(\bm{W}_1,\dots, \bm{W}_K)\in \mathcal{A}_p(M_{nest})$. 

We concentrate now on the post-cAIC CI for a mixed parameter $\mu_{i}$. Since $\bm{u}\in\mathbbm{R}^{r}$ is not subject to the selection process, no geometrical restrictions are imposed on the support of the asymptotic normal distribution of random effects. Hence, 
to construct $\mathcal{A}^{\mu}_{p}(\mathcal{M}_{nest})$ we need to enlarge the dimensionality of $\bm{w}$ to account for random effects, that is $\bm{w}\in \mathbbm{R}^{a+K+r}$. 
We have $\hat{\mu}_{i}(p)-\mu_{i}(p)=\tilde{\mu}_{i}(p)-\mu_{i}(p) + O_p(n^{-1/2})$ and the first term on the right hand side converges to a normally distributed random variable. Denote $\mathcal{E}=\{p_0,\dots,a+K\}$ and consider the selection of model $p$:
\begin{eqnarray}
&&\lim\limits_{n\rightarrow\infty}\frac{P\left[\left(n^{1/2}\left[\{\bm{c}^{s}_i(p)\}^t
\{ \hat{\bm{\varrho}}(p)-\bm{\varrho}(p)\}\right]
<t
\right)\cap
\left\{ 2(l'^{c}_{n,p}-l'^{c}_{n,j})\geqslant2(\rho_{pj}+b_{pj}), j \in \mathcal{E}
\right\}
\right]  }{P\left\{ 2(l'^{c}_{n,p}-l'^{c}_{n,j})\geqslant 2(\rho_{pj}+b_{pj}), j \in \mathcal{E}
\right\}} \nonumber\\
&&\approx
\frac{P\left(
\left[\{\bm{c}^s_{i}(p)\}^t\bm{K}^{-1/2}(p) \bm{W}^{s}_{\mu}(p)\leqslant t
\right]\cap
\left\{\bm{W}_{\mu}\in\mathcal{A}^{\mu}_{p}(\mathcal{M}_{nest})
\right\}
\right)  }{P\left\{\bm{W}_{\mu}\in\mathcal{A}^{\mu}_{p}(\mathcal{M}_{nest})
\right\}}\label{eq:cond_nest}\\
&&=
P\left[  \{\bm{c}^{s}_{i}(p)\}^t\bm{K}^{-1/2}(p) \bm{W}^{s}_{\mu}(p)\leqslant t|\bm{W}_{\mu}\in \mathcal{A}^{\mu}_{p}(\mathcal{M}_{nest})   \right],\nonumber
\end{eqnarray}
where the second line is a consequence of $\{ \hat{\bm{\varrho}}(p)-\bm{\varrho}(p)\}\approx N\left\{ \bm{0},\bm{K}^{-1}(p)\right\}$.
\vspace{-0.4cm}

\subsubsection{Proof of Proposition \ref{prop:prop_overparam}}

Similarly as for $\mathcal{M}_{nest}$, we calculate the set with constraints by pairwise comparisons of the $\mathrm{cAIC}'$ values. Therefore we slightly rewrite the expression in \eqref{eq:l_c_T2}. Consider $2l'^{c}_{n,M_i}=-2\{\ell_n^c(\bm{\beta}_0(M_i))-\ell_n^c(\hat{\bm{\beta}}_{m})\}$, $\mathcal{M}_i\in\mathcal{M}_{o}$. It follows that $\mathrm{cAIC}'(M_{\mathrm{cAIC}})-\mathrm{cAIC}'(M_i)\geqslant0$ which implies $2(l'^{c}_{n,M_{\mathrm{cAIC}}}-l'^{c}_{n,{M}_i})\geqslant2(\rho_{M_{\mathrm{cAIC}}}-\rho_{M_{i}})+2(b_{M_{\mathrm{cAIC}}}-b_{M_{i}})$. The region in Proposition \ref{prop:prop_overparam} (I) is defined using the extended selection matrix and an analogous analysis as in \eqref{eq:cond_nest}. Regarding Proposition \ref{prop:prop_overparam} (II), observe that
\begin{equation}\label{eq:region_overparm}
	n(\hat{\bm{\beta}}_{m,\mathcal{M}_o}-\bm{\beta}_{0,\mathcal{M}_o})^t\bm{B}_{\mathrm{cAIC},i}(\hat{\bm{\beta}}_{m,\mathcal{M}_o}-\bm{\beta}_{0,\mathcal{M}_o})+o_p(1)\geqslant 2(\rho_{M_{\mathrm{cAIC}}}-\rho_{M_{i}} + b_{M_{\mathrm{cAIC}}}-b_{M_{i}}),
\end{equation}
where $\bm{B}_{\mathrm{cAIC},i}$ is a diagonal matrix with blocks $\mathcal{J}^c_{M_{\mathrm{cAIC}}}$ and $-\mathcal{J}^{c}_{M_{i}}$ corresponding to models $M_{cAIC}$ and $M_{i}$. $\bm{J}^m_{o}$ is a diagonal matrix with the same structure as $\bm{B}_{\mathrm{cAIC},i}$ with blocks $\mathcal{I}^m_{M_{\mathrm{cAIC}}}$, $\mathcal{I}^m_{M_i}$. Using Lemma \ref{lemma:joint_con} and a continuous mapping theorem, equation \eqref{eq:region_overparm} equals asymptotically $\bm{W}^t(\bm{J}^m_{o})^{-1/2}\bm{B}_{\mathrm{cAIC},i}(\bm{J}^{m}_{o})^{-1/2}\bm{W}\geqslant2(\rho_{M_{\mathrm{cAIC}}}-\rho_{M_{i}})+2(b_{M_{\mathrm{cAIC}}}-b_{M_{i}})$.

To obtain a selection region for a mixed parameter, denote by $\bm{A}_{\mathrm{cAIC},i}$ a diagonal matrix with two blocks, that is
\begin{equation}\label{eq:matrix_A_gen}
	\bm{A}_{\mathrm{cAIC},i}=
	\begin{pmatrix}
		\bm{A}_{1\mathrm{cAIC},i} & \bm{0}\\
		\bm{0} & \bm{A}_{\mathrm{2cAIC},i}
	\end{pmatrix}
	=
	\begin{pmatrix}
		(\bm{J}^m_{o})^{-1/2}\bm{B}_{\mathrm{cAIC},i}(\bm{J}^{m}_{o})^{-1/2} & \bm{0}\\
		\bm{0} & \bm{I}_r
	\end{pmatrix}.
\end{equation}
Finally, we need to account for the lack of orthogonality multiplying $\bm{A}_{\mathrm{cAIC},i}$ by $\bm{E}$ which leads to $\bm{W}_{\mu}^t\bm{E}^{1/2}\bm{A}_{\mathrm{cAIC},i}\bm{E}^{1/2}\bm{W}_{\mu} \geqslant 2(\rho_{M_{\mathrm{cAIC}}}-\rho_{M_i})+2(b_{M_{\mathrm{cAIC}}}-b_{M_{i}})$.
\vspace{-0.4cm}

\subsubsection{Proof of Lemma \ref{lemma: misspec}}
The result follows from a uniform version of the Lindeberg-Feller central limit theorem and the continuous mapping theorem demonstrated by \cite{kasy} for a general vector and applied by \cite{charkhi2018asymptotic} to a likelihood based model. The same arguments outlined in the latter are valid within our 
settings with misspecified models.
\vspace{-0.5cm}

\subsubsection{Proof of Proposition \ref{prop:uniform_conv}}
The proof of Proposition \ref{prop:uniform_conv} proceeds along identical steps as the proof of Proposition 4 in \cite{charkhi2018asymptotic} changing the types of likelihoods and parameters of interest. Consider $\alpha=[n^{1/2}\{\bm{c}^s_{i}(M_{cAIC})\}^t\{\hat{\bm{\varrho}}_e(M_{cAIC})-\bm{\varrho}'_{0}(M_{cAIC})\}]$ and $\gamma=\bigcap_{M\in\mathcal{M}}\{n(\hat{\bm{\varrho}}_{e,\mathcal{M}}-\bm{\varrho}'_{0,\mathcal{M}})^t(\bm{A}_{M_{cAIC},M_s}-\bm{A}_{M_{i},M_s})(\hat{\bm{\varrho}}_{e,\mathcal{M}}-\bm{\varrho}'_{0,\mathcal{M}})\} +o_P(1)$,where $\bm{A}_{M_{cAIC},M_s}$ as defined in Proposition \ref{prop:prop_overparam}. It follows that
\begin{equation}\label{eq:ost_eq}
	P\left[
	n^{1/2}\{\hat{\mu}_{i}(M_{cAIC})-\mu_{i}\}<t|M_{\mathrm{cAIC}}\right]=P(\alpha\cap\gamma)/P(\gamma).
\end{equation}
If we combine equation \eqref{eq:cond_nest} and Lemma \ref{lemma: misspec}, it follows that the difference between the expression in equation \eqref{eq:ost_eq} and
\begin{equation*}
\frac{P\left([  \{\bm{c}^s_{i}(M_{cAIC})\}^t\bm{K}^{-1/2}(M_{cAIC}) \bm{W}^{s}(M_{cAIC})<t] \cap\left\{
\bm{W}\in\mathcal{A}^{\mu}_{M_{\mathrm{cAIC}}}(\mathcal{M})    \right\}   \right)}{P\left\{
\bm{W}\in\mathcal{A}^{\mu}_{M_{\mathrm{cAIC}}}(\mathcal{M})    \right\} }
\end{equation*}
converges to 0, as stated in equation \eqref{eq:unif_conv_mu}.

\section{Supplementary material}\label{sec:sup_mat}
This section contains additional theoretical derivations and numerical results. Specifically, in Section \ref{sec:mse} and Section \ref{sec:cAIC} we present spelled-out formulas of the mean squared error and the cAIC of \cite{kubokawa2011conditional}. The selection properties of cAIC are presented in Section \ref{sec:selection_properties}. Afterwards, we present more numerical results of the simulations in Section \ref{sec:simulations_sm} and additional details on the data example in Section \ref{sec:data_example_sm}. Finally in Section \ref{sec:assump_sm} we provide an extended list of assumptions and further technical proofs. To facilitate the readability of additional results, all tables are included at the end of this document.

\subsection{MSE of mixed parameter}\label{sec:mse}

In this section we present explicitly the constituents of matrices $\bm{K}$ and $\bm{K}^{-1}$. The former follows immediately if we rewrite the mixed model equations in \eqref{eq:mom_eq_mat} using the strategy of \cite{gilmour1995average}. 
\begin{equation*}
	\begin{split}
		\bm{K}\tilde{\bm{\varrho}}=\bm{C}^t\bm{R}^{-1}\bm{y},\quad\tilde{\bm{\varrho}}&=\left(\tilde{\bm{\beta}}^t,\tilde{\bm{u}}^t\right)^t,\quad\bm{C}=\left[\bm{X}\:\bm{Z}\right],\quad\bm{K}=\bm{C}^t\bm{R}^{-1}\bm{C}+\bm{G}^{+},\\
		\bm{G}^{+}&=\begin{pmatrix}\bm{0}_{(a+K)\times (a+K)} & \bm{0}_{(p+1)\times {n}}\\
			\bm{0}_{{n}\times (p+1)} & \bm{G}^{-1}_{{n}\times {n}}\end{pmatrix}.
	\end{split}
\end{equation*}
Regarding $\bm{K}^{-1}$, we have:
\begin{equation*}
	\begin{split}
		\bm{K}^{-1}&=
		\begin{pmatrix}
			\bm{K}^{-1}_{11} & \bm{K}^{-1}_{12}  \\
			\bm{K}^{-1}_{21} & \bm{K}^{-1}_{22}
		\end{pmatrix}
		\\
		&=
		\begin{pmatrix}
			(\bm{X}^t\bm{V}^{-1}\bm{X})^{-1}\quad -(\bm{X}^t\bm{V}^{-1}\bm{X})^{-1}\bm{X}^t\bm{V}^{-1}\bm{Z}\bm{G} \\
			-(\bm{X}^t\bm{V}^{-1}\bm{X})^{-1}\bm{X}^t\bm{V}^{-1}\bm{Z}\bm{G} \quad \bm{F}+\bm{G}\bm{Z}^t\bm{V}^{-1}(\bm{X}^t\bm{V}^{-1}\bm{X})^{-1}\bm{X}^t\bm{V}^{-1}\bm{Z}\bm{G}
		\end{pmatrix},
	\end{split}
\end{equation*}
where $\bm{F}=(\bm{Z}^t\bm{R}\bm{Z}+\bm{G}^{-1})^{-1}$. The direct calculation under linear mixed models (LMM) can be found in \cite{gumedze2011parameter}.

On the other hand, the first order MSE estimator for a mixed parameter is given by
\begin{equation*}\label{eq:naive_mse_sm}
	mse_{1}(\hat{\mu}_{i})=\bm{c}^t_{i}\hat{\bm{K}}^{-1} \bm{c}_{i}=g_{1i}(\hat{\bm{\theta}})+g_{2i}(\hat{\bm{\theta}}),
\end{equation*}
where
\begin{equation*}\label{eq:MSE_b_first_term}
	\bm{m}^t_{i}(\bm{G}_{i}-\bm{G}_{i}\bm{Z}^t_{i} \bm{V}_{i}^{-1}\bm{Z}_{i}\bm{G}_{i})\bm{m}_{i}+\bm{d}_{i}^t\left(\sum_{i=1}^{n}\bm{X}^t_{i}\bm{V}^{-1}_{i}\bm{X}_{i}\right)^{-1}\bm{d}_{i} =: g_{1i}(\bm{\theta}) + g_{2i}(\bm{\theta}),
\end{equation*}
with  $\bm{d}_{i}^t=\bm{k}^t_{i}-\bm{m}^t_{i}\bm{G}\bm{Z}^t_{i}\bm{V}^{-1}_{i}\bm{X}_{i}$. In the small area estimation (SAE) literature, this estimator is called first-order correct, because $\mathbb{E}\{mse_{1}(\hat{\mu}_{i})\}=\mathrm{MSE}(\hat{\mu}_{i} )+O({n}^{-1})$. In addition, $g_{1i}$ accounts for the variability of $\tilde{\mu}_{i}$ once $\bm{\beta}$ is known, and $g_{2i}$ for the variability arising from the estimation of $\tilde{\bm{\beta}}$. An analytical second-order correct estimator is given by
\begin{equation}\label{eq:MSE_lin_sm}
	\mathrm{mse}_{2}(\hat{\mu}_{i} )= g_{1i}(\hat{\bm{{\theta}}}) + g_{2i}(\hat{\bm{\theta}}) + 2 g_{3i}(\hat{\bm{\theta}}),\quad g_{3i}(\bm{\theta})= \text{tr}\left\{ (\partial \bm{a}_{i}^t/\partial\bm{\theta}) \bm{V}_{i} (\partial \bm{a}_{i}^t/\partial\bm{\theta} )^t\bm{V}_A(\hat{\bm{\theta}}) \right\},
\end{equation}
where $\bm{a}^t_{i}=\bm{m}^t_{i}\bm{G}\bm{Z}^t_{i}\bm{V}^{-1}_{i}$ with $\bm{V}_A(\hat{\bm{\theta}})$ denoting the asymptotic covariance matrix, and  $\mathbb{E}\left\{\mathrm{mse}_{2}(\hat{\mu}_{i} )\right\}=\mathrm{MSE}(\hat{\mu}_{i} )+o({n}^{-1})$.

\subsection{cAIC of \cite{kubokawa2011conditional}}\label{sec:cAIC}

Before spelling out the exact form of cAIC, we define the derivatives with respect to $\bm{\theta}$ and differential operators with respect to $\bm{y}$
\begin{equation}\label{eq:diff_oper}
	\begin{split}
		\bm{A}_{(i)}(\bm{\theta})=\frac{\partial \bm{A}(\bm{\theta})}{\partial \theta_i}, &\quad \bm{A}_{(ij)}(\bm{\theta})=\frac{\partial^2 \bm{A}(\bm{\theta})}{\partial \theta_i \partial \theta_j}, \quad \bm{A}_{(ijk)}(\bm{\theta})=\frac{\partial^3 \bm{A}(\bm{\theta})}{\partial \theta_i \partial \theta_j \partial \theta_k},\\
		\nabla_{\bm{y}}&=\frac{\partial}{\partial \bm{y}}, \quad
		\nabla_{\bm{y}} \nabla_{\bm{y}}^t=\frac{\partial}{\partial \bm{y}}\frac{\partial}{\partial \bm{y}^t},
	\end{split}
\end{equation}	
where $\bm{A}(\bm{\theta})$ may denote a scalar, a vector or a matrix. In addition, the $i$th element of $\nabla_{\bm{y}}$ and $(i,j)$th of $\nabla_{\bm{y}} \nabla_{\bm{y}}^t$ are $\partial/\partial y_i$ and $\partial^2/\partial y_i \partial y_j$. cAIC of \cite{kubokawa2011conditional} defined in \eqref{eq:c_AIC_Kubo} includes a correction term $b(\hat{\bm{\theta}})$. We have 
\begin{equation}\label{eq:correction_cAIC}
	\begin{split}
		b(\bm{\theta})=&-\frac{1}{2}\sum_{i=1}^{h}\mathrm{\rm{tr}} \{ \bm{V}(\bm{V}^{-1}\bm{R}\bm{V}^{-1})_{(i)} \bm{V}\mathbb{E}(\nabla_{\bm{y}} \nabla_{\bm{y}}^t\hat{\theta}^*_i)\}-\sum_{i=1}^{h}\mathrm{\rm{tr}} \{ \bm{R}_{(i)}(\bm{R}^{-1}-\bm{V}^{-1})\} \mathbb{E}(\hat{\theta}_i^{**})\\
		&-\sum_{i=1}^{h}\sum_{j=1}^{h}\mathrm{\rm{tr}} \left[\frac{1}{2} \bm{R}_{(ij)}(\bm{R}^{-1}-\bm{V}^{-1})+\bm{R}_{(i)} \{(\bm{R}^{-1})_{(j)}-(\bm{V}^{-1})_{(j))}\}\right] \mathbb{E}(\hat{\theta}_i^{*}\hat{\theta}_j^{*}),
	\end{split}
\end{equation}
where $\hat{\bm{\theta}}^*$, $\hat{\bm{\theta}}^{**}$ are defined in Section \ref{sec:assump} and $\hat{b}=b(\hat{\bm{\theta}})$.

\subsection{Selection properties of cAIC}\label{sec:selection_properties}

\subsubsection{Nested models}\label{sec:sec_nested_sm}

In Figure \ref{fig:nest_single}  we presented the allowable domains for random variables $W_1$, $W_2$ and $W_3$ using the restrictions imposed by the cAIC selection. These figures were plotted based on our simulated dataset with $y_{ij}=\sum_{d=1}^{3}\beta_dx_{dij}+u_{i}+e_{ij}$, $n=30$, $m_i=5$, $\sigma^2_u=1$, $\sigma^2_e=0.5$ and $\bm{\beta}=(2.25, -1.1, 2.43)$. Under this model, the expression to estimate $b$ in equation (\ref{eq:correction_cAIC}) is substantially simplified and spelled out in \cite{kubokawa2011conditional}. We estimated $\bm{\Sigma}$ using empirical versions of $\bm{V}$ and $\bm{R}$ defined in Section \ref{sec:inference_LMM}, and we obtained (the numbers are rounded to 3 digits)
\begin{equation*}
	\hat{\bm{\Sigma}}=
	\begin{pmatrix}
		9.263 &-0.309 &-0.053\\
		-0.309 & 1.233 &-0.027\\
		-0.053 &-0.027 & 1.225
	\end{pmatrix},
\end{equation*}

This exemplary setting was chosen in a subjective way, and other choices are possible too. Table \ref{tab:cAIC_rho_examp_nest} displays estimated values of $\rho$, $b$ and cAIC for the models from sets $\mathcal{M}_{nest}$ defined above and $\mathcal{M}_{all}$ defined in Section \ref{sec:sec_overlap}. Figure \ref{fig:nest_single} depicts geometrical regions which restrict the domains of $W_1$, $W_2$ and $W_3$. The regions are defined by the appropriate equations from Section \ref{sec:sec_nested} applied to the selection between models $M_0$, $M_1$ and $M_2$ defined therein. Once $M_{cAIC}=M_0$, the exact sets of inequalities was derived in Section \ref{sec:sec_nested}. In addition, it was depicted in panel (1) of Figure \ref{fig:nest_single}. If $M_{cAIC}=M_1$, we have
\begin{equation*}
	\begin{split}
		&\mathcal{A}_{M_1}(\mathcal{M}_{nest})=\{\bm{w}\in \mathbb{R}^3:
		w_2^2\Sigma_{2}+2w_1w_2\Sigma_{12}\geqslant 2(\rho_{21}+b_{21}),\\
		&w_3^2\Sigma_{3}+2w_1w_3 \Sigma_{13}+2w_2w_3\Sigma_{23}<2(\rho_{32}+b_{32}) \},
	\end{split}
\end{equation*}
which is presented in panel (2) of Figure \ref{fig:nest_single}. Finally, for $M_{cAIC}=M_2$, we have
\begin{equation*}
	\begin{split}
		&\mathcal{A}_{M_2}(\mathcal{M}_{nest})=\{\bm{w}\in \mathbb{R}^3:
		w_3^2 \Sigma_{3}+ 2w_1w_3 \Sigma_{13}+ 2w_2w_3\Sigma_{23}  \geqslant 2(\rho_{32}+b_{32}),\\&w_2^2 \Sigma_{2}+w_3^2 \Sigma_{3} + 2w_1w_3 \Sigma_{13} + 2w_2w_3 \Sigma_{23}+2w_1w_2 \Sigma_{12}\geqslant 2(\rho_{31}+b_{31}) \},
	\end{split}
\end{equation*}
which is illustrated in panel (3) of Figure \ref{fig:nest_single}.

\begin{figure}[htb]
	\centering
	\includegraphics[width=0.6\textwidth]{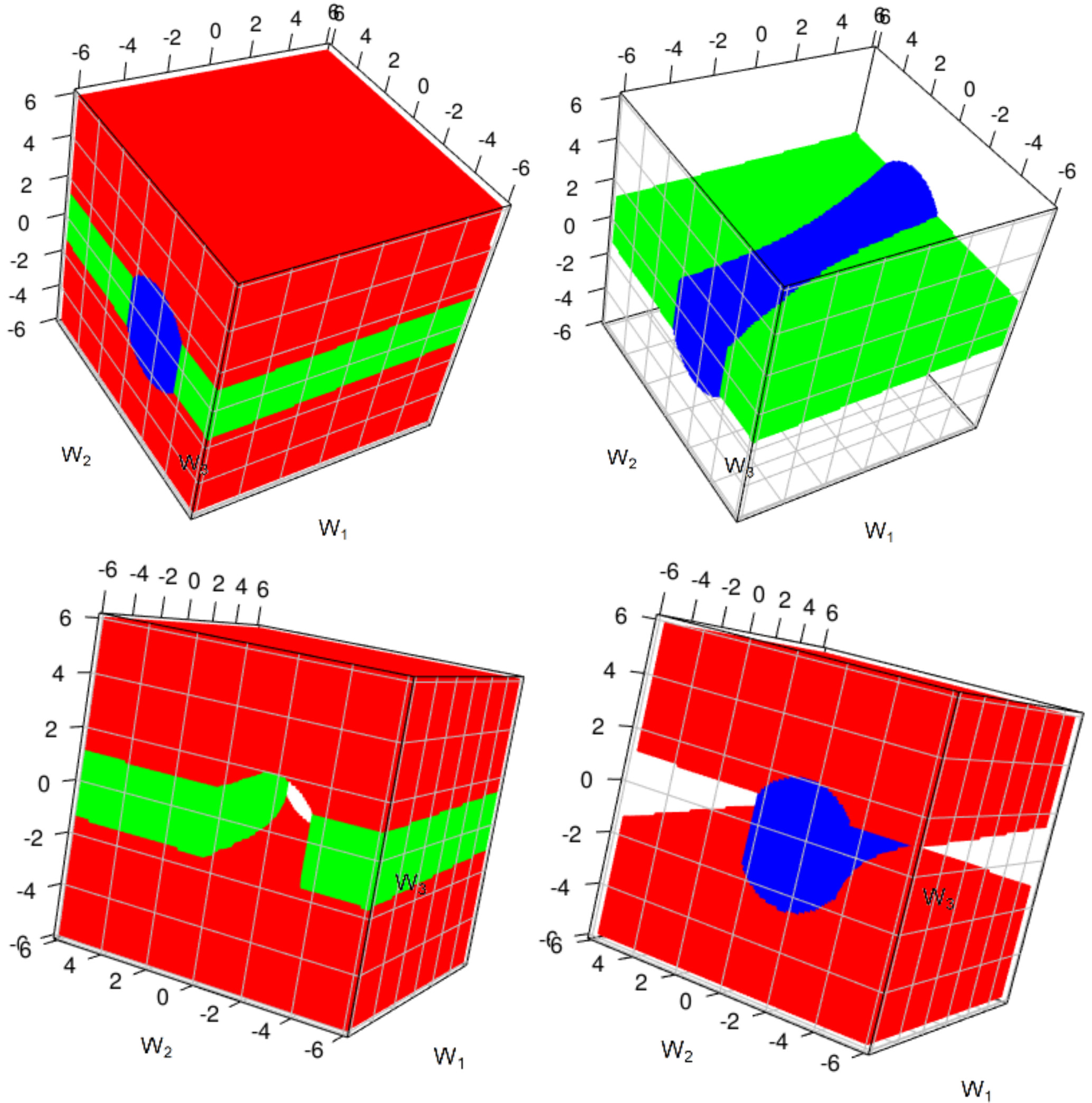}
	\caption{\label{fig:nest_mult} Allowable domains of $W_1$, $W_2$ and $W_3$ for nested model selection when cAIC selects: (top, left) $M_0$ or $M_1$ or $M_2$, (top, right) $M_1$ or $M_2$, (bottom, left) $M_2$ or $M_3$, (bottom, right) $M_1$ or $M_3$.}
	\vspace{-0.5cm}
\end{figure}

Figure \ref{fig:nest_mult} shows the partition of 3-dimensional space composed of $W_1$, $W_2$ and $W_3$. In the remaining three panels, we provide the  partition of the space made by two of selected models. The purpose of Figure \ref{fig:nest_mult} is to show that the restrictions  partition the 3-dimensional space and that there is no overlap. Of course, in practice we select only one model which would correspond to only one of these regions.

\begin{figure}[htb]
	\centering
	\makebox{\includegraphics[width=0.8\textwidth]{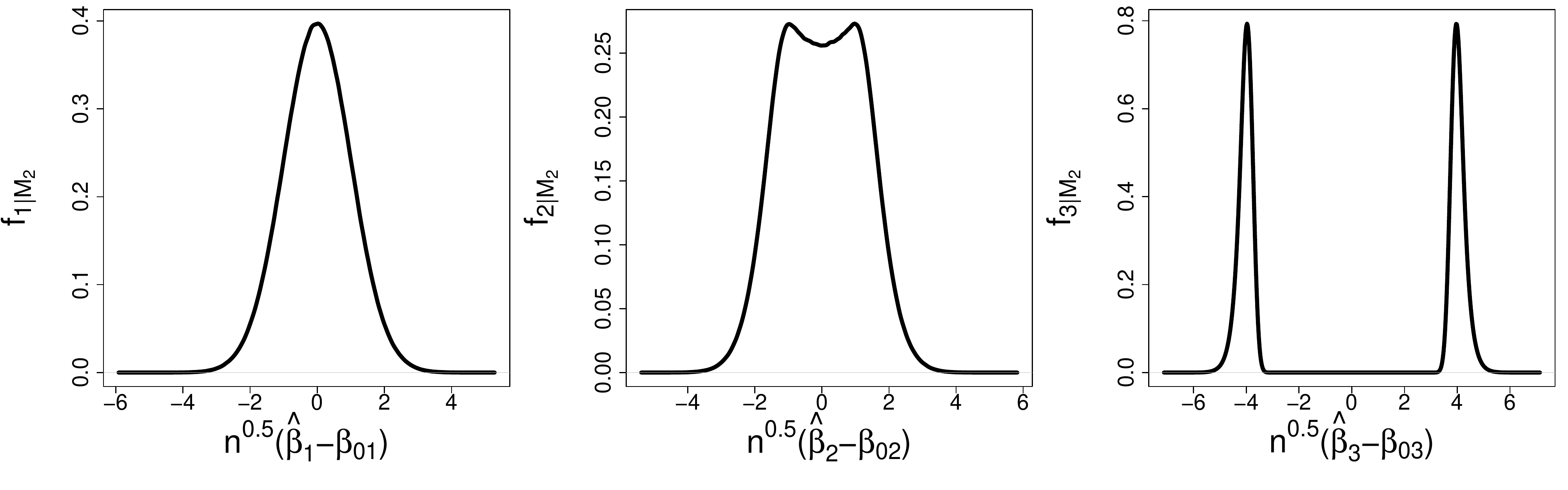}}
	\caption{\label{fig:post_sel_dens} Marginal post-selection densities $f_{j|M_2}$ of $n^{1/2}(\hat{\beta}_d-\beta_{0d})$, conditional on $\hat{p}_0=3$ when $p_0=1$ ($d=1,2,3$).}
\end{figure}

Finally, we illustrate the effect of the selection on the limiting densities. Figure \ref{fig:post_sel_dens} depicts post-selection densities of $\bm{\beta}_1$, $\bm{\beta}_2$ and $\bm{\beta}_3$ assuming that we select model $M_2$ with $\hat{p}_0=3$, whereas the true model is $M_0$ with $p_0=1$ \citep[see Figure 3 in][which depicts AIC post-selection densities]{charkhi2018asymptotic}. We can immediately notice several features. First of all, the asymptotic density of $n^{1/2}(\hat{\beta}_1-\beta_{01})$ does not seem to be affected, which is plausible since the true model includes $\beta_{01}$. Furthermore,  $n^{1/2}(\hat{\beta}_2-\beta_{02})$ seems to be affected only slightly in the centre of its density. In contrast, the post-selection density of $n^{1/2}(\hat{\beta}_3-\beta_{03})$ is heavily influenced by the cAIC selection procedure -- the density is bimodal with much larger high quantiles than in case of the normal distribution presented in the left panel of Figure \ref{fig:post_sel_dens}. This clearly shows that the application of standard quantiles might lead to wrong conclusions, for example while constructing confidence intervals or carrying out tests.

\subsubsection{General models}\label{sec:sec_overlap_sm}

The choice of $\mathcal{M}$ is of paramount importance -- it affects the distribution of all parameters, even those which are common to all models. We follow up with the example in Section \ref{sec:sec_overlap_sm}. In particular, we consider $\mathcal{M}_{all}=\{M_0, M_1, M_2, M_3 \}$ with $M_0$, $M_1$, $M_2$ as in the framework of the nested models, and $M_3=(\bm{\beta}_1,\bm{\beta}_3)$ (see Table \ref{tab:cAIC_rho_examp_nest} for specific values of estimated $\rho$, $b$ and cAIC). We derive a set of inequalities which impose the restrictions on the domain of random variables once $M_{cAIC}=M_1$, $M_2$ or $M_3$ (the inequalities under $M_{cAIC}=M_0$ were given in Section \ref{sec:sec_overlap_sm}). If $M_{cAIC}=M_1$, the allowable domain is presented in 
panel (2) of Figure \ref{fig:all_single} which was constructed using the following set of equations
\begin{equation*}
	\begin{split}
		&\mathcal{A}_{M_1}(\mathcal{M}_{all})=\{\bm{w}\in \mathbb{R}^3:
		w_2^2\Sigma_{2}+2w_1w_2\Sigma_{12}\geqslant 
		2(\rho_{M_1, M_0}+b_{M_1, M_0}),\\
		&w_3^2\Sigma_{3}+2w_1w_3 \Sigma_{13}+2w_2w_3\Sigma_{23}<
		2(\rho_{M_2, M_1}+b_{M_2, M_1}), \\
		&w_2^2\Sigma_{2}+2w_1w_2
		\Sigma_{12}-w_3^2\Sigma_{3}-2w_1w_3 \Sigma_{13}\geqslant 
		2(\rho_{M_1, M_3}+b_{M_1, M_3})\}.
	\end{split}
\end{equation*}
For $M_{cAIC}=M_2$, we have
\begin{equation*}
	\begin{split}
		&\mathcal{A}_{M_2}(\mathcal{M}_{all})=\{\bm{w}\in \mathbb{R}^3:
		w_3^2 \Sigma_{3}+ 2w_1w_3 \Sigma_{13}+ 2w_2w_3\Sigma_{23}  \geqslant 
		2(\rho_{M_2, M_1}+b_{M_2, M_1}),\\
		&w_2^2 \Sigma_{2}+w_3^2 \Sigma_{3} +
		2w_1w_3 \Sigma_{13} + 2w_2w_3 \Sigma_{23}+2w_1w_2 \Sigma_{12}\geqslant 
		2(\rho_{M_2, M_0}+b_{M_2, M_0}),\\
		&w_2^2\Sigma_{2}+ 2w_1w_2 \Sigma_{12}+ 2w_2w_3 \Sigma_{23}\geqslant 
		2(\rho_{M_2, M_3}+b_{M_2,M_3}) \},
	\end{split}
\end{equation*}
presented in panel (3) of Figure \ref{fig:all_single}. Finally, if our selection process chooses $M_{cAIC}=M_3$, it follows that
\begin{equation*}
	\begin{split}
		&\mathcal{A}_{M_3}(\mathcal{M}_{all})=\{\bm{w}\in \mathbb{R}^3:
		w_3^2 \Sigma_{3}+ 2w_1w_3 \Sigma_{13} \geqslant
		2(\rho_{M_3, M_0}+b_{M_3, M_0}),\\
		&w_2^2\Sigma_{2}+
		2w_1w_2 \Sigma_{12}+2w_2w_3 \Sigma_{23}<
		2(\rho_{M_2, M_3}+b_{M_2, M_3}), \\
		&w_3^2\Sigma_{3}+2w_1w_3 \Sigma_{13} -w_2^2\Sigma_{2}-2w_1w_2 \Sigma_{12}\geqslant 
		2(\rho_{M_3, M_1}+b_{M_3, M_1})\},
	\end{split}
\end{equation*}
which is presented in panel (4) of Figure \ref{fig:all_single}. 
\subsection{Additional results of our simulation study}\label{sec:simulations_sm}

In this section we provide the simulation results obtained for two extended selection matrices $\bm{\upsilon}^3_{all}$ and $\bm{\upsilon}^4_{all}$, i.e.\ when the first three (respectively four) covariates are included in all models. The simulation setting is the same as described in Section \ref{sec:simulations}. We consider three model sets. Selection matrix $\bm{\upsilon}^2_{all}$ correspond to a model set in which some models exclude truly nonzero covariates. In contrast, $\bm{\upsilon}^3_{all}$ describes a model set in which all models are forced to include truly nonzero covariates, whereas  $\bm{\upsilon}^4_{all}$ corresponds to a set in which we force all models to include truly nonzero parameters, but also an irrelevant covariate $\beta_4$.  
Before presenting additional simulations, we describe a practical procedure to obtain post-selected confidence intervals in a form of an algorithm:
\begin{enumerate}
	\item In a numerical study, generate a suitable dataset to fit NERM.
	\item Define the initial set of candidate models $\mathcal{M}$.
	\item Fit the model to the data and obtain consistent estimators $\hat{\bm{\beta}}_m$, $\hat{\sigma}^2_e$ and $\hat{\sigma}^2_u$ using maximum likelihood (or restricted maximum likelihood for $\hat{\sigma}^2_e$ and $\hat{\sigma}^2_u$).
	\item Estimate cAIC for all models in $\mathcal{M}$ and select model $M$ with the smallest value of the information criterion.
	\item Calculate matrices $\hat{\bm{\Sigma}}$, $\hat{\bm{K}}$ and $\hat{\mathcal{I}}^m$ which are the empirical counterparts of the matrices $\bm{\Sigma}$, $\bm{K}$ and $\mathcal{I}^m$ in the model with all parameters.
	\item Retrieve matrices $\hat{\bm{K}}(M)$ and $\hat{\mathcal{I}}^m(M)$, which are submatrices of $\hat{\bm{K}}$ and $\hat{\mathcal{I}}^m$, respectively, corresponding to model $M$.
	\item Calculate quadratic constrains that define selection regions $\mathcal{A}_M$  $\mathcal{A}_M^{\mu}$ for the set of general models in Section \ref{sec:sec_overlap}.
	\item Using for instance an R package \textit{tmg}, select $B=10000$ Monte Carlo samples from a truncated, multivariate normal distribution such that $\bm{W}^{s}_{(b)}\in \mathcal{A}_M$ and $\bm{W}^{s}_{\mu(b)}\in \mathcal{A}_M^{\mu}$, $b=1, \dots, B$.
	\item Retrieve high quantiles $c^{\mu}_i(\alpha/2)$ and $c_j(\alpha/2)$ from the empirical post-cAIC distributions of $\mu_i$ and $\bm{\beta}$, that is
	\begin{equation*}
		\begin{split}
			&\bar{\bm{c}}^s_{i}(M)^t\hat{\bm{K}}^{-1/2}(M) \bm{W}^{s}_{\mu(b)}(M),\quad i=1, \dots, {n},\quad b=1, \dots, B, \\
			&
			\{\hat{\mathcal{I}}^{m}(M)\}^{-1/2}\bm{W}^{s}_{(b)}(M),\quad b=1, \dots, B.
		\end{split}
	\end{equation*}
	\item Construct post-cAIC confidence intervals for $\mu_i$ and $\beta_j$, $j \in |M|$.
\end{enumerate}

It might be quite difficult to find starting values to sample from $(|M|+{n})$-dimensional truncated multivariate distribution which is necessary to construct post-cAIC intervals for mixed parameters. We suggest thus selecting them randomly. In addition, since the constraints are imposed only on the asymptotic distribution of the fixed effects, it seems to be more efficient to first sample from $|M|$-dimensional truncated distribution and afterwards from ${n}$-dimensional multivariate normal to mimic the asymptotic distribution of the random effects. The former serves in the construction of the post-cAIC intervals for fixed effects. These are not subject to restrictions because they are not involved in the selection procedure. Then we merge them into a matrix with the first $|M|$ columns from the truncated normal distribution and the next ${n}$ columns from the multivariate normal distribution. The aforementioned procedure is valid for all contexts we considered in Sections \ref{sec:sec_nested}, \ref{sec:sec_overlap} and \ref{sec:misspec}, that is, when assuming nested models, a general set of models and for misspecified models. 

We turn to the additional simulation results. Tables \ref{tab:PoSI_beta_1_sm}, \ref{tab:PoSI_beta_2} and \ref{tab:PoSI_beta_3} present coverage probabilities (CP) and lengths (L) for post-cAIC (p.-cAIC) and naive (N) confidence intervals for the components of fixed parameters $\beta_j$  within three different model sets which correspond to selection matrices $\bm{\upsilon}^2_{all}$, $\bm{\upsilon}^3_{all}$, and $\bm{\upsilon}^4_{all}$. 
Table \ref{tab:PoSI_beta_1_sm} completes the results in Table Table \ref{tab:PoSI_beta_1} under $\bm{\upsilon}^2_{all}$. The results for $\beta_2$ and $\beta_4$ resemble those we found for $\beta_1$ and $\beta_5$ respectively. In this case, covariate $\beta_3$ is relevant, i.e. truly nonzero, but not included in all models. Regarding Table \ref{tab:PoSI_beta_2} it is remarkable that although all truly nonzero covariates are included in each model, the naive CI fails to provide a good coverage not only for $\beta_4$ and $\beta_5$, but also for $\beta_3$. In contrast, when the irrelevant covariate $\beta_4$ is always included, Table \ref{tab:PoSI_beta_3} shows that the naive CI fails mainly for $\beta_5$ which is truly zero. In sum, while the post-cAIC intervals lead to a close to nominal coverage under all settings, the undercoverage of the naive method is the most striking feature in the tables. 
Tables \ref{tab:PoSI_betax_1}, \ref{tab:PoSI_betax_2} and \ref{tab:PoSI_betax_3} present coverage probabilities and lengths of CI for linear combinations of the components of fixed parameters under the same three selection matrices $\bm{\upsilon}^2_{all}$, $\bm{\upsilon}^3_{all}$, and $\bm{\upsilon}^4_{all}$. The performance of post-cAIC intervals is better overall than the one of the naive intervals except for the small sample sizes. Tables~\ref{tab:PoSI_2} and~\ref{tab:PoSI_3} show coverage probabilities and lengths for post-cAIC (p.-cAIC), post-OBSP (p.-OBSP) and naive intervals constructed using the first-order (N1) and the second-order (N2) correct MSE estimators for a mixed parameter under two different selection matrices $\bm{\upsilon}^3_{all}$ and $\bm{\upsilon}^4_{all}$. We can draw the same conclusions as in case of Table \ref{tab:PoSI_1}.

\subsection{Post-cAIC inference with income data from Galicia}\label{sec:data_example_sm}

In this section, we provide additional details about the post-cAIC inference applied to study the average household income in counties of Galicia. First, we focus on the estimation of the parameters of interest. Second, we complete the model selection analysis.

In Section \ref{sec:data_example}, we calculate the EBLUP using the survey estimates of covariate means $\hat{\bar{\bm{X}}}^{dir}_{i}$ and the means of transformed household income. The SSHG does include the official estimates of total $X^{dir}_{di}$ and mean $\bar{X}^{dir}_{di}$ at the county level, but we retrieved them using the standard formulas:
\begin{equation}\label{eq:covariates}
	\hat{X}^{dir}_{ik}=\sum_{j\in \mathcal{R}_i} w_{j}x_{jk},\quad
	\hat{\bar{X}}^{dir}_{ik}=\hat{X}^{dir}_{ik}/\hat{N}^{dir}_i
	\quad
	\text{and}\quad
	\hat{N}^{dir}_d=\sum_{j\in \mathcal{R}_i} w_{j},
\end{equation}
where $\hat{N}^{dir}_i$ refers to the estimated county size $N^{dir}_i$, $\mathcal{R}_i$ to the sample in county $i$ and $w_j$ to the sample weight.

In Section \ref{sec:data_example} we admitted not to having used all covariates available form the SSHG. On the basis of the previous, related studies \citep{boubeta2016empirical,reluga2021simultaneous}, we selected a set of 16 covariates which included those describing characteristics of the household and a member of this household who was considered as a main person. We analysed five binary variables describing the type of the household: household with 1 person (Typ1), household with more than one person (Typ2), household with a couple with children (Typ3), household with a couple without children (Typ4) and household with a single parent (Typ5). Furthermore we considered variables regarding the status of the property: without mortgage (Ten1) and with mortgage (Ten2), and the difficulties of the household at the end of the month: some difficulties (Dif2) and a lot of difficulties (Dif3). When it comes to the covariates describing the main person, we analysed a variable indicating the place of birth: Galicia (Birth1) and Spain except for Galicia (Birth2), and the eduction: primary (Edu1) and secondary (Edu2). We have also analysed a covariate indicating if the size of the municipality was smaller than ten thousands inhabitants (Size), a biological gender (Sex) as well as age: less than forty four years old (Age1) and between forty five and sixty four years old (Age2).
\begin{figure}[hbt]
	\vspace{-0.4cm}
	\centering
	\makebox{\includegraphics[width=0.9\textwidth]{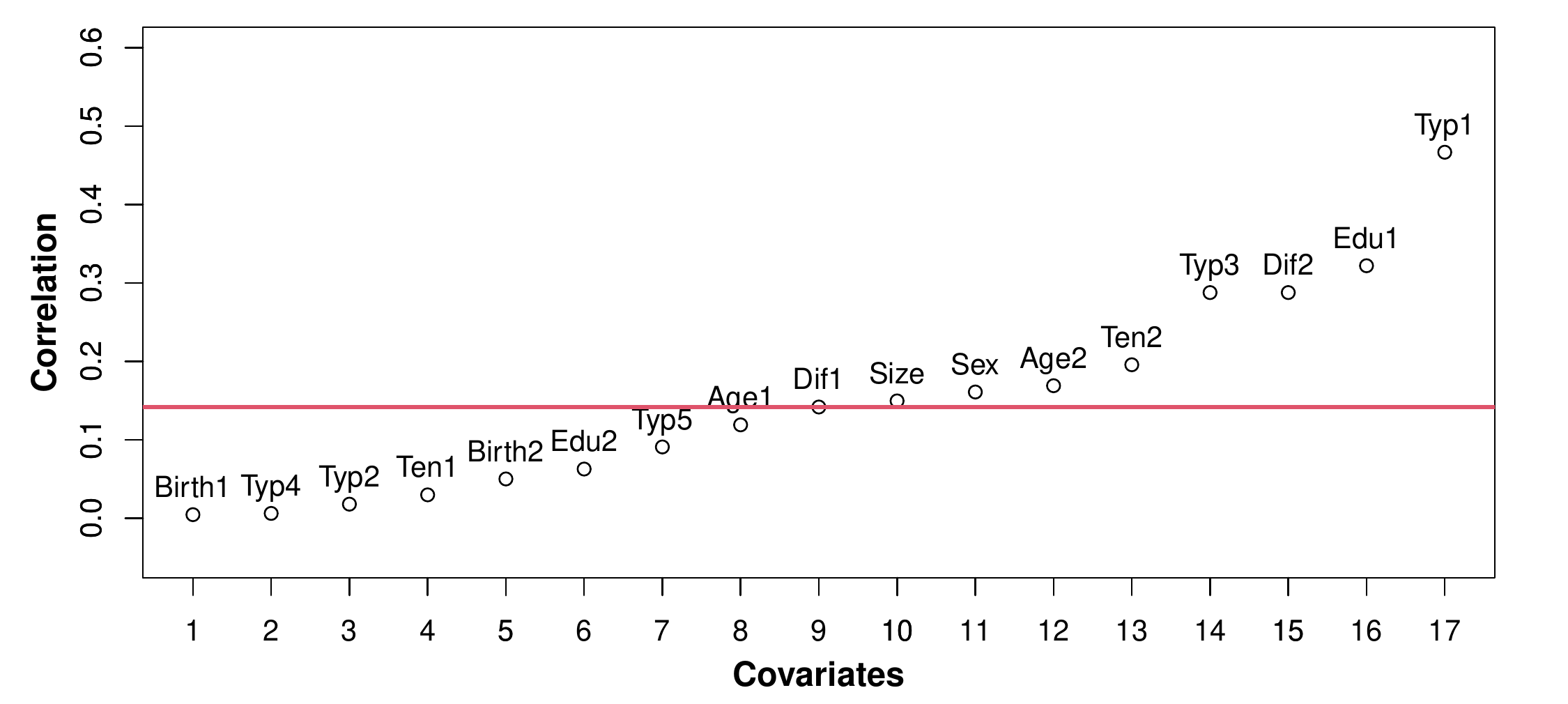}}
	\vspace{-0.2cm}
	\caption{\label{fig:correlation} Spearman's correlation coefficients between outcome variable and covariates}
	\vspace{-0.2cm}
\end{figure}
Figure \ref{fig:correlation} shows the Spearman's correlation coefficients between transformed income and covariates. We can see that the correlation between the transformed income and variables Typ1, Edu1, Dif2 and Typ3 is the strongest. We thus included them into each model. After that we inserted to the final set of models only those covariates with the correlation coefficients higher than the median value. We ended up with the set of eight variables and $2^4=16$ models. Table \ref{tab:ch4_models_data_example2} presents the inclusion of covariates into different models, whereas the selection criteria are outlined in Table \ref{tab:cAIC_data}. In the left part of Table \ref{tab:int_descript} we can see the descriptive statistics of the lengths of the post-cAIC, post-OBSP and naive intervals for a mixed parameter in the left. The descriptive statistics for the regression-synthetic estimates are presented in the right part  of Table \ref{tab:int_descript}.

\subsection{Additional assumptions}\label{sec:assump_sm}

The derivation of the extended cAIC of \cite{kubokawa2011conditional} as well as the first- and second-order correct MSE estimators requires some additional regularity conditions. Let $\lambda_1(\bm{V})\leqslant\dots \leqslant\lambda_{m}(\bm{V})$ be the eigenvalues of $\bm{V}$, and  $\lambda^{i}_l(\bm{V})$, $\lambda^{ij}_l(\bm{V})$, $\lambda^{ijk}_l(\bm{V})$ the eigenvalues of $\bm{V}_{(i)}$, $\bm{V}_{(ij)}$ and $\bm{V}_{(ijk)}$ defined in \eqref{eq:diff_oper}, $0\leqslant i,j,k\leqslant h$, $l=1, \dots, {m}$ ordered such that $|\lambda^{i}_1(\bm{V})|\leqslant \dots \leqslant |\lambda^{i}_n(\bm{V})|$, $|\lambda^{ij}_1(\bm{V})|\leqslant \dots \leqslant |\lambda^{ij}_n(\bm{V})|$, $|\lambda^{ijk}_1(\bm{V})|\leqslant \dots \leqslant |\lambda^{ijk}_{m}(\bm{V})|$. Moreover we assume
\begin{enumerate}
	\item[R.1] Rate of convergence: $\sup_{{i}\geqslant1}m_i<<\infty$, $\sup_{{i}\geqslant1}q_{i}<<\infty$, $n\rightarrow\infty$, i.e., $m$ and $n$ are of the same asymptotic order.	
	\item[R.2] $\bm{X}_{i}$, $\bm{Z}_{i}$, $\bm{R}(\bm{\theta})>0$ and $\bm{V}(\bm{\theta})>0$, $i=1,\dots, n$ contain only finite values.
	\item[R.3] Covariance matrices $\bm{G}_{i}$ and $\bm{R}_{i}$ have a linear structure with respect to $\bm{\theta}$.
	\item[R.4] $\bm{e}_{i}^t=\bm{k}^t_{i}-\bm{m}^t_{i}\bm{G}\bm{Z}^t_{i}\bm{V}^{-1}_{i}\bm{X}_{i}$ with $e_{di}=O(1)$ for $i=1,\dots, a+K$.
	\item[R.5] $\{\frac{\partial}{\partial \theta_j}\bm{m}^t_{i}\bm{G}\bm{Z}^t_{i}\bm{V}^{-1}_{i}\bm{X}_{i}\}_i=O(1)$ for $j=1,\dots,h$ and $i=1,\dots, a+K$.
	\item[R.6]
	$\hat{\bm{\theta}}$ satisfies: $(i)$ $\hat{\bm{\theta}}-\bm{\theta}=O_p({n}^{-1/2})$, $(ii)$ $\hat{\bm{\theta}}(\bm{y})=\hat{\bm{\theta}}(-\bm{y})$ and $(iii)$ $\hat{\bm{\theta}}(\bm{y}+\bm{Xr})=\hat{\bm{\theta}}(\bm{y})$ for any $\bm{r}\in \mathbb{R}^{a+K}$ .
	\item[R.7] $\hat{\bm{\theta}}-\bm{\theta}$ can be expanded as $\hat{\bm{\theta}}-\bm{\theta}=\hat{\bm{\theta}}^*+\hat{\bm{\theta}}^{**}+O_p(n^{-3/2})$,
	where $\hat{\bm{\theta}}^{*}=O_p(n^{-1/2})$, $\hat{\bm{\theta}}^{**}=O_p(n^{-1})$ and $\mathbb{E}(\hat{\bm{\theta}}^{*})=\bm{0}$.
	\item[R.8]
	$\hat{\bm{\theta}}^{*}$ and $\hat{\bm{\theta}}^{**}$ satisfy that $\mathbb{E}\left\{\text{tr}(\nabla_{\bm{y}} \nabla_{\bm{y}}^t\hat{\theta}_i^{**})\right\}=O(n^{-1})$, $\mathbb{E}\left[\{\text{tr}(\nabla_{\bm{y}} \nabla_{\bm{y}}^t\hat{\theta}_i^{*})\hat{\theta}_j^{*}\}\right]=O(n^{-1})$ and  $\mathbb{E}\left[\text{tr}\left\{(\nabla_{\bm{y}}\hat{\theta}_i^{*})(\nabla_{\bm{y}}\hat{\theta}_j^{*})^t\right\}\right]=O(n^{-1})$, where $\hat{\bm{\theta}}^{*}=(\hat{\theta}_1^{*}, \dots, \hat{\theta}_h^{*})^t$, $\hat{\bm{\theta}}^{**}=(\hat{\theta}_1^{**}, \dots, \hat{\theta}_h^{**})^t$.
\end{enumerate}

\subsection{Additional derivations and proofs}\label{sec:ch4_App_PoSI_sm}

First, we provide a proof of Lemma \ref{lemma:oversel}. Second, we present two algebraic properties.

\subsubsection{Proof of Lemma 1}
To prove the stated overselection property of cAIC, we proceed along similar steps as in the proof of Lemma 1 in \cite{charkhi2018asymptotic}. Let  $M_{pars}$ be the smallest true model. For all $M'\not\in\mathcal{M}_{o}$ it holds
\begin{equation*}
	\begin{split}
		&\mathrm{P}(M_{\mathrm{cAIC}}=M')\\
		&\leqslant\mathrm{P}\left\{
		\mathrm{cAIC}(M')\geqslant \max_{M\in \mathcal{M}_{O}} \mathrm{cAIC}(M) \right\}
		\leqslant
		\mathrm{P}\left\{ \mathrm{cAIC}(M')\geqslant \mathrm{cAIC}(M_{pars})\right\}\\
		&=\mathrm{P}\left[
		\ell_n^c\{\hat{\bm{\beta}}_{m}(M')\}-\hat{\rho}_{M'}-\hat{b}_{M'}
		\geqslant \ell_n^c\{\hat{\bm{\beta}}_{m}(M_{pars})\}-\hat{\rho}_{M_{pars}}-\hat{b}_{M_{pars}}
		\right]
		\\
		&=\mathrm{P}\left[
		\ell_n^c\{\hat{\bm{\beta}}_{m}(M')\}-\ell_n^c\left\{\bm{\beta}_{0m}(M_{pars})\right\}-\hat{\rho}_{M'}-\hat{b}_{M'}\right.\\
		&\left.\geqslant \ell_n^c\{\hat{\bm{\beta}}_{0m}(M_{pars})\}-\ell_n^c\left\{\bm{\beta}_{0m}(M_{pars})\right\}-\hat{\rho}_{M_{pars}}
		-\hat{b}_{M_{pars}}\right]\\
		&\rightarrow 0
	\end{split}
\end{equation*}
where the last line follows from
\begin{equation*}
	\ell_n^c\left\{\hat{\bm{\beta}}_{0m}(M_{pars})\right\}-\ell_n^c\left\{\bm{\beta}_{0m}(M_{pars})\right\}\geqslant \ell_n^c\left\{\hat{\bm{\beta}}_{m}(M')\right\}-\ell_n^c\left\{\bm{\beta}_{0m}(M_{pars})\right\}
\end{equation*}
as well as $\hat{\rho}_{M_{pars}} \geqslant \hat{\rho}_{M'}$ and $\hat{b}_{M_{pars}} \geqslant \hat{b}_{M'}$.

\subsubsection{Algebraic derivations}\label{sec:ch4_App_algebra}

The purpose of this section is to show the equivalence between the marginal and the conditional fixed parameters, that is $\hat{\bm{\beta}}_{m}=\hat{\bm{\beta}}_{c}$. Recall that $\hat{\bm{\beta}}_{m}$ is the estimated EBLUE defined as a solution of minimisation of equation \ref{eq:marginal_loglik}, whereas $\hat{\bm{\beta}}_c$ is an empirical counterpart of $\bm{\beta}_c=(\bm{X}^t\bm{R}^{-1}\bm{X})^{-1}(\bm{X}^t\bm{R}^{-1}\bm{y}-\bm{X}^t\bm{R}^{-1}\bm{Zu})$ derived from the equation for $\mathcal{R}_{n, K}^c$ below equation \eqref{eq:l_c_T}. In addition, we have $\bm{u}=\bm{GZ}^t\bm{V}^{-1}(\bm{y}-\bm{X\beta})$, and define  $\bm{L}=\bm{X}^t\bm{R}^{-1}\bm{X}$, $\bm{L}_Z=\bm{X}^t\bm{R}^{-1}\bm{Z}$, $\bm{G}_X=\bm{GZ}^t\bm{V}^{-1} \bm{X}$, $\bm{K}_y=\bm{X}^t\bm{R}^{-1}\bm{y}$, $\bm{G}_y=\bm{GZ}^t\bm{V}^{-1}\bm{y}$. Recall that $\bm{u}$ is EBLUP, which might be estimated using a two-stage procedure or the extended likelihood. We make use of Properties \ref{prop:prop1} and \ref{prop:prop2} to obtain the equivalence between $\hat{\bm{\beta}}_{m}$ and $\hat{\bm{\beta}}_{c}$.
\begin{property}\label{prop:prop1}
	For matrices $\bm{B}^{p\times m}$, $\bm{D}^{m\times p}$ and for non-singular matrices $\bm{C}^{n\times n}$, $\bm{A}^{p\times p}$, \cite{rao1973linear} showed that
	\begin{equation}\label{eq:prop_inverse}
		(\bm{A}+\bm{BCD})^{-1}=\bm{A}^{-1}-
		\bm{A}^{-1}\bm{B}(\bm{C}^{-1}+\bm{D}\bm{A}^{-1}\bm{B})^{-1}\bm{DA}^{-1}.
	\end{equation}
\end{property}

\begin{property}\label{prop:prop2}
	\begin{equation*}
		\begin{split}
			&\bm{V}^{-1}=\bm{R}^{-1}-\bm{R}^{-1}\bm{Z}(\bm{Z}^t\bm{R}^{-1}\bm{Z}+\bm{G}^{-1})^{-1}\bm{Z}^t\bm{R}^{-1},\\
			&(\bm{Z}^t\bm{R}^{-1}\bm{Z}+\bm{G}^{-1})^{-1}\bm{Z}^t\bm{R}^{-1}=\bm{GZV}^{-1},\\
			&(\bm{I}_q-\bm{G}_X\bm{L}^{-1}\bm{L}_Z)^{-1}\bm{G}_X\bm{L}^{-1} = \bm{G}_X(\bm{L}-\bm{L}_Z\bm{G}_X)^{-1}.
		\end{split}
	\end{equation*}
\end{property}
Property \ref{prop:prop2} is left without proof, because it only consists of simple but tedious algebraic transformations. In what follows, we show the equivalence between $\tilde{\bm{\beta}}_m$ and $\tilde{\bm{\beta}}_c$. We have  
\begin{eqnarray*}
	\tilde{\bm{\beta}}_{m} &=& 
	(\bm{X}^t\bm{V}^{-1}\bm{X})^{-1}(\bm{X}^t\bm{V}^{-1}\bm{y})
	=\left\{\bm{X}^t\bm{R}^{-1}\bm{X}-\bm{X}^t\bm{R}^{-1}\bm{Z} \left(\bm{Z}^t\bm{R}^{-1}\bm{Z} \right. \right.\\
	&&\left. \left. +\bm{G}^{-1}\right)^{-1}\bm{Z}^t\bm{R}^{-1}\bm{X}\right\}^{-1} \left\{\bm{X}^t\bm{R}^{-1}\bm{y}-\bm{X}^t\bm{R}^{-1}\bm{Z}(\bm{Z}^t\bm{R}^{-1}\bm{Z}+\bm{G}^{-1})^{-1}\bm{Z}^t\bm{R}^{-1}\bm{y}\right\}\\
	&=& \left(\bm{X}^t\bm{R}^{-1}\bm{X}-\bm{X}^t\bm{R}^{-1}\bm{Z}\bm{GZ}^t\bm{V}^{-1} \bm{X}\right)^{-1}
	(\bm{X}^t\bm{R}^{-1}\bm{y}-\bm{X}^t\bm{R}^{-1}\bm{Z}\bm{GZ}^t\bm{V}^{-1}\bm{y})\\
	&=& (\bm{L}-\bm{L}_Z\bm{G}_X)^{-1}(\bm{K}_y-\bm{L}_Z\bm{G}_y)\\
	&= &\bm{L}^{-1}\bm{K}_{y}-\bm{L}^{-1}\bm{L}_Z\bm{G}_y+ \bm{L}^{-1}\bm{L}_{Z}(\bm{I}_q-\bm{G}_X\bm{L}^{-1}\bm{L}_Z)^{-1}\bm{G}_X\bm{L}^{-1}\bm{K}_y\\
	&&-\bm{L}^{-1}\bm{L}_{Z}(\bm{I}_q-\bm{G}_X\bm{L}^{-1}\bm{L}_Z)^{-1}\bm{G}_X\bm{L}^{-1}\bm{L}_Z\bm{G}_y,
\end{eqnarray*}
where we used Property \ref{prop:prop2} in the first equation and Property \ref{prop:prop1} in the last equation.
On the other hand
\begin{eqnarray*}
	\tilde{\bm{\beta}}_c &=& (\bm{X}^t\bm{R}^{-1}\bm{X})^{-1}(\bm{X}^t\bm{R}^{-1}\bm{y}-\bm{X}^t\bm{R}^{-1}\bm{Z}\bm{GZ}^t\bm{V}^{-1}\bm{y}+\bm{X}^t\bm{R}^{-1}\bm{Z}\bm{GZ}^t\bm{V}^{-1}\bm{X\beta})\\
	&=& (\bm{X}^t\bm{R}^{-1}\bm{X})^{-1}\bm{X}^t\bm{R}^{-1}\bm{y}		-(\bm{X}^t\bm{R}^{-1}\bm{X})^{-1}\bm{X}^t\bm{R}^{-1}\bm{Z}\bm{GZ}^t\bm{V}^{-1}\bm{y}\\
	&& +(\bm{X}^t\bm{R}^{-1}\bm{X})^{-1}\bm{X}^t\bm{R}^{-1}\bm{Z}\bm{GZ}^t\bm{V}^{-1}\bm{X\beta}\\
	&=& \bm{L}^{-1}\bm{K}_{y}-\bm{L}^{-1}\bm{L}_Z\bm{G}_y
	+\bm{L}^{-1}\bm{L}_{Z}\bm{G}_X(\bm{L}-\bm{L}_Z\bm{G}_X)^{-1}\bm{K}_y \\ &&
	-\bm{L}^{-1}\bm{L}_{Z}\bm{G}_X(\bm{L}-\bm{L}_Z\bm{G}_X)^{-1}\bm{L}_{Z}\bm{G}_y
\end{eqnarray*}
The desired result follows applying the third line of Property \ref{prop:prop2} and replacing $\bm{R}$, $\bm{G}$ and $\bm{V}$ with $\hat{\bm{R}}$, $\hat{\bm{G}}$ and $\hat{\bm{V}}$.

\bibliographystyle{rss}
\bibliography{literature_review}

\begin{thebibliography}{48}
\expandafter\ifx\csname natexlab\endcsname\relax\def\natexlab#1{#1}\fi
\expandafter\ifx\csname url\endcsname\relax
  \def\url#1{\texttt{#1}}\fi
\expandafter\ifx\csname urlprefix\endcsname\relax\def\urlprefix{URL: }\fi

\bibitem[{Akaike(1973)}]{akaike1973information}
Akaike, H. (1973) Information theory and an extension of the maximum likelihood
  principle.
\newblock In \textit{Proc. 2nd Int. Symp. Info. Theory, Tsahkadsor, Armenia,
  USSR, Septemebr 2-8, 1971}, 267--281. B.Petrov \& F.Cski, eds. Budapest:
  Akad{\'e}miai Kiad{\'o}.

\bibitem[{Bachoc et~al.(2019)Bachoc, Leeb and P{\"o}tscher}]{bachoc2019valid}
Bachoc, F., Leeb, H. and P{\"o}tscher, B.~M. (2019) Valid confidence intervals
  for post-model-selection predictors.
\newblock \textit{Ann. Statist.}, \textbf{47}, 1475--1504.

\bibitem[{Belloni et~al.(2015)Belloni, Chernozhukov and
  Kato}]{belloni2015uniform}
Belloni, A., Chernozhukov, V. and Kato, K. (2015) Uniform post-selection
  inference for least absolute deviation regression and other z-estimation
  problems.
\newblock \textit{Biometrika}, \textbf{102}, 77--94.

\bibitem[{Berk et~al.(2013)Berk, Brown, Buja, Zhang and Zhao}]{berk2013valid}
Berk, R., Brown, L., Buja, A., Zhang, K. and Zhao, L. (2013) Valid
  post-selection inference.
\newblock \textit{Ann. Statist.}, \textbf{41}, 802--837.

\bibitem[{Bolker et~al.(2009)Bolker, Brooks, Clark, Geange, Poulsen, Stevens
  and White}]{bolker2009generalized}
Bolker, B.~M., Brooks, M.~E., Clark, C.~J., Geange, S.~W., Poulsen, J.~R.,
  Stevens, M. H.~H. and White, J.-S.~S. (2009) Generalized linear mixed models:
  a practical guide for ecology and evolution.
\newblock \textit{Trends Ecol. Evol.}, \textbf{24}, 127--135.

\bibitem[{Boubeta et~al.(2016)Boubeta, Lombard{\'\i}a and
  Morales}]{boubeta2016empirical}
Boubeta, M., Lombard{\'\i}a, M.~J. and Morales, D. (2016) Empirical best
  prediction under area-level poisson mixed models.
\newblock \textit{Test}, \textbf{25}, 548--569.

\bibitem[{Charkhi and Claeskens(2018)}]{charkhi2018asymptotic}
Charkhi, A. and Claeskens, G. (2018) Asymptotic post-selection inference for
  the {A}kaike information criterion.
\newblock \textit{Biometrika}, \textbf{105}, 645--664.

\bibitem[{Cunen et~al.(2020)Cunen, Walløe and Hjort}]{cunen2020}
Cunen, C., Walløe, L. and Hjort, N.~L. (2020) Focused model selection for
  linear mixed models with an application to whale ecology.
\newblock \textit{Ann. Appl. Stat.}, \textbf{14}, 872--904.

\bibitem[{Fay and Herriot(1979)}]{fay_herriot}
Fay, R.~E. and Herriot, R.~A. (1979) Estimates of income for small places: An
  application of {J}ames-{S}tein procedures to census data.
\newblock \textit{J. Am. Statist. Assoc.}, \textbf{74}, 269--277.

\bibitem[{Ferrari and Yang(2015)}]{ferrari2015confidence}
Ferrari, D. and Yang, Y. (2015) Confidence sets for model selection by
  {F}-testing.
\newblock \textit{Stat. Sin.}, \textbf{25}, 1637--1658.

\bibitem[{Francq et~al.(2019)Francq, Lin and Hoyer}]{FrancqEtal2019}
Francq, B.~G., Lin, D. and Hoyer, W. (2019) Confidence, prediction, and
  tolerance in linear mixed models.
\newblock \textit{Stat. Med.}, \textbf{38}, 5603--5622.

\bibitem[{Gallant and White(1988)}]{gallant1988unified}
Gallant, A. and White, H. (1988) \textit{A Unified Theory of Estimation and
  Inference for Nonlinear Dynamic Models}.
\newblock Wiley–Blackwell, New Jersey.

\bibitem[{Gilmour et~al.(1995)Gilmour, Thompson and
  Cullis}]{gilmour1995average}
Gilmour, A.~R., Thompson, R. and Cullis, B.~R. (1995) Average information
  {REML}: an efficient algorithm for variance parameter estimation in linear
  mixed models.
\newblock \textit{Biometrics}, \textbf{51}, 1440--1450.

\bibitem[{Greven and Kneib(2010)}]{greven2010behaviour}
Greven, S. and Kneib, T. (2010) On the behaviour of marginal and conditional
  {AIC} in linear mixed models.
\newblock \textit{Biometrika}, \textbf{97}, 773--789.

\bibitem[{Gumedze and Dunne(2011)}]{gumedze2011parameter}
Gumedze, F. and Dunne, T. (2011) Parameter estimation and inference in the
  linear mixed model.
\newblock \textit{Linear Algebra Appl.}, \textbf{435}, 1920--1944.

\bibitem[{Harville(1977)}]{harville1977maximum}
Harville, D.~A. (1977) Maximum likelihood approaches to variance component
  estimation and to related problems.
\newblock \textit{J. Am. Statist. Assoc.}, \textbf{72}, 320--338.

\bibitem[{Henderson(1950)}]{henderson1950}
Henderson, C.~R. (1950) Estimation of genetic parameters.
\newblock \textit{Ann. Math. Statist.}, \textbf{21}, 226--252.

\bibitem[{Henderson(1975)}]{henderson}
--- (1975) Best linear unbiased estimation and prediction under a selection
  model.
\newblock \textit{Biometrics}, \textbf{31}, 423--447.

\bibitem[{Hjort and Claeskens(2003)}]{hjort2003frequentist}
Hjort, N.~L. and Claeskens, G. (2003) Frequentist model average estimators.
\newblock \textit{J. Am. Statist. Assoc.}, \textbf{98}, 879--899.

\bibitem[{Hodges and Sargent(2001)}]{hodges2001counting}
Hodges, J.~S. and Sargent, D.~J. (2001) Counting degrees of freedom in
  hierarchical and other richly-parameterised models.
\newblock \textit{Biometrika}, \textbf{88}, 367--379.

\bibitem[{Jiang(2007)}]{jiang2007linear}
Jiang, J. (2007) \textit{Linear and generalized linear mixed models and their
  applications}.
\newblock Springer, New York.

\bibitem[{Kasy(2018)}]{kasy}
Kasy, M. (2018) Uniformity and the delta method.
\newblock \textit{J. Econom. Methods}, \textbf{8}, 1--19.

\bibitem[{Kawakubo and Kubokawa(2014)}]{kawakubo2014modified}
Kawakubo, Y. and Kubokawa, T. (2014) Modified conditional {AIC} in linear mixed
  models.
\newblock \textit{J. Multiv. Anal.}, \textbf{129}, 44--56.

\bibitem[{Kubokawa(2011)}]{kubokawa2011conditional}
Kubokawa, T. (2011) Conditional and unconditional methods for selecting
  variables in linear mixed models.
\newblock \textit{J. Multiv. Anal.}, \textbf{102}, 641--660.

\bibitem[{Laird and Ware(1982)}]{laird1982random}
Laird, N.~M. and Ware, J.~H. (1982) Random-effects models for longitudinal
  data.
\newblock \textit{Biometrics}, \textbf{38}, 963--974.

\bibitem[{Lee et~al.(2016)Lee, Sun, Sun and Taylor}]{lee2016exact}
Lee, J.~D., Sun, D.~L., Sun, Y. and Taylor, J.~E. (2016) Exact post-selection
  inference, with application to the lasso.
\newblock \textit{Ann. Statist.}, \textbf{44}, 907--927.

\bibitem[{Leeb and P{\"o}tscher(2003)}]{leeb2003finite}
Leeb, H. and P{\"o}tscher, B.~M. (2003) The finite-sample distribution of
  post-model-selection estimators and uniform versus nonuniform approximations.
\newblock \textit{Econ. Theory}, \textbf{19}, 100--142.

\bibitem[{Leeb and P{\"o}tscher(2006)}]{leeb2006can}
--- (2006) Can one estimate the conditional distribution of
  post-model-selection estimators?
\newblock \textit{Ann. Statist.}, \textbf{34}, 2554--2591.

\bibitem[{Leeb and P{\"o}tscher(2008)}]{leeb2008can}
--- (2008) Can one estimate the unconditional distribution of
  post-model-selection estimators?
\newblock \textit{Econ. Theory}, \textbf{24}, 338--376.

\bibitem[{Liang et~al.(2008)Liang, Wu and Zou}]{liang2008}
Liang, H., Wu, H. and Zou, G. (2008) A note on conditional {AIC} for linear
  mixed-effects models.
\newblock \textit{Biometrika}, \textbf{95}, 773--778.

\bibitem[{Lombard{\'\i}a et~al.(2017)Lombard{\'\i}a, L{\'o}pez-Vizca{\'\i}no
  and Rueda}]{lombardia2017mixed}
Lombard{\'\i}a, M.~J., L{\'o}pez-Vizca{\'\i}no, E. and Rueda, C. (2017) Mixed
  generalized {A}kaike information criterion for small area models.
\newblock \textit{J. R. Statist. Soc. A}, \textbf{180}, 1229--1252.

\bibitem[{Marhuenda et~al.(2017)Marhuenda, Molina, Morales and
  Rao}]{marhuenda2017poverty}
Marhuenda, Y., Molina, I., Morales, D. and Rao, J. (2017) Poverty mapping in
  small areas under a twofold nested error regression model.
\newblock \textit{J. R. Statist. Soc. A}, \textbf{180}, 1111--1136.

\bibitem[{McCulloch and Neuhaus(2011)}]{mcculloch2011misspecifying}
McCulloch, C.~E. and Neuhaus, J.~M. (2011) Misspecifying the shape of a random
  effects distribution: why getting it wrong may not matter.
\newblock \textit{Stat. Sci}, \textbf{26}, 388--402.

\bibitem[{Morales et~al.(2021)Morales, Esteban~Lefler, Perez and
  Hobza}]{morales2021sae}
Morales, D., Esteban~Lefler, M., Perez, A. and Hobza, T. (2021) \textit{A
  Course on Small Area Estimation and Mixed Models: Methods, Theory and
  Applications in R}.
\newblock Springer.

\bibitem[{M{\"u}ller et~al.(2013)M{\"u}ller, Scealy and
  Welsh}]{muller2013model}
M{\"u}ller, S., Scealy, J.~L. and Welsh, A.~H. (2013) Model selection in linear
  mixed models.
\newblock \textit{Stat. Sci.}, \textbf{28}, 135--167.

\bibitem[{Rao(1973)}]{rao1973linear}
Rao, C.~R. (1973) \textit{Linear statistical inference and its applications}.
\newblock Wiley New York.

\bibitem[{Rao and Molina(2015)}]{rao2015small}
Rao, J. N.~K. and Molina, I. (2015) \textit{Small area estimation}.
\newblock John Wiley \& Sons.

\bibitem[{Reluga et~al.(2021)Reluga, Lombardía and
  Sperlich}]{reluga2021simultaneous}
Reluga, K., Lombardía, M.~J. and Sperlich, S.~A. (2021) Simultaneous inference
  for linear mixed model parameters with an application to small area
  estimation.
\newblock \textit{arXiv:1903.02774}.

\bibitem[{Srivastava and Kubokawa(2010)}]{srivastava2010conditional}
Srivastava, M.~S. and Kubokawa, T. (2010) Conditional information criteria for
  selecting variables in linear mixed models.
\newblock \textit{J. Multiv. Anal.}, \textbf{101}, 1970--1980.

\bibitem[{Sugasawa et~al.(2019)Sugasawa, Kawakubo and
  Datta}]{sugasawa2019observed}
Sugasawa, S., Kawakubo, Y. and Datta, G.~S. (2019) Observed best selective
  prediction in small area estimation.
\newblock \textit{J. Multiv. Anal.}, \textbf{173}, 383--392.

\bibitem[{Tibshirani et~al.(2018)Tibshirani, Rinaldo, Tibshirani and
  Wasserman}]{tibshirani2018uniform}
Tibshirani, R.~J., Rinaldo, A., Tibshirani, R. and Wasserman, L. (2018) Uniform
  asymptotic inference and the bootstrap after model selection.
\newblock \textit{Ann. Statist.}, \textbf{46}, 1255--1287.

\bibitem[{Tibshirani et~al.(2016)Tibshirani, Taylor, Lockhart and
  Tibshirani}]{tibshirani2016exact}
Tibshirani, R.~J., Taylor, J., Lockhart, R. and Tibshirani, R. (2016) Exact
  post-selection inference for sequential regression procedures.
\newblock \textit{J. Am. Statist. Assoc.}, \textbf{111}, 600--620.

\bibitem[{Vaida and Blanchard(2005)}]{vaida2005conditional}
Vaida, F. and Blanchard, S. (2005) Conditional {A}kaike information for
  mixed-effects models.
\newblock \textit{Biometrika}, \textbf{92}, 351--370.

\bibitem[{Verbeke and Molenberghs(2000)}]{verbekeMolenberghs2000}
Verbeke, G. and Molenberghs, G. (2000) \textit{Linear Mixed Models for
  Longitudinal Data}.
\newblock Springer.

\bibitem[{Vuong(1989)}]{vuong1989}
Vuong, Q.~H. (1989) Likelihood ratio tests for model selection and non-nested
  hypotheses.
\newblock \textit{Econometrica}, \textbf{57}, 307--333.

\bibitem[{White(1994)}]{white_1994}
White, H. (1994) \textit{Estimation, Inference and Specification Analysis}.
\newblock Econometric Society Monographs. Cambridge University Press.

\bibitem[{Woodroofe(1982)}]{woodroofe1982model}
Woodroofe, M. (1982) On model selection and the arc sine laws.
\newblock \textit{Ann. Statist.}, \textbf{10}, 1182--1194.

\bibitem[{You et~al.(2016)You, M{\"u}ller and Ormerod}]{you2016generalized}
You, C., M{\"u}ller, S. and Ormerod, J.~T. (2016) On generalized degrees of
  freedom with application in linear mixed models selection.
\newblock \textit{Stat. Comput.}, \textbf{26}, 199--210.

\end{thebibliography}

\newpage
\begin{table}
	\caption{\label{tab:cAIC_rho_examp_nest} Estimated values of $\rho$, $b$ and cAIC for the models from sets $\mathcal{M}_{nest}$ and $\mathcal{M}_{all}$.}
	\centering
	\fbox{%
			\vspace{-0.5cm}
		\begin{tabular}{c c cccc}
			\hline
			Model & & $M_0$ & $M_1$   & $M_2$ & $M_3$  \\\hline
			$\hat{\rho}+\hat{b}$  & &24.386 & 25.449 & 26.450 & 25.369 \\
			cAIC & &429.245 & 430.323 & 431.334  & 430.236 \\\hline
	\end{tabular}}
\vspace{-0.1cm}
\end{table}
\begin{table}
	\caption{\label{tab:PoSI_beta_1_sm} Coverage probabilities and average lengths (in parenthesis) of post-cAIC and naive confidence intervals for $\beta_{j}$. Nominal coverage probability: 95\%, selection matrix: $\bm{\upsilon}^2_{all}$.}
	\centering
	\setlength{\tabcolsep}{1.6pt}
	\fbox{%
		\begin{tabular}{ccccccccccc}
			&       &           &  & $(15:5)$     & $(30:5)$     & $(60:5)$     & $(90:5)$     &  & $(30:10)$    & $(30:20)$    \\
			S & Meth. & $\beta_j$ &  & CP (L)    & CP (L)    & CP (L)    & CP (L)    &  & CP (L)    & CP (L)    \\
			\hline
			S1  & p.-aAIC  & $\beta_2$ &  & 94.9 (0.500) & 94.1 (0.401) & 94.6 (0.270)  & 95.2 (0.229) &  & 93.5 (0.262) & 94.1 (0.179) \\
			&       & $\beta_3$ &  & 98.7 (0.819) & 98.9 (0.649) & 99.9 (0.441) & 99.6 (0.384) &  & 99.3 (0.369) & 99.6 (0.245) \\
			&       & $\beta_4$ &  & 87.4 (0.742) & 97.1 (0.570)  & 94.3 (0.388) & 92.2 (0.310) &  & 92.2 (0.318) & 88.3 (0.217) \\[2mm]
			& N & $\beta_2$ &  & 94.6 (0.487) & 93.3 (0.389) & 94.3 (0.265) & 94.4 (0.222) &  & 92.6 (0.258) & 93.9 (0.176) \\
			&       & $\beta_3$ &  & 93.2 (0.517) & 93.0 (0.366) & 94.0 (0.252)   & 93.8 (0.207) &  & 93.6 (0.244) & 94.2 (0.173) \\
			&       & $\beta_4$ &  & 70.0 (0.577) & 72.7 (0.361) & 70.2 (0.264) & 66.5 (0.220) &  & 71.5 (0.249) & 67.2 (0.178) \\[2mm]
			S2  & p.-cAIC  & $\beta_2$ &  & 94.0 (0.495)   & 94.8 (0.399) & 94.2 (0.270)  & 95.5 (0.228) &  & 93.4 (0.263) & 94.1 (0.179) \\
			&       & $\beta_3$ &  & 99.1 (0.854) & 99.6 (0.691) & 99.7 (0.466) & 99.8 (0.402) &  & 99.6 (0.400) & 99.5 (0.245) \\
			&       & $\beta_4$ &  & 89.5 (0.759) & 97.5 (0.616) & 95.1 (0.419) & 92.6 (0.321) &  & 93.7 (0.341) & 86.2 (0.214) \\[2mm]
			& N & $\beta_2$ &  & 93.6 (0.483) & 94.2 (0.384) & 93.3 (0.262) & 94.1 (0.220) &  & 92.6 (0.257) & 94.0 (0.175) \\
			&       & $\beta_3$ &  & 93.9 (0.510)  & 92.5 (0.361) & 94.1 (0.249) & 93.1 (0.205) &  & 93.5 (0.243) & 94.1 (0.173) \\
			&       & $\beta_4$ &  & 70.4 (0.571) & 74.3 (0.354) & 72.0 (0.261)   & 69.5 (0.217) &  & 69.1 (0.248) & 67.4 (0.177)
		\end{tabular}
	}	
	\vspace{-0.1cm}
\end{table}
\begin{table}
	\caption{\label{tab:PoSI_beta_2} Coverage probabilities and average lengths (in parenthesis) of post-cAIC and naive confidence intervals for $\beta_{j}$. Nominal coverage probability: 95\%, selection matrix: $\bm{\upsilon}^3_{all}$.}
	\centering
	\setlength{\tabcolsep}{1.6pt}
	\fbox{%
		\begin{tabular}{ccccccccccc}
			&       &           &  & $(15:5)$     & $(30:5)$     & $(60:5)$     & $(90:5)$     &  & $(30:10)$    & $(30:20)$    \\
			S & Meth. & $\beta_j$ &  & CP (L)    & CP (L)    & CP (L)    & CP (L)    &  & CP (L)    & CP (L)    \\\hline
			S1  & p.-cAIC  & $\beta_1$ &  & 92.9 (1.131) & 94.1 (0.786) & 94.3 (0.561) & 94.1 (0.454) &  & 94.7 (0.782) & 94.0 (0.744) \\
			&       & $\beta_2$ &  & 94.2 (0.499) & 93.5 (0.396) & 94.8 (0.269) & 95.5 (0.225) &  & 93.3 (0.263) & 94.2 (0.177) \\
			&       & $\beta_3$ &  & 94.3 (0.526) & 93.4 (0.385) & 95.2 (0.259) & 95.3 (0.211) &  & 94.4 (0.253) & 93.7 (0.333) \\
			&       & $\beta_4$ &  & 92.2 (0.758) & 94.5 (0.471) & 93.4 (0.349) & 92.7 (0.292) &  & 93.7 (0.333) & 93.9 (0.339) \\
			&       & $\beta_5$ &  & 91.3 (0.746) & 93.1 (0.497) & 93.6 (0.353) & 92.3 (0.290) &  & 93.9 (0.339) & 95.2 (0.237) \\[2mm]
			& N & $\beta_1$ &  & 92.6 (1.103) & 94.1 (0.779) & 94.3 (0.555) & 94.1 (0.453) &  & 93.5 (0.745) & 93.4 (0.727) \\
			&       & $\beta_2$ &  & 93.9 (0.488) & 92.7 (0.388) & 94.6 (0.265) & 95.1 (0.222) &  & 92.1 (0.258) & 93.9 (0.176) \\
			&       & $\beta_3$ &  & 93.6 (0.517) & 91.6 (0.365) & 94.4 (0.252) & 94.4 (0.207) &  & 93.7 (0.244) & 69.8 (0.249) \\
			&       & $\beta_4$ &  & 69.4 (0.578) & 71.9 (0.360) & 68.5 (0.265) & 67.5 (0.220)  &  & 69.8 (0.249) & 67.8 (0.256) \\
			&       & $\beta_5$ &  & 66.4 (0.569) & 72.5 (0.377) & 69.8 (0.267) & 68.9 (0.218) &  & 67.8 (0.256) & 67.1 (0.175) \\[2mm]
			S2  & p.-cAIC  & $\beta_1$ &  & 92.2 (0.844) & 94.2 (0.598) & 94.1 (0.425) & 94.1 (0.346) &  & 93.6 (0.562) & 94.1 (0.532) \\
			&       & $\beta_2$ &  & 94.4 (0.493) & 94.2 (0.392) & 94.9 (0.267) & 95.2 (0.222) &  & 93.6 (0.263) & 94.6 (0.177) \\
			&       & $\beta_3$ &  & 93.2 (0.519) & 93.8 (0.380) & 94.9 (0.256) & 93.3 (0.209) &  & 95.0 (0.252) & 93.7 (0.332) \\
			&       & $\beta_4$ &  & 92.5 (0.751) & 96.4 (0.466) & 93.5 (0.345) & 92.6 (0.290) &  & 93.7 (0.332) & 93.5 (0.341) \\
			&       & $\beta_5$ &  & 88.7 (0.739) & 93.0 (0.495) & 92.2 (0.352) & 91.0 (0.288) &  & 93.5 (0.341) & 95.1 (0.237) \\[2mm]
			& N & $\beta_1$ &  & 92.2 (0.839) & 94.2 (0.598) & 93.9 (0.425) & 94.3 (0.346) &  & 92.9 (0.552) & 94.0 (0.527) \\
			&       & $\beta_2$ &  & 94.0 (0.483) & 94.0 (0.384) & 94.5 (0.262) & 95.0 (0.219) &  & 92.7 (0.257) & 94.5 (0.175) \\
			&       & $\beta_3$ &  & 92.8 (0.510) & 91.6 (0.361) & 94.2 (0.249) & 92.7 (0.205) &  & 93.8 (0.243) & 68.2 (0.248) \\
			&       & $\beta_4$ &  & 69.9 (0.570) & 74.6 (0.354) & 68.5 (0.260) & 67.4 (0.217) &  & 68.2 (0.248) & 66.7 (0.255) \\
			&       & $\beta_5$ &  & 65.5 (0.561) & 72.0 (0.373) & 67.8 (0.264) & 66.7 (0.215) &  & 66.7 (0.255) & 66.7 (0.175)
		\end{tabular}
	}	
		\vspace{-0.1cm}
\end{table}
\newpage
\begin{table}
	\caption{\label{tab:PoSI_beta_3} Coverage probabilities and average lengths (in parenthesis) of post-cAIC and naive confidence intervals for $\beta_{j}$. Nominal coverage probability: 95\%, selection matrix: $\bm{\upsilon}^4_{all}$.}
	\centering
	\setlength{\tabcolsep}{1.6pt}
	\fbox{%
		\begin{tabular}{ccccccccccc}
			&         &           &  & $(15:5)$     & $(30:5)$     & $(60:5)$     & $(90:5)$     &  & $(30:10)$    & $(30:20)$    \\
			S & Meth.   & $\beta_j$ &  & CP (L)    & CP (L)    & CP (L)    & CP (L)    &  & CP (L)    & CP (L)    \\
			\hline
			S1  & p.-cAIC & $\beta_1$ &  & 92.4 (1.118) & 94.5 (0.787) & 93.8 (0.559) & 95.3 (0.453) &  & 94.6 (0.780) & 93.1 (0.742) \\
			&         & $\beta_2$ &  & 94.1 (0.496) & 94.8 (0.394) & 94.5 (0.268) & 94.3 (0.223) &  & 94.7 (0.261) & 94.0 (0.176) \\
			&         & $\beta_3$ &  & 94.7 (0.521) & 93.4 (0.372) & 95.0 (0.255) & 95.5 (0.209) &  & 95.4 (0.247) & 95.0 (0.175) \\
			&         & $\beta_4$ &  & 94.8 (0.586) & 95.1 (0.364) & 94.6 (0.268) & 93.2 (0.223) &  & 95.3 (0.250) & 94.7 (0.179) \\
			&         & $\beta_5$ &  & 91.4 (0.754) & 93.2 (0.502) & 93.9 (0.354) & 94.3 (0.288) &  & 94.4 (0.340) & 94.5 (0.236) \\[2mm]
			& N      & $\beta_1$ &  & 92.0 (1.106) & 94.5 (0.781) & 93.7 (0.554) & 95.2 (0.452) &  & 94.1 (0.746) & 92.7 (0.726) \\
			&         & $\beta_2$ &  & 94.1 (0.491) & 94.7 (0.392) & 94.3 (0.266) & 94.3 (0.222) &  & 94.4 (0.259) & 94.0 (0.176) \\
			&         & $\beta_3$ &  & 94.6 (0.520)  & 93.1 (0.369) & 94.8 (0.254) & 95.5 (0.208) &  & 95.2 (0.245) & 94.9 (0.173) \\
			&         & $\beta_4$ &  & 94.6 (0.581) & 95.0 (0.363) & 94.4 (0.266) & 93.1 (0.220) &  & 95.2 (0.249) & 94.6 (0.178) \\
			&         & $\beta_5$ &  & 69.1 (0.573) & 71.4 (0.380) & 71.3 (0.268) & 70.5 (0.218) &  & 70.6 (0.257) & 70.2 (0.176) \\[2mm]
			S2  & p.-cAIC & $\beta_1$ &  & 92.7 (0.852) & 94.8 (0.597) & 93.3 (0.426) & 95.3 (0.346) &  & 94.4 (0.561) & 94.4 (0.561) \\
			&         & $\beta_2$ &  & 93.7 (0.492) & 95.8 (0.388) & 94.9 (0.266) & 93.7 (0.221) &  & 94.1 (0.260) & 94.1 (0.260)  \\
			&         & $\beta_3$ &  & 94.7 (0.515) & 92.7 (0.367) & 94.6 (0.252) & 94.7 (0.207) &  & 95.7 (0.246) & 95.7 (0.246) \\
			&         & $\beta_4$ &  & 95.2 (0.579) & 94.9 (0.356) & 94.2 (0.263) & 93.6 (0.220) &  & 95.7 (0.249) & 95.7 (0.249) \\
			&         & $\beta_5$ &  & 91.0 (0.754) & 93.6 (0.499) & 93.3 (0.352) & 93.8 (0.286) &  & 94.3 (0.342) & 94.3 (0.342) \\[2mm]
			& N      & $\beta_1$ &  & 92.6 (0.850)  & 95.0 (0.597) & 93.3 (0.426) & 95.3 (0.346) &  & 94.0 (0.552) & 94.0 (0.552) \\
			&         & $\beta_2$ &  & 93.5 (0.486) & 95.8 (0.387) & 94.8 (0.263) & 93.7 (0.220)  &  & 94.0 (0.258) & 94.0 (0.258) \\
			&         & $\beta_3$ &  & 94.5 (0.514) & 92.4 (0.363) & 94.0 (0.250) & 94.4 (0.205) &  & 96.0 (0.244) & 95.6 (0.244) \\
			&         & $\beta_4$ &  & 95.1 (0.574) & 94.9 (0.356) & 93.9 (0.261) & 93.4 (0.217) &  & 95.3 (0.248) & 95.3 (0.248) \\
			&         & $\beta_5$ &  & 69.7 (0.565) & 73.5 (0.375) & 71.0 (0.265) & 70.7 (0.216) &  & 70.1 (0.256) & 70.2 (0.256)
		\end{tabular}
	}	
	\vspace{-0.5cm}
\end{table}
\begin{table}
	\caption{\label{tab:PoSI_betax_1} Coverage probabilities and average lengths (in parenthesis) of post-cAIC and naive confidence intervals for $\bm{k}_i^t\hat{\bm{\beta}}$ and $\bar{\bm{k}}^t\hat{\bm{\beta}}$. Nominal coverage probability: 95\%, selection matrix: $\bm{\upsilon}^2_{all}$.}
	\centering
	\setlength{\tabcolsep}{1.6pt}
	\fbox{%
		\begin{tabular}{ccccccccccc}
			&         &                  &  & $(15:5)$     & $(30:5)$     & $(60:5)$     & $(90:5)$     &  & $(30:10)$    & $(30:20)$    \\
			S & Meth.   & $\beta_j$        &  & CP (L)    & CP (L)    & CP (L)    & CP (L)    &  & CP (L)    & CP (L)    \\\hline
			S1  & p.-cAIC & $\bm{k}_1^t\hat{\bm{\beta}}$                &  & 91.5 (1.152) & 94.3 (1.018) & 97.7 (0.699) & 97.2 (0.515) &  & 96.6 (0.891) & 93.9 (0.751) \\
			&         & $\bm{k}_2^t\hat{\bm{\beta}}$                &  & 91.4 (1.305) & 97.9 (1.118) & 94.0 (0.572) & 96.3 (0.500) &  & 96.2 (0.870) & 94.5 (0.793) \\
			&         & $\bm{k}_3^t\hat{\bm{\beta}}$                &  & 97.8 (1.571) & 96.2 (0.967) & 97.0 (0.690) & 95.7 (0.464) &  & 97.1 (0.904) & 96.0 (0.864) \\
			&         & $\bar{\bm{k}}^t\hat{\bm{\beta}}$ &  & 95.3 (1.374) & 95.8 (0.998) & 96.7 (0.711) & 97.1 (0.572) &  & 96.5 (0.891) & 94.5 (0.786) \\[2mm]
			& N       & $\bm{k}_1^t\hat{\bm{\beta}}$                &  & 91.8 (1.171) & 90.7 (0.886) & 94.7 (0.574) & 96.0 (0.469) &  & 92.9 (0.765) & 93.6 (0.731) \\
			&         & $\bm{k}_2^t\hat{\bm{\beta}}$                &  & 88.9 (1.211) & 93.7 (0.868) & 94.5 (0.577) & 95.1 (0.470) &  & 93.3 (0.752) & 93.3 (0.728) \\
			&         & $\bm{k}_3^t\hat{\bm{\beta}}$                &  & 92.7 (1.179) & 93.4 (0.835) & 93.9 (0.572) & 95.6 (0.459) &  & 93.4 (0.750) & 93.6 (0.737) \\
			&         & $\bar{\bm{k}}^t\hat{\bm{\beta}}$ &  & 92.0 (1.159) & 92.3 (0.844) & 93.4 (0.594) & 94.1 (0.487) &  & 93.5 (0.758) & 93.3 (0.731) \\[2mm]
			S2  & p.-cAIC & $\bm{k}_1^t\hat{\bm{\beta}}$                &  & 96.1 (1.071) & 97.5 (0.723) & 97.9 (0.584) & 96.9 (0.389) &  & 96.3 (0.721) & 94.4 (0.548) \\
			&         & $\bm{k}_2^t\hat{\bm{\beta}}$                &  & 96.1 (1.329) & 98.0 (0.904)   & 99.2 (0.568) & 96.9 (0.451) &  & 97.6 (0.728) & 93.7 (0.538) \\
			&         & $\bm{k}_3^t\hat{\bm{\beta}}$                &  & 98.3 (1.334) & 97.5 (0.785) & 95.2 (0.618) & 98.7 (0.495) &  & 95.1 (0.625) & 96.2 (0.585) \\
			&         & $\bar{\bm{k}}^t\hat{\bm{\beta}}$ &  & 95.2 (1.117) & 97.0 (0.844)   & 97.0 (0.597) & 97.6 (0.485) &  & 96.0 (0.661) & 94.5 (0.560) \\[2mm]
			& N       & $\bm{k}_1^t\hat{\bm{\beta}}$                &  & 93.6 (0.934) & 94.9 (0.623) & 92.4 (0.476) & 95.7 (0.366) &  & 91.8 (0.581) & 93.5 (0.529) \\
			&         & $\bm{k}_2^t\hat{\bm{\beta}}$                &  & 86.4 (0.946) & 89.5 (0.675) & 95.7 (0.453) & 94.0 (0.393) &  & 93.4 (0.572) & 93.5 (0.529) \\
			&         & $\bm{k}_3^t\hat{\bm{\beta}}$                &  & 92.4 (0.937) & 94.7 (0.670)  & 90.1 (0.493) & 96.4 (0.379) &  & 93.3 (0.569) & 94.1 (0.535) \\
			&         & $\bar{\bm{k}}^t\hat{\bm{\beta}}$ &  & 91.0 (0.912) & 92.3 (0.677) & 92.9 (0.477) & 93.9 (0.390) &  & 93.1 (0.571) & 93.5 (0.532)
		\end{tabular}
	}	
		\vspace{-0.5cm}
\end{table}

\begin{table}
	\caption{\label{tab:PoSI_betax_2} Coverage probabilities and average lengths (in parenthesis) of post-cAIC and naive confidence intervals for $\bm{k}_i^t\hat{\bm{\beta}}$ and $\bar{\bm{k}}^t\hat{\bm{\beta}}$. Nominal coverage probability: 95\%, selection matrix: $\bm{\upsilon}^3_{all}$.}
	\centering
	\setlength{\tabcolsep}{1.6pt}
	\fbox{%
		\begin{tabular}{ccccccccccc}
			&         &                  &  & $(15:5)$     & $(30:5)$     & $(60:5)$     & $(90:5)$     &  & $(30:10)$    & $(30:20)$    \\
			S & Meth.   & $\beta_j$        &  & CP (L)    & CP (L)    & CP (L)    & CP (L)    &  & CP (L)    & CP (L)    \\\hline
			S1  & p.-cAIC & $\bm{k}_1^t\hat{\bm{\beta}}$                &  & 95.4 (1.287) & 92.5 (1.030) & 92.1 (0.618) & 94.6 (0.496) &  & 95.9 (0.833) & 94.7 (0.772) \\
			&         & $\bm{k}_2^t\hat{\bm{\beta}}$                &  & 96.8 (1.552) & 94.7 (0.895) & 96.3 (0.714) & 94.2 (0.551) &  & 96.3 (0.865) & 93.8 (0.748) \\
			&         & $\bm{k}_3^t\hat{\bm{\beta}}$                &  & 93.1 (1.151) & 96.4 (0.989) & 95.3 (0.618) & 94.6 (0.505) &  & 94.7 (0.817) & 94.3 (0.760) \\
			&         & $\bm{k}_4^t\hat{\bm{\beta}}$                &  & 96.1 (1.330) & 95.2 (0.849) & 95.8 (0.623) & 94.4 (0.516) &  & 94.3 (0.760) & 95.4 (0.869) \\
			&         & $\bm{k}_5^t\hat{\bm{\beta}}$                &  & 93.9 (1.295) & 93.4 (0.948) & 95.2 (0.582) & 94.3 (0.567) &  & 95.4 (0.869) & 94.0 (0.740) \\
			&         & $\bar{\bm{k}}^t\hat{\bm{\beta}}$ &  & 93.7 (1.285) & 94.3 (0.898) & 94.7 (0.643) & 94.7 (0.519) &  & 95.1 (0.823) & 94.3 (0.756) \\[2mm]
			& N       & $\bm{k}_1^t\hat{\bm{\beta}}$                &  & 92.9 (1.109) & 86.1 (0.913) & 90.3 (0.575) & 94.0 (0.488) &  & 94.4 (0.757) & 93.4 (0.730) \\
			&         & $\bm{k}_2^t\hat{\bm{\beta}}$                &  & 92.9 (1.232) & 93.0 (0.819) & 90.9 (0.599) & 93.2 (0.516) &  & 93.0 (0.749) & 93.2 (0.731) \\
			&         & $\bm{k}_3^t\hat{\bm{\beta}}$                &  & 93.1 (1.133) & 91.7 (0.820) & 93.8 (0.559) & 93.3 (0.475) &  & 92.2 (0.754) & 94.0 (0.750) \\
			&         & $\bm{k}_4^t\hat{\bm{\beta}}$                &  & 93.5 (1.120)  & 93.1 (0.783) & 93.9 (0.559) & 92.7 (0.472) &  & 94.0 (0.750) & 92.3 (0.762) \\
			&         & $\bm{k}_5^t\hat{\bm{\beta}}$                &  & 91.0 (1.139) & 91.6 (0.870) & 94.5 (0.569) & 90.7 (0.503) &  & 92.3 (0.762) & 93.8 (0.730) \\
			&         & $\bar{\bm{k}}^t\hat{\bm{\beta}}$   &  & 91.7 (1.163) & 92.3 (0.828) & 92.5 (0.589) & 92.9 (0.484) &  & 93.2 (0.755) & 93.6 (0.731) \\[2mm]
			S2  & p.-cAIC & $\bm{k}_1^t\hat{\bm{\beta}}$                &  & 93.3 (0.998) & 95.4 (0.846) & 94.5 (0.496) & 95.9 (0.515) &  & 93.6 (0.588) & 94.6 (0.565) \\
			&         & $\bm{k}_2^t\hat{\bm{\beta}}$                &  & 92.1 (0.968) & 96.0 (0.772) & 95.7 (0.545) & 94.7 (0.528) &  & 95.6 (0.630) & 94.2 (0.535) \\
			&         & $\bm{k}_3^t\hat{\bm{\beta}}$                &  & 95.0 (0.971) & 96.0 (0.685) & 95.2 (0.527) & 94.9 (0.448) &  & 95.6 (0.679) & 94.1 (0.596) \\
			&         & $\bm{k}_4^t\hat{\bm{\beta}}$                &  & 93.9 (0.943) & 93.6 (0.634) & 95.6 (0.523) & 95.3 (0.625) &  & 94.1 (0.596) & 93.0 (0.583) \\
			&         & $\bm{k}_5^t\hat{\bm{\beta}}$                &  & 92.5 (0.890)  & 96.0 (0.673)   & 95.6 (0.451) & 94.4 (0.424) &  & 93.0 (0.583) & 95.3 (0.575) \\
			&         & $\bar{\bm{k}}^t\hat{\bm{\beta}}$ &  & 93.2 (1.013) & 94.6 (0.727) & 94.6 (0.518) & 94.7 (0.422) &  & 94.4 (0.607) & 94.4 (0.546) \\[2mm]
			& N       & $\bm{k}_1^t\hat{\bm{\beta}}$               &  & 90.6 (0.885) & 87.4 (0.681) & 91.4 (0.440) & 89.8 (0.404) &  & 92.6 (0.568) & 93.8 (0.537) \\
			&         & $\bm{k}_2^t\hat{\bm{\beta}}$                &  & 91.5 (0.951) & 90.6 (0.653) & 94.4 (0.506) & 86.3 (0.430) &  & 93.5 (0.576) & 93.7 (0.529) \\
			&         & $\bm{k}_3^t\hat{\bm{\beta}}$                &  & 91.4 (0.844) & 91.8 (0.610) & 91.8 (0.460) & 91.6 (0.401) &  & 92.4 (0.575) & 92.9 (0.559) \\
			&         & $\bm{k}_4^t\hat{\bm{\beta}}$                &  & 93.9 (0.936) & 93.6 (0.629) & 94.9 (0.477) & 86.0 (0.503) &  & 92.9 (0.559) & 92.2 (0.568) \\
			&         & $\bm{k}_5^t\hat{\bm{\beta}}$                &  & 91.7 (0.843) & 92.4 (0.607) & 94.8 (0.427) & 91.2 (0.379) &  & 92.2 (0.568) & 94.1 (0.540) \\
			&         & $\bar{\bm{k}}^t\hat{\bm{\beta}}$   &  & 90.5 (0.914) & 91.5 (0.658) & 91.8 (0.467) & 92.1 (0.385) &  & 92.7 (0.565) & 93.9 (0.532)
		\end{tabular}
	}	
\end{table}

\begin{table}
	\caption{\label{tab:PoSI_betax_3} Coverage probabilities and average lengths (in parenthesis) of post-cAIC and naive confidence intervals for $\bm{k}_i^t\hat{\bm{\beta}}$ and $\bar{\bm{k}}^t\hat{\bm{\beta}}$. Nominal coverage probability: 95\%, selection matrix: $\bm{\upsilon}^4_{all}$.}
	\centering
	\setlength{\tabcolsep}{1.6pt}
	\fbox{%
		\begin{tabular}{ccccccccccc}
			&         &                  &  & $(15:5)$      & $(30:5)$     & $(60:5)$     & $(90:5)$     &  & $(30:10)$    & $(30:20)$    \\
			S & Meth.   & $\beta_j$        &  & Cov. (L.)     & Cov. (L.)    & Cov. (L.)    & Cov. (L.)    &  & Cov. (L.)    & Cov. (L.)    \\
			\hline
			S1 & p.-cAIC & 	$\bm{k}_1^t\hat{\bm{\beta}}$                &  & 95.7 (1.463)  & 93.6 (0.843) & 96.7 (0.613) & 95.2 (0.479) &  & 94.7 (0.790) & 93.3 (0.742) \\
			&         & $\bm{k}_2^t\hat{\bm{\beta}}$                &  & 93.6 (1.239)  & 94.2 (0.904) & 96.3 (0.640) & 96.2 (0.490) &  & 94.2 (0.793) & 94.2 (0.762) \\
			&         & $\bm{k}_3^t\hat{\bm{\beta}}$                &  & 95.1 (1.284)  & 96.1 (0.921) & 95.8 (0.594) & 95.3 (0.483) &  & 96.2 (0.818) & 93.1 (0.743) \\
			&         & $\bm{k}_4^t\hat{\bm{\beta}}$                &  & 94.1 (1.193)  & 94.2 (0.871) & 93.1 (0.588) & 95.1 (0.469) &  & 95.7 (0.829) & 93.7 (0.746) \\
			&         & $\bm{k}_5^t\hat{\bm{\beta}}$                &  & 94.8 (1.210)  & 95.3 (0.914) & 96.2 (0.679) & 95.2 (0.476) &  & 96.4 (0.874) & 93.6 (0.742) \\
			&         & $\bar{\bm{k}}^t\hat{\bm{\beta}}$  &  & 93.4 (1.228)  & 94.7 (0.871) & 94.6 (0.622) & 95.5 (0.503) &  & 95.2 (0.815) & 93.6 (0.751) \\[2mm]
			& N       & $\bm{k}_1^t\hat{\bm{\beta}}$                 &  & 90.2 (1.193) & 93.8 (0.846) & 94.8 (0.577) & 95.3 (0.483) &  & 94.1 (0.758) & 93.1 (0.729) \\
			&         & $\bm{k}_2^t\hat{\bm{\beta}}$                 &  & 93.0 (1.196)  & 94.4 (0.917) & 95.1 (0.596) & 95.3 (0.466) &  & 93.4 (0.754) & 93.1 (0.728) \\
			&         & $\bm{k}_3^t\hat{\bm{\beta}}$                 &  & 93.2 (1.141) & 93.6 (0.824) & 94.2 (0.556) & 95.5 (0.487) &  & 94.3 (0.758) & 93.0 (0.727) \\
			&         & $\bm{k}_4^t\hat{\bm{\beta}}$                 &  & 92.4 (1.122) & 93.6 (0.847) & 93.2 (0.590) & 95.3 (0.471) &  & 94.3 (0.764) & 93.2 (0.726) \\
			&         & $\bm{k}_5^t\hat{\bm{\beta}}$                 &  & 92.9 (1.112) & 95.1 (0.893) & 93.4 (0.598) & 95.4 (0.479) &  & 93.3 (0.763) & 92.7 (0.728) \\
			&         & $\bar{\bm{k}}^t\hat{\bm{\beta}}$  &  & 92.3 (1.166)  & 93.8 (0.831) & 93.5 (0.589) & 94.6 (0.483) &  & 93.5 (0.755) & 92.9 (0.730) \\[2mm]
			S2  & p.-cAIC & $\bm{k}_1^t\hat{\bm{\beta}}$                 &  & 93.6 (0.878)  & 95.6 (0.638) & 94.7 (0.450) & 95.6 (0.356) &  & 95.9 (0.597) & 95.9 (0.597) \\
			&         & $\bm{k}_2^t\hat{\bm{\beta}}$                 &  & 93.2 (0.866)  & 96.6 (0.761) & 94.0 (0.449) & 95.9 (0.374) &  & 96.3 (0.657) & 96.3 (0.657) \\
			&         & $\bm{k}_3^t\hat{\bm{\beta}}$                 &  & 93.5 (0.857)  & 93.8 (0.654) & 93.0 (0.440) & 96.7 (0.448) &  & 95.1 (0.615) & 95.1 (0.615) \\
			&         & $\bm{k}_4^t\hat{\bm{\beta}}$                 &  & 93.0 (0.948)  & 95.9 (0.653) & 93.6 (0.478) & 95.0 (0.364) &  & 94.8 (0.588) & 94.8 (0.588) \\
			&         & $\bm{k}_5^t\hat{\bm{\beta}}$                 &  & 94.1 (1.020)  & 94.5 (0.603) & 95.2 (0.463) & 96.5 (0.434) &  & 95.9 (0.589) & 95.9 (0.589) \\
			&         & $\bar{\bm{k}}^t\hat{\bm{\beta}}$  &  & 93.2 (0.916)  & 95.0 (0.693) & 94.2 (0.497) & 95.2 (0.405) &  & 95.1 (0.598) & 95.1 (0.598) \\[2mm]
			& N       & $\bm{k}_1^t\hat{\bm{\beta}}$                &  & 93.3 (0.872)  & 94.5 (0.604) & 93.5 (0.427) & 95.7 (0.356) &  & 94.4 (0.560) & 94.4 (0.560) \\
			&         & $\bm{k}_2^t\hat{\bm{\beta}}$                 &  & 93.1 (0.856)  & 92.5 (0.648) & 93.5 (0.444) & 95.7 (0.361) &  & 92.5 (0.566) & 92.5 (0.566) \\
			&         & $\bm{k}_3^t\hat{\bm{\beta}}$                 &  & 93.2 (0.849)  & 93.9 (0.656) & 93.3 (0.443) & 91.3 (0.382) &  & 93.5 (0.577) & 93.5 (0.576) \\
			&         & $\bm{k}_4^t\hat{\bm{\beta}}$                &  & 92.7 (0.950)  & 94.1 (0.607) & 93.9 (0.482) & 95.0 (0.364) &  & 93.4 (0.554) & 93.4 (0.554) \\
			&         & $\bm{k}_5^t\hat{\bm{\beta}}$                 &  & 93.7 (0.997)  & 94.6 (0.606) & 93.9 (0.446) & 93.1 (0.382) &  & 94.7 (0.556) & 94.7 (0.556) \\
			&         & $\bar{\bm{k}}^t\hat{\bm{\beta}}$  &  & 93.0 (0.909)  & 93.7 (0.658) & 93.0 (0.469) & 94.0 (0.384) &  & 93.9 (0.565) & 93.9 (0.565)
		\end{tabular}
	}	
\end{table}


\begin{table}
	\caption{\label{tab:PoSI_2} Coverage probabilities and average lengths (in parenthesis) of post-cAIC, post-OBSP and naive confidence intervals for $\mu_{i}$. Nominal coverage probability: 95\%, selection matrix: $\bm{\upsilon}^3_{all}$.}
	\setlength{\tabcolsep}{2pt}
	\fbox{%
		\begin{tabular}{cccccccccc}
			&                    &            & $(15:5)$     & $(30:5)$     & $(60:5)$     & $(90:5)$     &  & $(30:10)$    & $(30:20)$    \\
			S & Meth.              &            & CP (L)    & CP (L)    & CP (L)    & CP (L)    &  & CP (L)    & CP (L)    \\ \hline
			S1 & p.-cAIC               &            & 94.3 (1.609) & 95.0 (1.621)   & 95.2 (1.626) & 95.3 (1.624) &  & 95.0 (1.202) & 95.0 (0.868) \\
			& \multicolumn{2}{c}{N1} & 93.7 (1.567) & 94.6 (1.597) & 95.0 (1.613) & 95.2 (1.616) &  & 94.8 (1.191) & 94.9 (0.864) \\
			& \multicolumn{2}{c}{N2} & 95.0 (1.650) & 95.1 (1.629) & 95.2 (1.628) & 95.3 (1.626) &  & 95.1 (1.202) & 95.0 (0.868) \\
			& p.-OBSP           &            & 94.7 (1.631) & 95.0 (1.623) & 95.2 (1.625) & 95.3 (1.624) &  & 95.1 (1.200) & 94.9 (0.868) \\[2mm]
			S2 &  p.-cAIC                &            & 92.5 (1.465) & 94.6 (1.499) & 94.9 (1.500) & 95.0 (1.501) &  & 94.8 (1.146) & 95.0 (0.847) \\
			& \multicolumn{2}{c}{N1} & 91.6 (1.422) & 94.3 (1.473) & 94.7 (1.486) & 94.9 (1.492) &  & 94.6 (1.136) & 95.0 (0.843) \\
			& \multicolumn{2}{c}{N2} & 96.3 (1.667) & 95.7 (1.556) & 95.3 (1.524) & 95.3 (1.516) &  & 95.1 (1.163) & 95.1 (0.851) \\
			&  p.-OBSP            &            & 93.5 (1.517) & 95.1 (1.520) & 95.0 (1.508) & 95.1 (1.506) &  & 94.8 (1.156) & 95.4 (0.851)
	\end{tabular}}
\end{table}

\begin{table}
	\caption{\label{tab:PoSI_3} Coverage probabilities and average lengths (in parenthesis) of post-cAIC, post-OBSP and naive confidence intervals for $\mu_{i}$. Nominal coverage probability: 95\%, selection matrix: $\bm{\upsilon}^4_{all}$.}
	\setlength{\tabcolsep}{2pt}
	\fbox{
		\begin{tabular}{cccccccccc}
			&                    &            & $(15:5)$     & $(30:5)$     & $(60:5)$     & $(90:5)$     &  & $(30:10)$    & $(30:20)$    \\
			S & Meth.             &            & CP (L)    & CP (L)    & CP (L)    & CP (L)    &  & CP (L)    & CP (L)    \\ \hline
			S1                   & p.-cAIC                &            & 94.0 (1.599) & 94.8 (1.622) & 95.1 (1.626) & 95.2 (1.622) &  & 95.2 (1.203) & 95.1 (0.868) \\
			& \multicolumn{2}{c}{N1} & 93.7 (1.578) & 94.6 (1.611) & 95.0 (1.620) & 95.1 (1.617) &  & 95.0 (1.195) & 95.0 (0.866) \\
			& \multicolumn{2}{c}{N2} & 95.1 (1.656) & 95.1 (1.644) & 95.2 (1.635) & 95.3 (1.627) &  & 95.2 (1.206) & 95.2 (0.869) \\
			&  p.-OBSP          &            & 94.8 (1.642) & 95.0 (1.638) & 95.2 (1.632) & 95.2 (1.625) &  & 95.2 (1.204) & 95.1 (0.868) \\[2mm]
			S2                 &  p.-cAIC                &            & 92.5 (1.451) & 94.6 (1.489) & 94.8 (1.498) & 95.1 (1.496) &  & 94.9 (1.146) & 94.8 (0.845) \\
			& \multicolumn{2}{c}{N1} & 92.2 (1.438) & 94.4 (1.476) & 94.7 (1.491) & 95.0 (1.492) &  & 94.8 (1.140) & 94.8 (0.843) \\
			& \multicolumn{2}{c}{N2} & 96.2 (1.654) & 95.8 (1.563) & 95.3 (1.530) & 95.4 (1.516) &  & 95.4 (1.166) & 95.0 (0.852) \\
			& p.-OBSP            &            & 94.4 (1.545) & 95.2 (1.525) & 95.0 (1.513) & 95.2 (1.506) &  & 95.1 (1.157) & 94.9 (0.849)
		\end{tabular}
	}
\end{table}
\begin{table}
	\caption{	\label{tab:ch4_models_data_example2} Inclusion of covariates into different models.}
	\setlength{\tabcolsep}{2.5pt}
	\raggedleft
	\fbox{%
		\begin{tabular}{lcccccccccccccccc}
			Covariate & M1 & M2 & M3 & M4 & M5 & M6 & M7 & M8 & M9 & M10 & M11 & M12 & M13 & M14 & M15 & M16 \\
			\hline
			Intercept     & \checkmark   & \checkmark  & \checkmark    & \checkmark   & \checkmark  &   \checkmark    & \checkmark   & \checkmark  & \checkmark    & \checkmark   & \checkmark  &  \checkmark    & \checkmark   & \checkmark  & \checkmark    & \checkmark    \\
			Typ1        & \checkmark  & \checkmark  & \checkmark   & \checkmark  & \checkmark  &  \checkmark   & \checkmark  & \checkmark  & \checkmark   & \checkmark  & \checkmark  & \checkmark   & \checkmark  & \checkmark  & \checkmark   & \checkmark    \\
			Edu1         & \checkmark   & \checkmark  & \checkmark   & \checkmark   & \checkmark  &  \checkmark   & \checkmark   & \checkmark  &  \checkmark   & \checkmark   & \checkmark  &  \checkmark   & \checkmark   & \checkmark  &  \checkmark   & \checkmark   \\
			Dif3         & \checkmark   & \checkmark  & \checkmark   & \checkmark   & \checkmark  &  \checkmark   & \checkmark   & \checkmark  &   \checkmark   & \checkmark   & \checkmark  & \checkmark   & \checkmark   & \checkmark  & \checkmark   & \checkmark   \\
			Typ3        & \checkmark  & \checkmark  & \checkmark   & \checkmark  & \checkmark  &   \checkmark   & \checkmark  & \checkmark  & \checkmark   & \checkmark  & \checkmark  & \checkmark   & \checkmark  & \checkmark  & \checkmark   & \checkmark    \\
			Ten2         &             & \checkmark   &
			& \checkmark  &    &
			\checkmark   &             & \checkmark   &
			&    \checkmark         &             &   \checkmark           &
			&   \checkmark           &
			&  \checkmark   \\
			Age2         &             &              &  \checkmark   & \checkmark  &              &
			& \checkmark  & \checkmark   &     &             &   \checkmark            &     \checkmark &             &               &  \checkmark   &   \checkmark  \\
			Sex       &             &              &
			&  &   \checkmark &
			\checkmark   & \checkmark  &   \checkmark                                                &            &             &            &            &     \checkmark         &    \checkmark           &   \checkmark   &   \checkmark   \\
			Size        &             &              &
			&             &              &    &             &              &   \checkmark &       \checkmark      &  \checkmark           &    \checkmark  &     \checkmark         &    \checkmark           &      \checkmark&   \checkmark
	\end{tabular}}
\end{table}

\begin{table}
	\caption{\label{tab:cAIC_data} Selection criteria.}
	\centering
	\setlength{\tabcolsep}{3pt}
	\fbox{%
		\begin{tabular}{cc cc cc ccc}
			Model & M1       & M2      & M3       & M4       & M5       & M6       & M7       & M8       \\
			cAIC  & \textbf{26097.45} & 26098.00   & 26098.40 & 26098.92 & 26104.15 & 26102.78 & 26103.18 & 26101.71 \\
			OBSP  & 995.658  & 983.52  & 990.50    & 977.97   & 995.76   & 983.58   & 990.61   & 978.05   \\ \hline
			Model & M9       & M10     & M11      & M12      & M13      & M14      & M5       & M16      \\
			cAIC  & 26103.39 & 26102.20 & 26102.24 & 26101.04 & 26102.42 & 26101.21 & 26101.26 & 26100.06 \\
			OBSP  & 993.55   & 982.06  & 988.42   & \textbf{976.55}   & 993.61   & 982.09   & 988.48   & 976.58
	\end{tabular}}	
\end{table}

\begin{table}
	\caption{\label{tab:int_descript} Descriptive statistics of lengths of naive and post-selection CI for mixed and regression-synthetic estimates of the log of the county-level averages of household income.}
	\centering
	\setlength{\tabcolsep}{4pt}
	\fbox{%
		\begin{tabular}{|c c ccccc c ccccc|}
			& &	\multicolumn{5}{c}{Mixed parameter} & &  \multicolumn{5}{c}{Linear combination} \\
			Meth.& & Min   & Max   & Median & Mean  & SD& &   Min   & Max   & Median & Mean  & SD\\
			\hline
			p.-cAIC     && 0.042 & 0.169 & 0.110  & 0.111 & 0.035 && 0.040 & 0.043 & 0.041 & 0.041& 0.001\\
			N1   && 0.040 & 0.176 & 0.108  & 0.112 & 0.038 && 0.034 & 0.041 & 0.038 & 0.037& 0.002\\
			N2   && 0.040 & 0.178 & 0.109  & 0.113 & 0.039 && - & - & -  & - & -\\
			p.-OBSP && 0.028 & 0.215 & 0.113 & 0.119& 0.053&& - & - & - & -& - \\
	\end{tabular}}
\end{table}

\end{document}